\def\ref{\par\noindent\hang}

\def\spose#1{\hbox to 0pt{#1\hss}}
\def\approxlt{\mathrel{\spose{\lower 3pt\hbox{$\sim$}}
	\raise 2.0pt\hbox{$<$}}}
\def\approxgt{\mathrel{\spose{\lower 3pt\hbox{$\sim$}}
	\raise 2.0pt\hbox{$>$}}}

\def\multleft#1{\hbox to size{\vbox {\halign {\lft{##}\cr #1}}\hfill}\par}
\def\multright#1{\hbox to size{\vbox {\halign {\rt{##}\cr #1}}\hfill}\par}
\def\Mdot{\hbox{$\dot M$}}
\def\degmark{^\circ}
\def\today{\ifcase\month\or January\or February\or March\or April\or May\or
      June\or July\or August\or September\or October\or November\or December\fi
      \space\number\day, \number\year}
\def\$<${\thinspace}
\def\s{\hbox{\phantom{5}}}	

\def\boxit#1{\vbox{\hrule\hbox{\vrule\kern3pt\vbox{\kern3pt
          #1 \kern3pt}\kern3pt\vrule}\hrule}}

\def\cm{{\rm\thinspace cm}}

\def\erg{{\rm\thinspace erg}}
\def\eV{{\rm\thinspace eV}}

\def\K{{\rm\thinspace K}}
\def\keV{{\rm\thinspace keV}}
\def\km{{\rm\thinspace km}}

\def\Msun{\hbox{$\rm\thinspace M_{\odot}$}}
\def\pc{{\rm\thinspace pc}}

\def\s{{\rm\thinspace s}}
\def\yr{{\rm\thinspace yr}}

\def\cmcu{\hbox{$\cm^3\,$}}
\def\pcmcu{\hbox{$\cm^{-3}\,$}}

\def\ergpcmsqps{\hbox{$\erg\cm^{-2}\s^{-1}\,$}}

\def\ergps{\hbox{$\erg\s^{-1}\,$}}

\def\kmps{\hbox{$\km\s^{-1}\,$}}

\def\Msunpyr{\hbox{$\Msun\yr^{-1}\,$}}

\def\pcmsq{\hbox{$\cm^{-2}\,$}}

\def\ps{\hbox{$\s^{-1}\,$}}

\documentstyle[psfig]{mn}

\title[Spectral study of type-1 AGN]
{An X-ray spectral study of 24 type-1 active galactic nuclei}

\author[C.~S.~Reynolds]
{C.~S.~Reynolds\thanks{Present address: JILA, University of Colorado, Boulder, CO 80309-0440, USA}\\
Institute of Astronomy, Madingley Road, Cambridge CB3 0HA}

\date{}

\begin{document}

\maketitle

\begin{abstract}
I present a study of the X-ray spectral properties of a sample
containing 24 type-1 active galactic nuclei using the medium spectral
resolution of {\it ASCA}.  The sample consists of 20 radio-quiet
objects (18 Seyfert 1 galaxies and 2 radio-quiet quasars) and 4
radio-loud objects (3 broad-line radio galaxies and 1 radio-loud
quasar).  A simple power-law continuum absorbed by Galactic material
provides a very poor description of the spectra of most objects.
Deviations from the power-law form are interpreted in terms of X-ray
reprocessing/absorption processes.  In particular, at least half of
the objects show K-shell absorption edges of warm oxygen ({\sc O\,vii}
and {\sc O\,viii}) characteristic of optically-thin, photoionized
material along the line-of-sight to the central engine, the so-called
warm absorber.  The amount and presence of this absorption is found to
depend on either the luminosity or radio-properties of the objects:
luminous and/or radio-loud objects are found to possess less ionized
absorption.  This ambiguity exists because the radio-loud objects are
also amongst the most luminous of the sample.  It is also found that
objects with significant optical reddening display deep {\sc O\,vii}
edges.  The converse is true with two possible exceptions (NGC~3783
and NGC~3516).  Coupled with other evidence resulting from detailed
study of particular objects, this suggests the existence of dusty warm
plasma.  A radiatively driven outflow originating from the molecular
torus is probably the source of this plasma.  Rapid variability of the
warm absorber also points to there being another component closer to
the central source and probably situated within the broad line region
(BLR).  Independent evidence for such an optically-thin, highly ionized
BLR component comes from detailed optical/UV studies.

Spectral features at energies characteristic of cold iron K$\alpha$
emission are common.  Such emission is expected to arise from the
fluorescence of cold iron in optically-thick material when illuminated
by the primary X-ray continuum.  Radio-quiet objects have iron
emission well described as originating from either the inner regions of
an accretion disk or, in a small number of cases, from the molecular
torus.  Two of the radio-loud objects (3C~120 and 3C~382) have a much
broader feature which presents problems for the relativistic disk
model.  The presence of radio-jets may be important in forming this
spectral feature.
\end{abstract}


\section{Introduction}

The high-energy emission of active galactic nuclei (AGN) originate
from the innermost regions of an accretion flow onto a putative
supermassive black hole.  It is widely believed that much of the
accretion energy is radiated as X-rays and $\gamma$-rays and that
reprocessing of this energy by circumnuclear material leads to much of
the observed UV, optical and IR luminosity.  The X-ray spectrum ($\sim
0.1$--100\,keV) of type-1 AGN (i.e. those displaying broad optical/UV
emission lines) can be well approximated by a power-law with photon
index $\Gamma\sim 2$.  Deviations from a power-law form can usually be
interpreted as the effects of X-ray reprocessing.  A detailed study of
this reprocessing yields direct information on the geometry and state
of matter in the central engines of these objects.

The first clear sign of atomic features in the X-ray spectra of AGN
came with {\it EXOSAT} and {\it Ginga} observations of nearby Seyfert
1 galaxies which found evidence for the fluorescent K$\alpha$ emission
line (at $6.4\keV$) of cold iron (Nandra et al. 1989; Matsuoka et
al. 1990).  {\it Ginga} also discovered a spectral flattening above
$\sim 10\keV$ (Nandra, Pounds \& Stewart 1990; Nandra \& Pounds 1994).
These features were explained as due to `reflection' of the primary
(power-law) X-ray continuum by cold optically-thick material out of
the line-of-sight to the observer (Guilbert \& Rees 1988; Lightman \&
White 1988; George \& Fabian 1991; Matt, Perola \& Piro 1991).  The
superior spectral resolution of the solid-state imaging spectrometers
(SIS) on board {\it ASCA} (Tanaka, Inoue \& Holt 1994) allowed
dramatic advances to be made.  Many Seyfert 1 galaxies were found to
have broad iron lines (with FWHM$\sim 0.1$--0.2$c$; Mushotzky et
al. 1995).  In particular, a long (4 day) {\it ASCA} observation of
MCG$-6-30-15$ found the line to be broad and skewed in exactly the
sense expected if the reprocessing material were the innermost parts
of an accretion disk around a black hole (Tanaka et al. 1995; Fabian
et al. 1995; Iwasawa et al. 1996).  The characteristic line profile
results from the combined effects of Doppler shifts, relativistic
aberration and gravitational redshifts.  Alternative interpretations
for the observed line profile do not seem viable (Fabian et al. 1995).
This was the first direct evidence for strong gravitational effects in
the vicinity of a putative black hole.

X-ray absorption by partially-ionized, optically-thin material, along
the line-of-sight to the central engine, the so-called warm absorber,
is another prominent feature in the X-ray spectrum of many AGN.  The
presence of such gas was first postulated in order to explain the
unusual form of the X-ray spectrum of the QSO MR~$2251-178$ (Halpern
1984; Pan, Stewart \& Pounds 1990).  Further evidence for such
material came from {\it Ginga} observations of nearby Seyfert 1
galaxies which suggested the presence of K-shell absorption edges of
highly ionized iron (Nandra, Pounds \& Stewart 1990; Nandra et
al. 1991; Nandra \& Pounds 1994).  However, it is unclear how these
conclusions were affected by an over-simplified treatment of the
fluorescent iron emission line discussed above.  The {\it ROSAT}
position sensitive proportional counter (PSPC) also discovered
probable K-shell absorption edges due to {\sc O\,vii} and {\sc
O\,viii} (Nandra \& Pounds 1992; Nandra et al. 1993; Fiore et
al. 1993; Turner et al. 1993a).  Unfortunately, due to the limited
bandpass and spectral resolution of {\it ROSAT} other explanations for
the observed features, such as a complex soft excess or partial
covering by a cold absorber, could not be firmly ruled out.

{\it ASCA} confirmed the presence of prominent {\sc O\,vii} and {\sc
O\,viii} edges (at rest energies of 0.74\,keV and 0.87\,keV
respectively) in the X-ray spectra of several bright Seyfert 1
galaxies.  The high ionization state strongly suggests it to be
photoionized material situated within $\sim 10\pc$ of the primary
continuum source.  Furthermore, {\it ASCA} performance verification
(PV) observations of MCG$-6-30-15$ found the warm absorber to be
variable on timescales down to several hours (Fabian et al. 1994a;
Reynolds et al. 1995).  The long {\it ASCA} observation of this object
clarified the nature of the variations: the optical depth of the {\sc
O\,vii} edge was found to remain constant on timescales of months to 
years whereas the depth of the {\sc O\,viii} edge was found to
anti-correlate with the primary flux on timescales less than $\sim
10^4$\,s (Otani et al. 1996).  These observations lead to a two-zone
warm absorber model for MCG$-6-30-15$: an inner absorber is
responsible for much of the {\sc O\,viii} edge and responds to changes
in the ionizing continuum on timescales $\approxlt 10^4\s$ whereas the
{\sc O\,vii} edge largely results from a tenuous outer absorber.  The
long recombination timescale in the tenuous outer absorber produces
the observed constancy of the absorption edge depth.

Despite these observational advances, the origin of the warm absorbing
material and its relation to other known structures within the central
engine remains unknown.  The variability results for MCG$-6-30-15$
suggest the inner absorber to be at radii characteristic of the broad
line region (BLR; $r\approxlt 10^{17}\cm$) and the outer absorber to
be at scales associated with the obscuring torus and Seyfert 2
scattering medium ($r\approxgt 1\pc$).  Both of these structures had
been previously considered as possible candidates for producing the
warm material (Krolik \& Kallman 1987; Netzer 1993; Reynolds \& Fabian
1995).  However, the general applicability of these results to other
AGN is uncertain.  An important step would be the analysis of an
unbiased sample of AGN with the medium spectral resolution offered by
{\it ASCA} in order to assess the frequency of occurence of warm
absorbers and their relation to other observable properties.

In this paper I present an analysis of the {\it ASCA} data for a
sample of 24 bright type-1 AGN including
18 Seyfert 1 galaxies, 3 broad-line radio galaxies (BLRG), 2
radio-quiet quasars (RQQ) and 1 radio-loud quasar (RLQ).  In
particular I focus on the nature of the warm absorber within this
sample.  The primary aim of this work is to present a uniform analysis
of the {\it ASCA} data for a moderately large sample of type-1 AGN.
Section 2 defines the sample and briefly describes the initial data
reduction.  The spectral fitting of the time-averaged {\it ASCA}
spectra is detailed in Section 3 and the results are discussed in
Section 4.  It is found that all objects display some sign of spectral
complexity indicating X-ray reprocessing phenomena are common.  In
particular, ionized absorption is detected with a high level of
significance in 12 of the 24 objects.  Since the dominant signatures
of this absorption across the {\it ASCA} band are K-shell absorption
edges due to {\sc O\,vii} and {\sc O\,viii}, the absorption can be
parameterised by the maximum optical depths of these edges.  This
provides the simpliest parametrization of what is clearly a complex
physical phenomenon.  Relationships between these edge depths and
other AGN properties (i.e. luminosity and optical reddening) are
found.  Section 5 presents the results of fitting the data with
one-zone photoionization models as well as a comparison of these
models with the simple two-edge fits.  Spectral variability is briefly
addressed in Section 6.  After some notes on individual objects
(Section 7), the general implications of these results on our
understanding of the central engines of AGN are discussed (Section 8).
My conclusions are presented in Section 9.

Throughout this paper it is assumed that $H_{\rm 0}=50\,{\rm km}\,{\rm
s}^{-1}\,{\rm Mpc}^{-1}$ and $q_{\rm 0}=0$.  Unless otherwise stated,
errors on physical quantities are quoted at the 90 per cent confidence
level for one interesting parameter ($\Delta\chi^2=2.7$).  Error bars
on plots are shown at the 1-$\sigma$ level for one interesting
parameter.

\section{The sample and basic data reduction}

\begin{table*}
\caption{The {\it ASCA} AGN sample.  
Column 2 indicates the type of nuclear activity (Sy1=Seyfert 1 galaxy;
BLRG=broad-line radio galaxy; RLQ=radio-loud quasar; RQQ=radio-quiet
quasar).  Columns 3, 4 and 5 give the co-ordinates (J2000.0) and
redshift of the source (from Veron-Cetty \& Veron 1993).  Column 6 gives
the Galactic {\sc H\,i} column density towards the source as
determined by 21-cm measurements [`a' indicates values obtained from
Elvis, Lockman \& Wilkes (1989) which are accurate to within 5 per
cent; otherwise quoted values are interpolations from the measurements
of Stark et al. (1992) and have uncertainties of $\sim 1\times
10^{20}\pcmsq$].  Column 7 gives the start date of the {\it ASCA}
observation used and column 8 gives the good exposure time for SIS0.}
\begin{center}
\begin{tabular}{lllccccc}\hline
common source & classification&  RA & Decl & redshift & Galactic $N_{\rm H}$ & start & SIS0 Exp. \\
name & of activity & (J2000.0) & (J2000.0) & ($z$) & ($10^{20}\pcmsq$) & date & (ks) \\\hline
Mrk~335 & Sy1 & 00 06 19.4 & 20 12 11 & 0.025 & 4.0 & 1993-Dec-9 & 19 \\
Fairall~9 & Sy1 & 01 23 45.7 & $-$58 48 21 & 0.046 & 3.0 & 1993-Nov-21 & 22 \\
Mrk~1040 & Sy1 & 02 28 14.4 & 31 18 41 & 0.016 & 7.07$^{\rm a}$ & 1994-Aug-19 &19 \\
3C~120 & BLRG & 04 33 11.0 & 05 21 15 & 0.033 & 12.32$^{\rm a}$ & 1994-Feb-17 & 45 \\
NGC~2992 & Sy1 & 09 45 41.9 & $-$14 19 35 & 0.008 & 5.56$^{\rm a}$ & 1994-May-6 & 28 \\
NGC~3227 & Sy1 & 10 23 30.5 & 19 51 55 & 0.003 & 2.2 & 1993-May-8 & 47 \\
NGC~3516 & Sy1 & 11 06 47.4 & 72 34 06 & 0.009 & 3.4 & 1994-Apr-2 & 28 \\
NGC~3783 & Sy1 & 11 39 01.7 & $-$37 44 18 & 0.009 & 3.7 & 1993-Dec-19 & 16 \\
NGC~4051 & Sy1 & 12 03 09.5 & 44 31 52 & 0.002 & 1.31$^{\rm a}$ & 1993-Apr-25 & 27 \\
3C~273 & RLQ & 12 29 06.6 & 02 03 08 & 0.158 & 3.0 & 1993-Dec-20 & 11 \\
NGC~4593 & Sy1 & 12 39 39.3 & $-$05 20 39 & 0.009 & 1.97$^{\rm a}$ & 1994-Jan-9 & 29 \\
MCG$-6-30-15$ & Sy1 & 13 35 53.3 & $-$34 17 48 & 0.008 & 4.06$^{\rm a}$ & 1994-Jul-23 & 147 \\
IC~4329a & Sy1 & 13 49 19.2 & $-$30 18 34 & 0.016 & 4.55$^{\rm a}$ & 1993-Aug-15 & 33  \\
NGC~5548 & Sy1 & 14 17 59.5 & 25 08 12 & 0.017 & 1.7 & 1993-Jul-27 & 27  \\
Mrk~841 & RQQ & 15 04 01.1 & 10 26 16 & 0.036 & 2.23$^{\rm a}$ & 1994-Feb-21 & 21 \\
Mrk~290 & Sy1 & 15 35 52.2 & 57 54 08 & 0.030 &  2.32$^{\rm a}$ & 1994-Jun-16 & 40 \\
3C~382 & BLRG & 18 35 03.3 & 32 41 46 & 0.059 & 7.4 & 1994-Apr-18 & 36 \\
3C~390.3 & BLRG & 18 42 08.7 & 79 46 16 & 0.057 & 4.1 & 1993-Nov-16 & 39 \\
ESO~141-G55 & Sy1 & 19 21 14.2 & $-$58 40 12 & 0.037 & 5.5 & 1994-Sep-21 & 10 \\
NGC~6814 & Sy1 & 19 42 40.5 & $-$10 19 24 & 0.006 & 9.80$^{\rm a}$ & 1993-May-4 & 45 \\
Mrk~509 & Sy1 & 20 44 09.6 & $-$10 43 23 & 0.035 & 4.2 & 1994-Apr-29 & 39 \\
MR~2251-178 & RQQ & 22 54 05.7 & $-$17 34 54 & 0.068 & 2.8 & 1993-Nov-6 & 7 \\
NGC~7469 & Sy1 & 23 03 15.5 & 08 52 26 & 0.017 & 4.82$^{\rm a}$ & 1993-Nov-24 & 12 \\
Mrk~926 & Sy1 & 23 04 43.4 &  $-$08 41 08 & 0.047 & 4.6 & 1993-May-25 & 30 \\\hline
\end{tabular}
\end{center}
\end{table*}

\noindent The sample consists of the 24 objects studied in the 
{\it EXOSAT} survey of Turner \& Pounds (1989) and the {\it Ginga}
survey of Nandra \& Pounds (1994, hereafter NP94) for which {\it ASCA}
data was publicly available at the time of the analysis (late 1995).
Table~1 defines the sample.  It contains 20 radio-quiet AGN (18
Seyfert 1 galaxies and 2 RQQ) and 4 radio-loud objects (3 BLRG and 1
RLQ).  This sample is not homogeneous or complete in any well defined
sense.  However, it should be unbiased with respect to any of the
spectral properties discussed in this paper.  Indeed, the two lists
from which the sample objects are drawn were compiled prior to any
detailed knowledge of AGN spectral complexity.

Data from both the SIS and gas imaging spectrometers (GIS) were used.
SIS data from both {\sc bright} and {\sc faint} mode were combined in
order to maximize the total signal and a standard {\sc grade}
selection was performed in order to reduce the effects of particle and
instrumental background.  Data from the SIS were further cleaned in
order to remove the effects of hot and flickering pixels and subjected
to the following data-selection criteria: 

i) the satellite should not be in the South Atlantic Anomaly (SAA),

ii) the object should be at least $5\degmark$ above the Earth's limb,

iii) the object should be at least $25\degmark$ above the day-time
Earth limb,

iv) the local geomagnetic cut-off rigidity (COR) should be greater
than 6\,GeV/$c$.

\noindent Data from the GIS were cleaned to remove the particle background and
subjected to the following data-selection criteria:

i) the satellite should not be in the SAA,

ii) the object should be at least $7\degmark$ above the Earth's limb
and

iii) the COR should be greater than 7\,GeV/$c$.

\noindent SIS and GIS data that satisfy these criteria shall
be referred to as `good' data.

Images, lightcurves and spectra were extracted from the good data
using a circular region centred on the source.  For the SIS, an
extraction radius of 3\,arcmins is used whereas a radius of 4\,arcmins
is used for the GIS.  These regions are sufficiently large to contain
all but a negligible portion of the source counts\footnote{See point
spread function in the `{\it ASCA} Technical Description\,',
Appendix~E of the {\it ASCA} Research Announcement}.  Background
spectra were extracted from source free regions of the same field of
view within each of the four {\it ASCA} instruments.  Background
regions for the SIS were taken to be rectangular regions along the
edges of the source chip whereas annular regions were used to extract
GIS background.

The data reduction described in the previous paragraphs was performed
using version 1.3 of the {\sc xselect} program contained within the
{\sc ftools} software package.

\section{Spectral analysis}

\subsection{Initial Spectral fitting}

\begin{figure*}
\hbox{
\psfig{figure=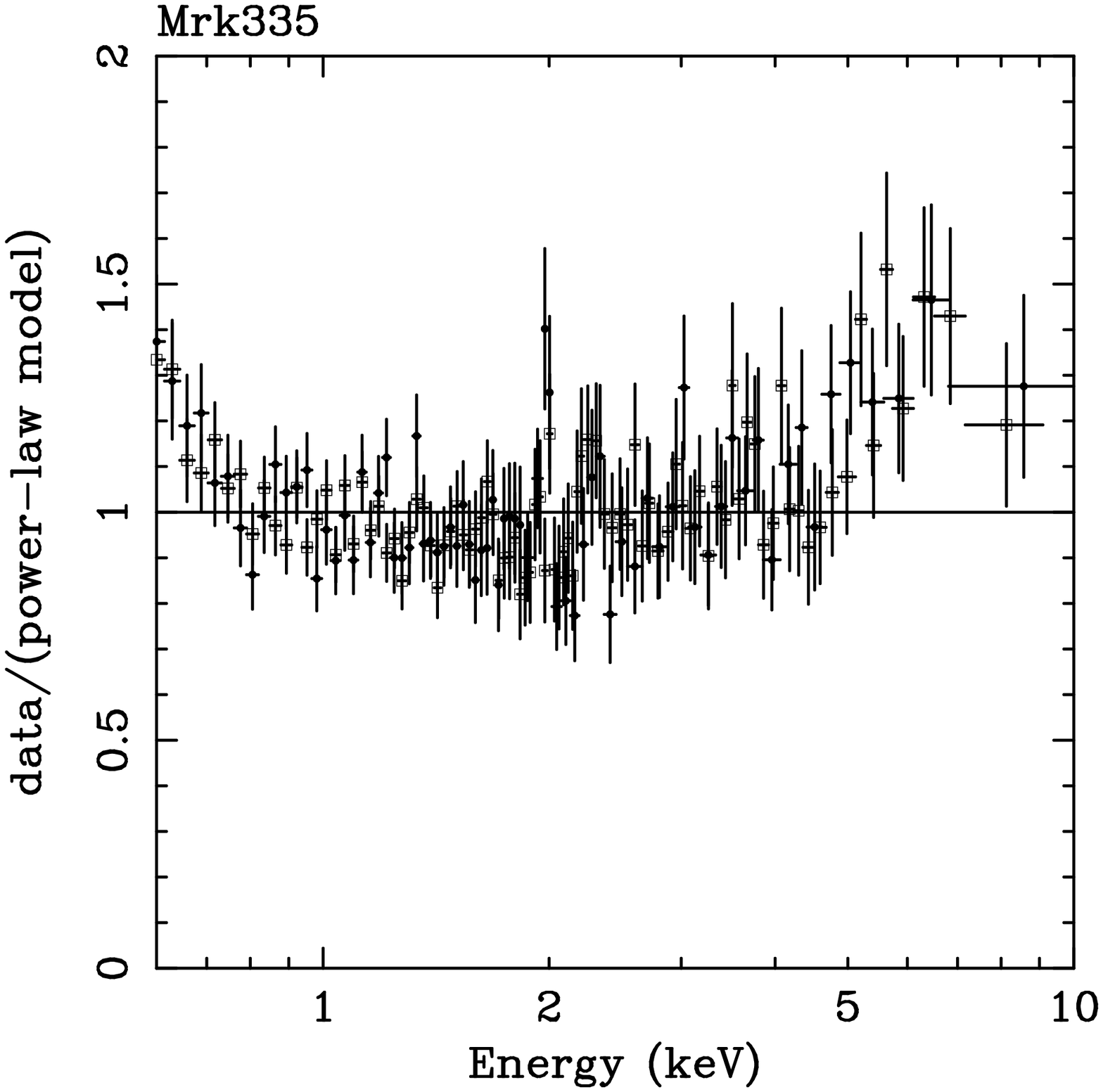,width=0.49\textwidth,height=0.23\textheight,angle=270}
\psfig{figure=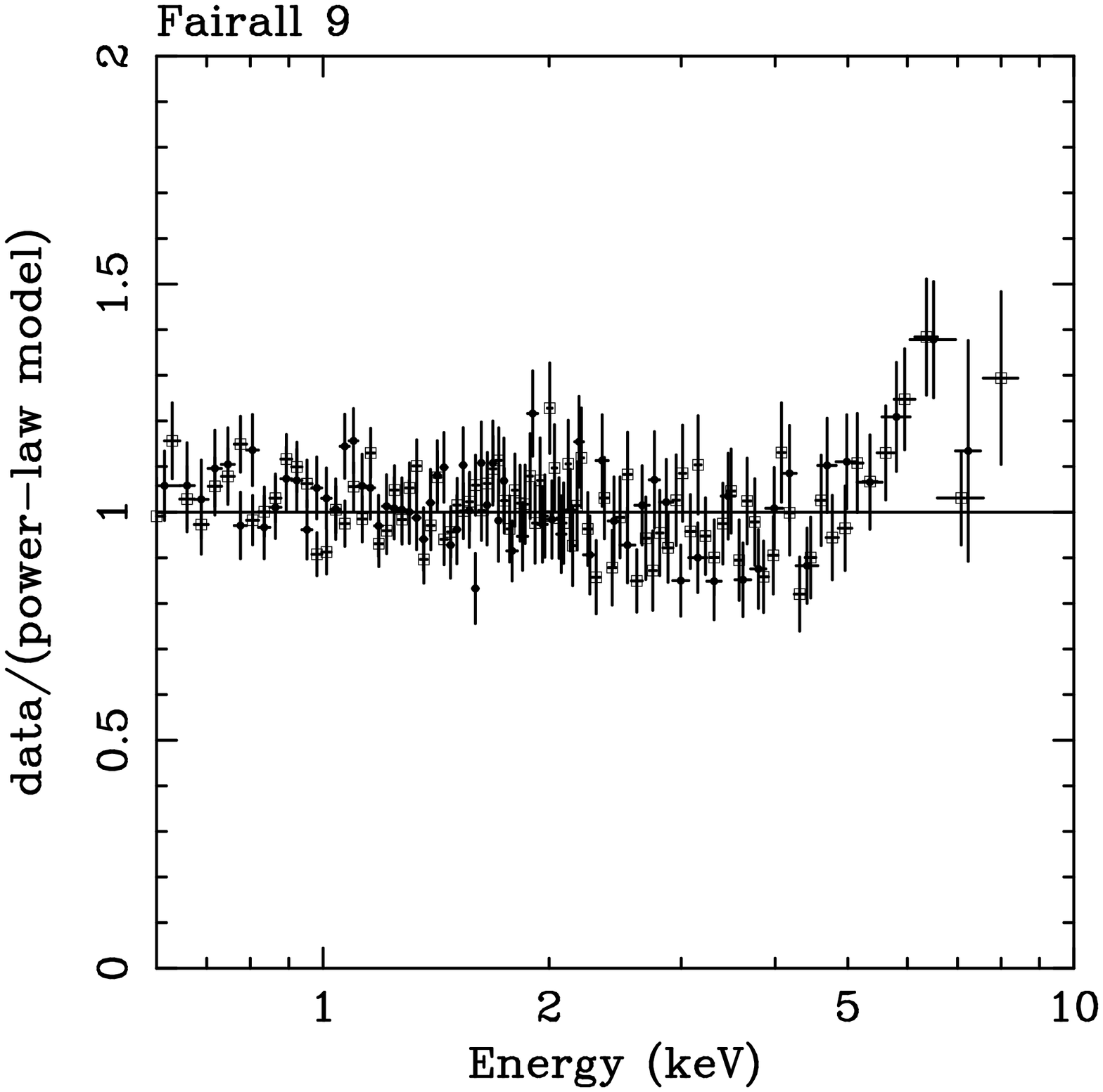,width=0.49\textwidth,height=0.23\textheight,angle=270}
}
\hbox{
\psfig{figure=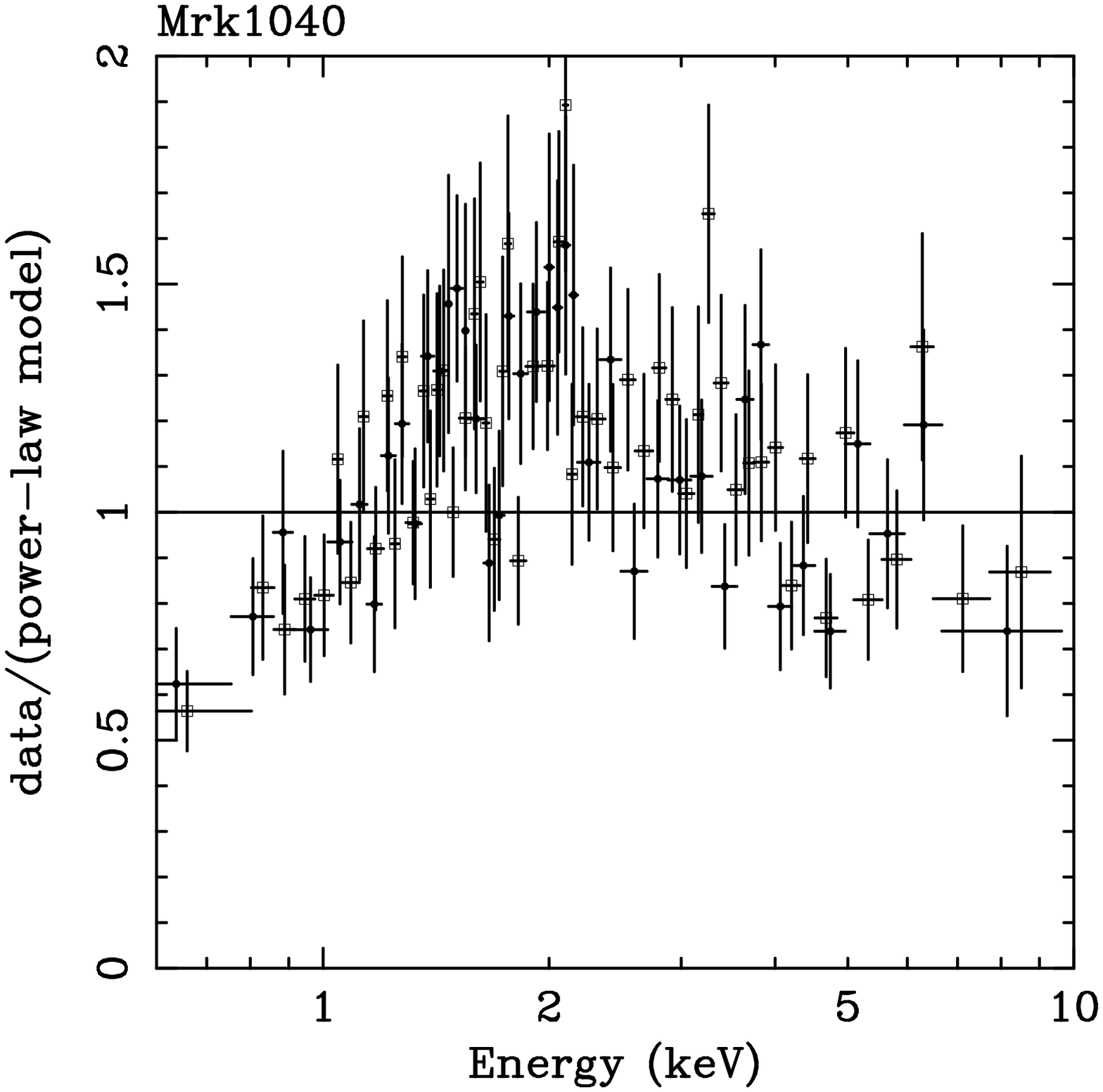,width=0.49\textwidth,height=0.23\textheight,angle=270}
\psfig{figure=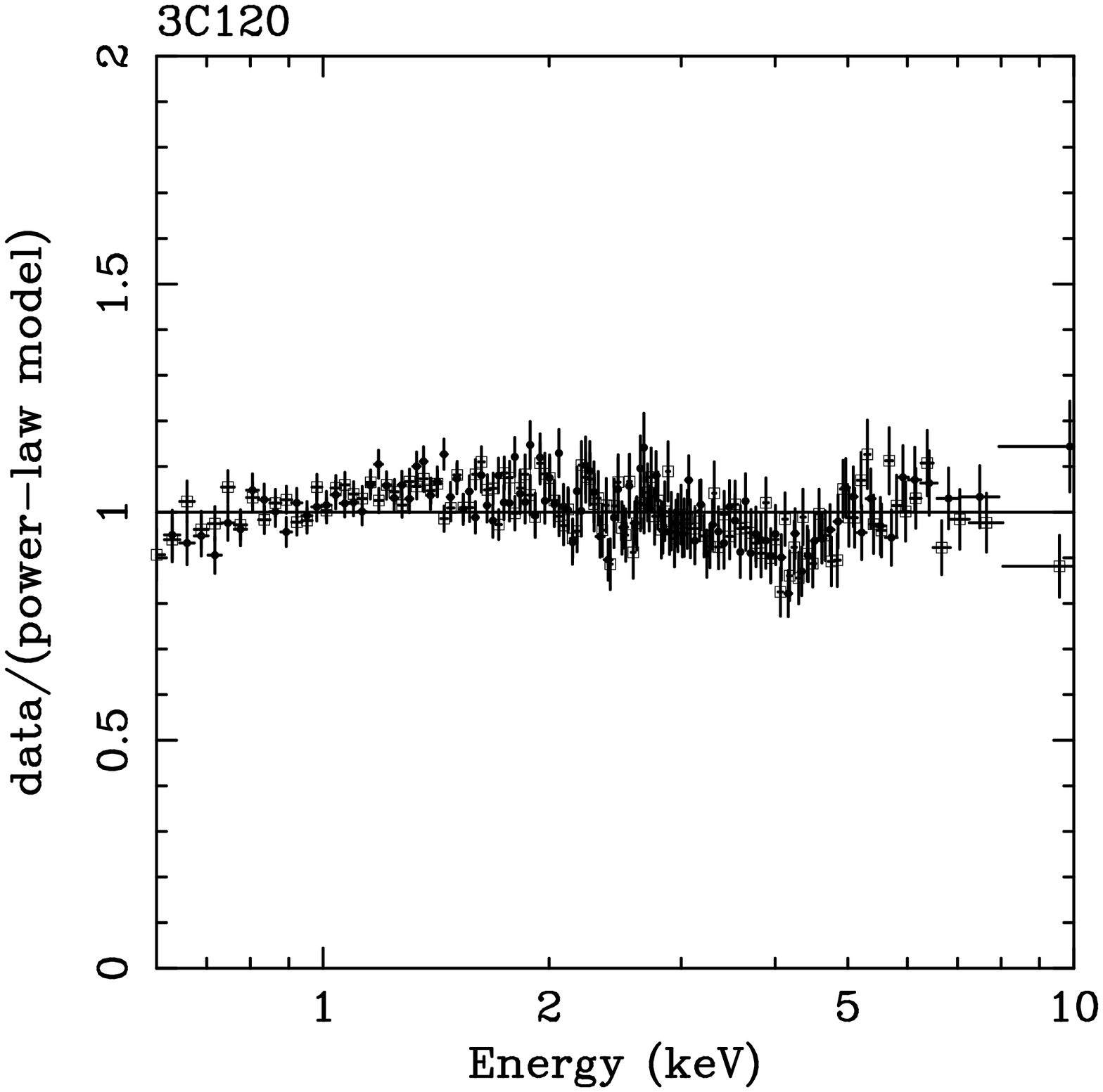,width=0.49\textwidth,height=0.23\textheight,angle=270}
}
\hbox{
\psfig{figure=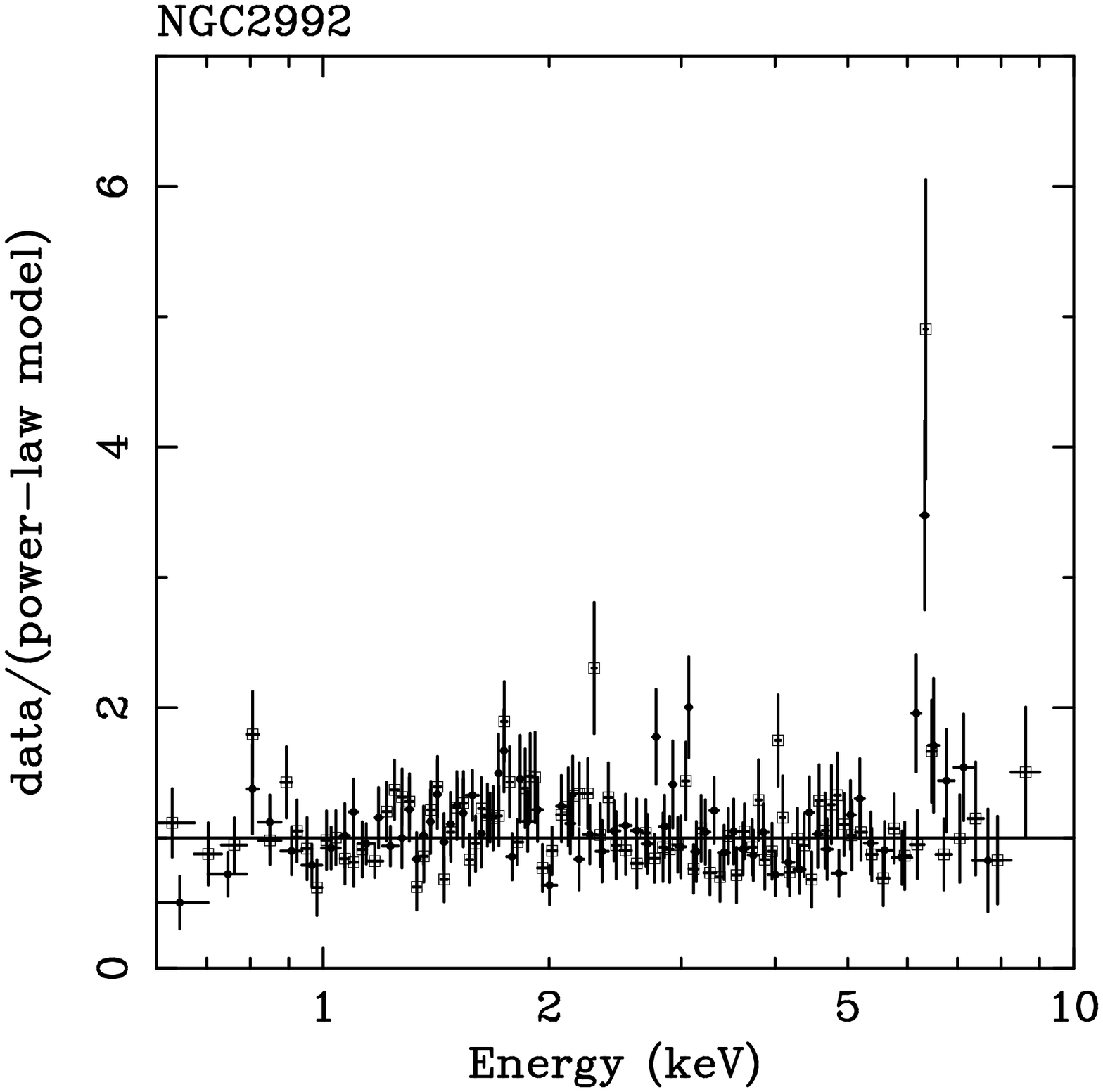,width=0.49\textwidth,height=0.23\textheight,angle=270}
\psfig{figure=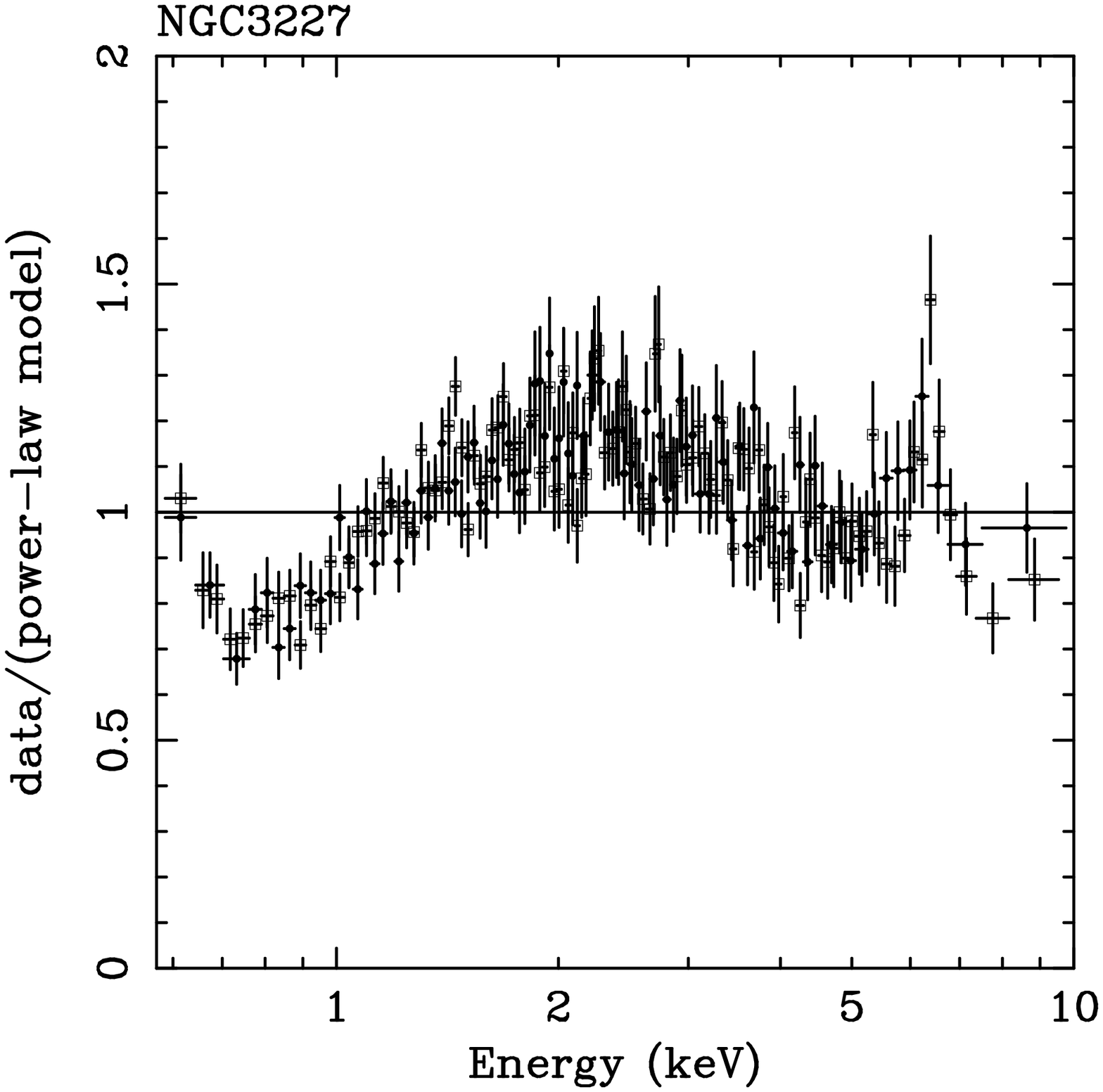,width=0.49\textwidth,height=0.23\textheight,angle=270}
}
\hbox{
\psfig{figure=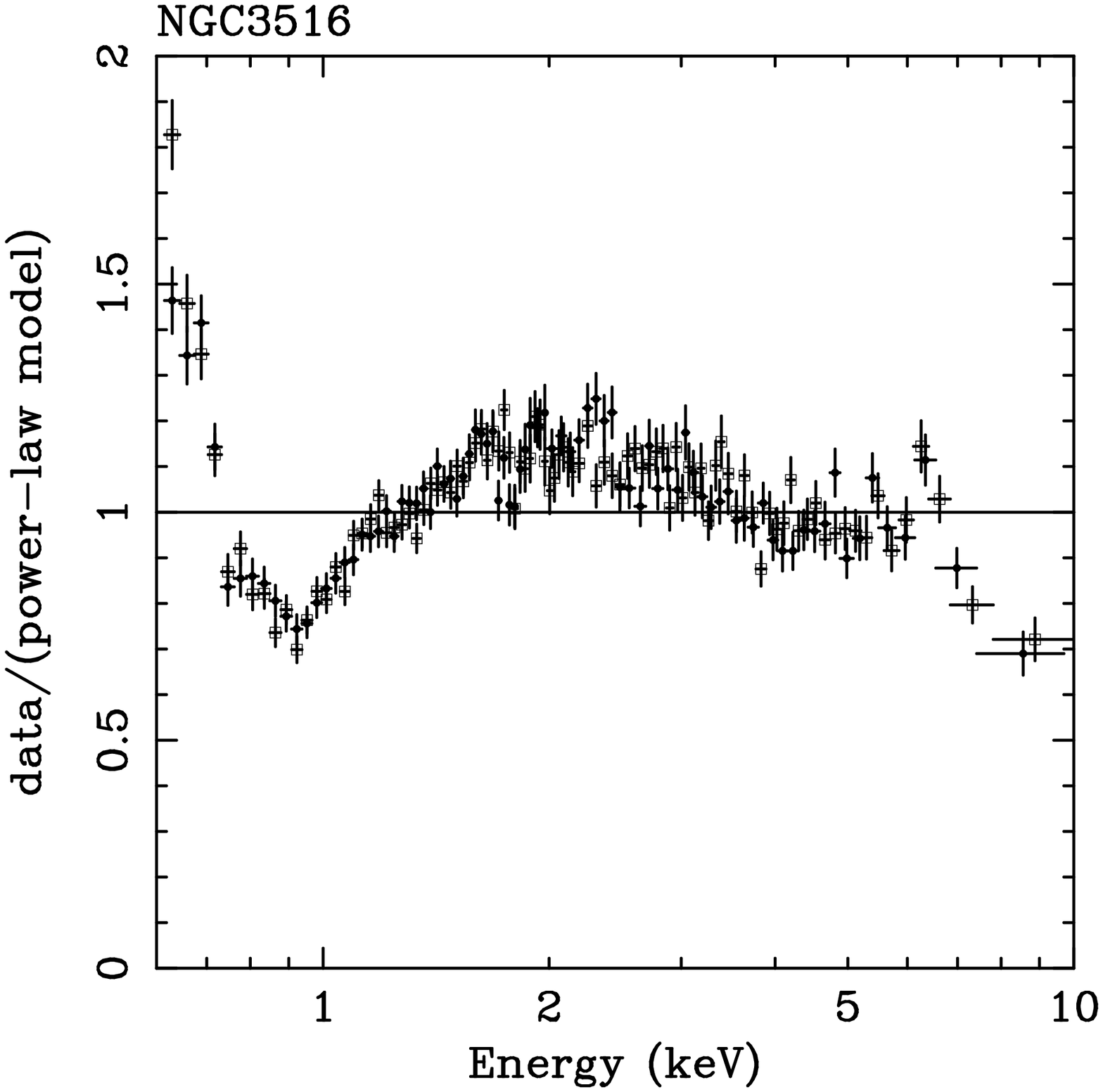,width=0.49\textwidth,height=0.23\textheight,angle=270}
\psfig{figure=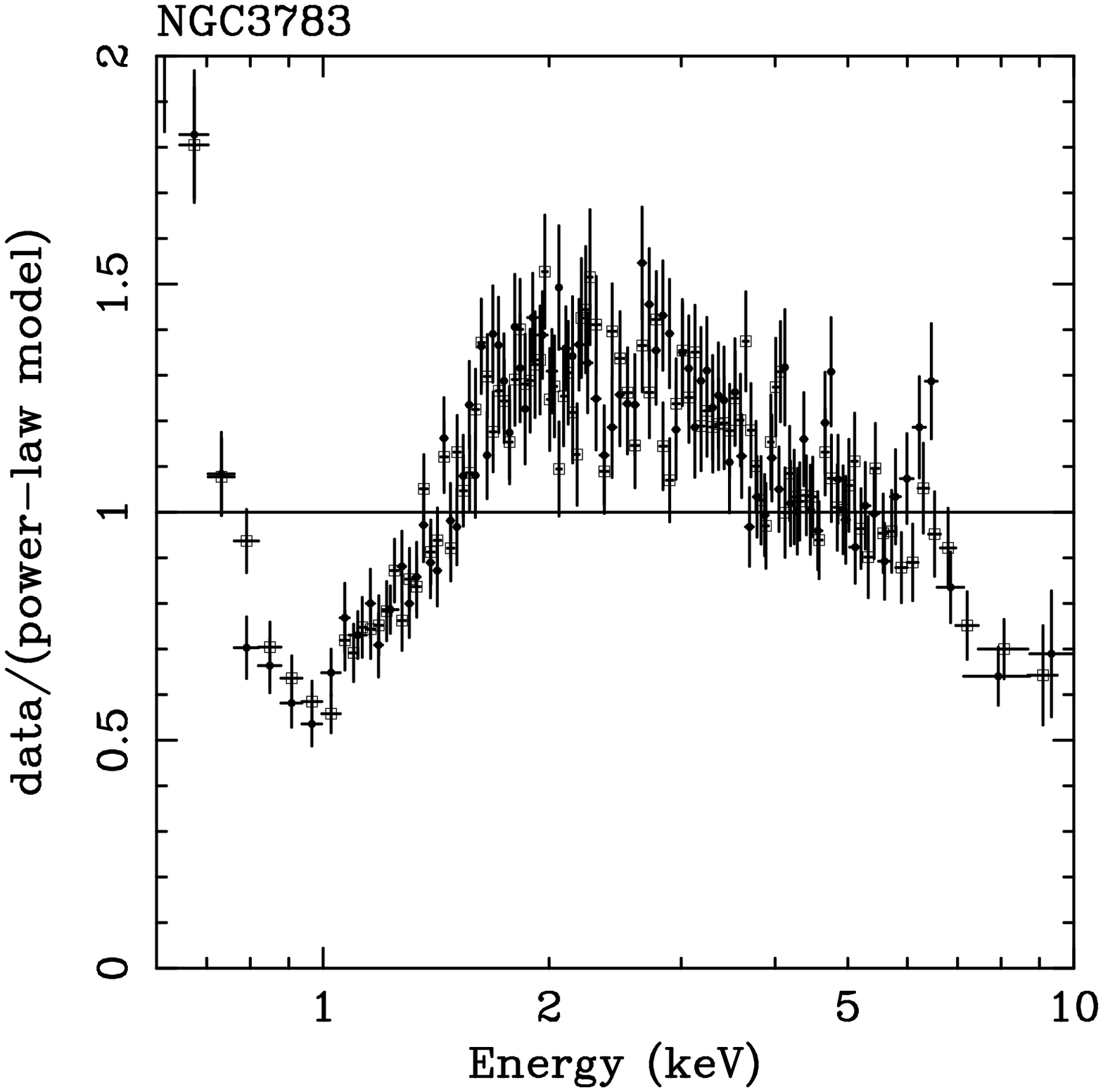,width=0.49\textwidth,height=0.23\textheight,angle=270}
}
\caption{}
\end{figure*}

\addtocounter{figure}{-1}

\begin{figure*}
\hbox{
\psfig{figure=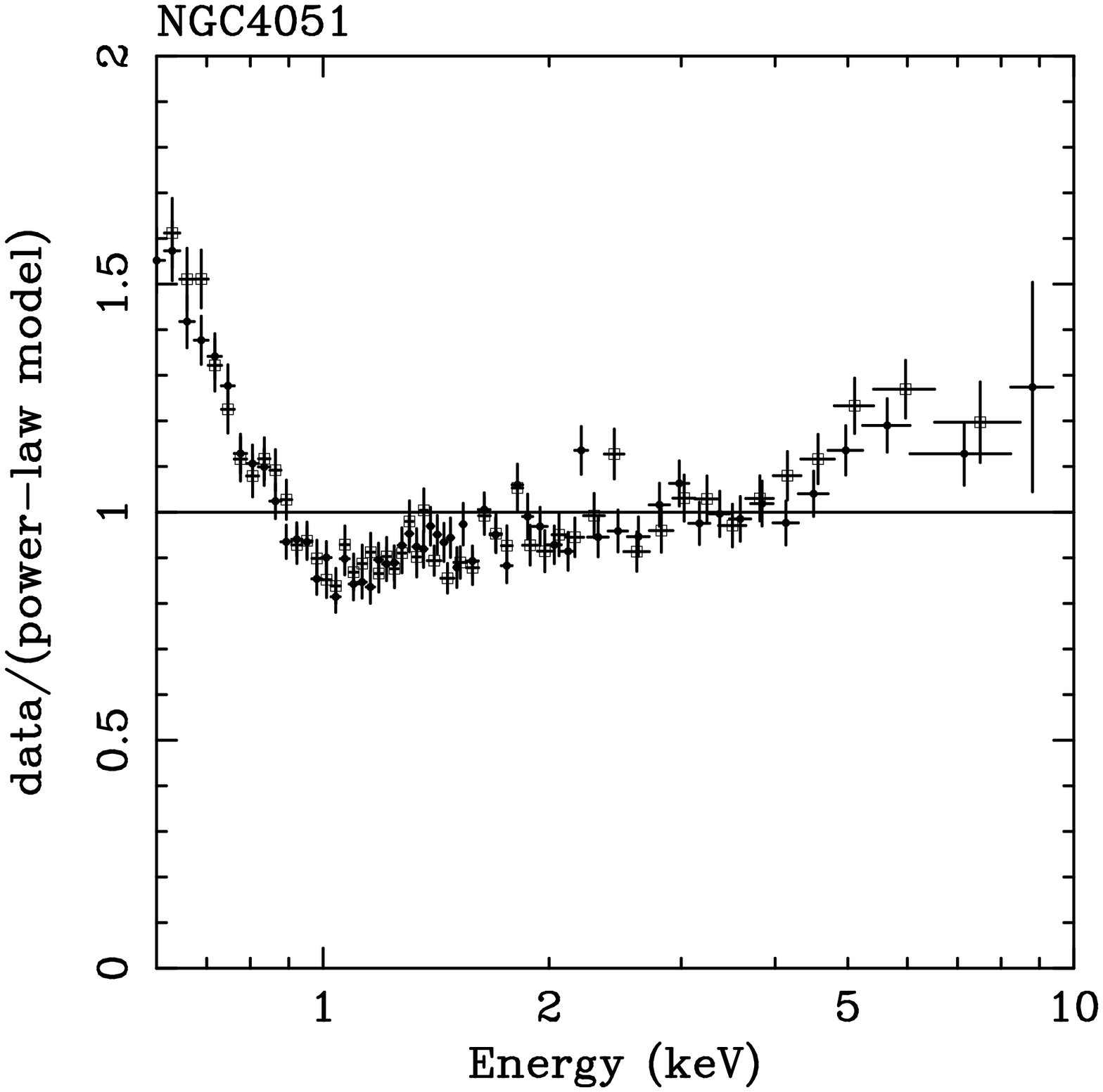,width=0.49\textwidth,height=0.23\textheight,angle=270}
\psfig{figure=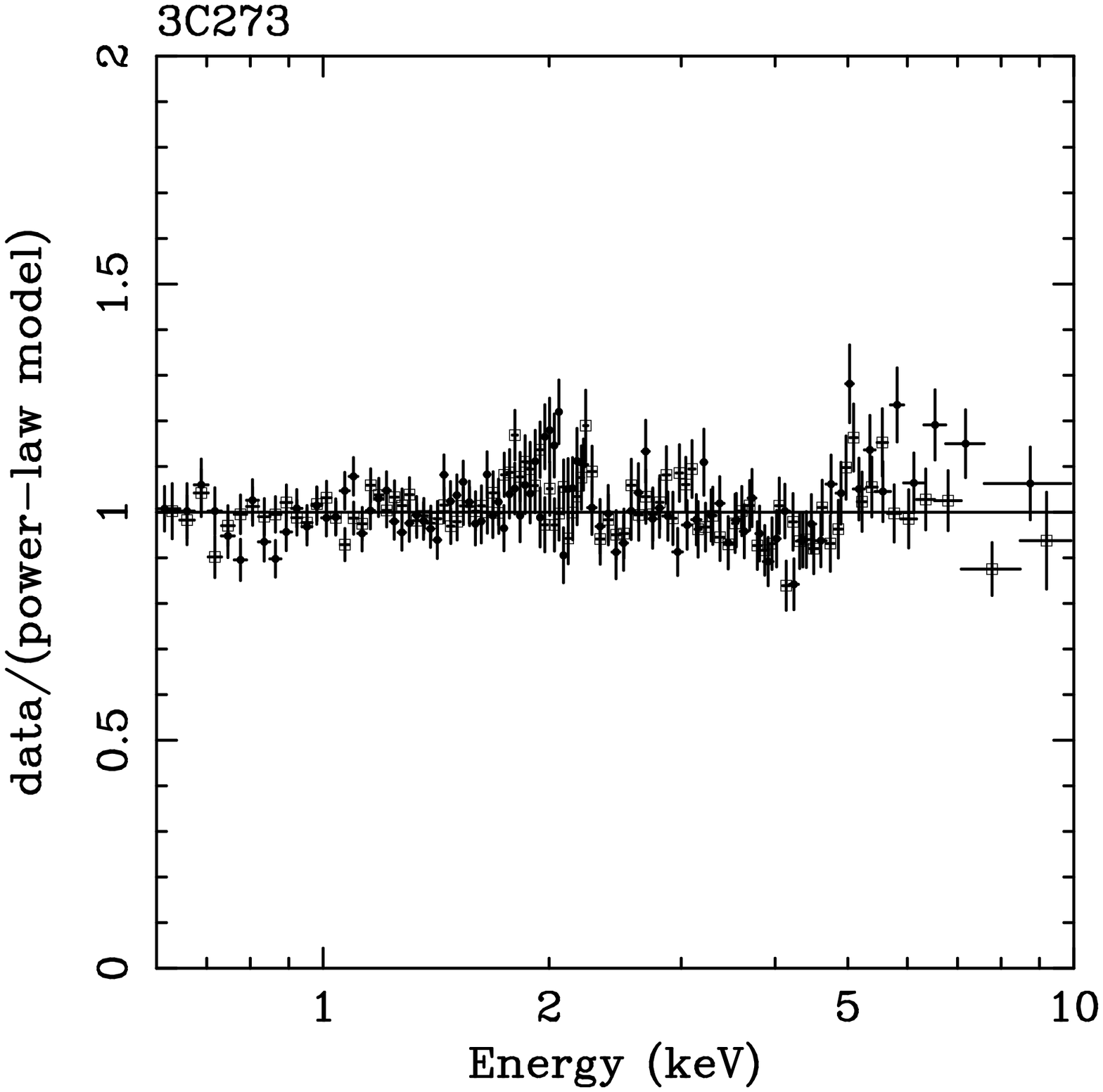,width=0.49\textwidth,height=0.23\textheight,angle=270}
}
\hbox{
\psfig{figure=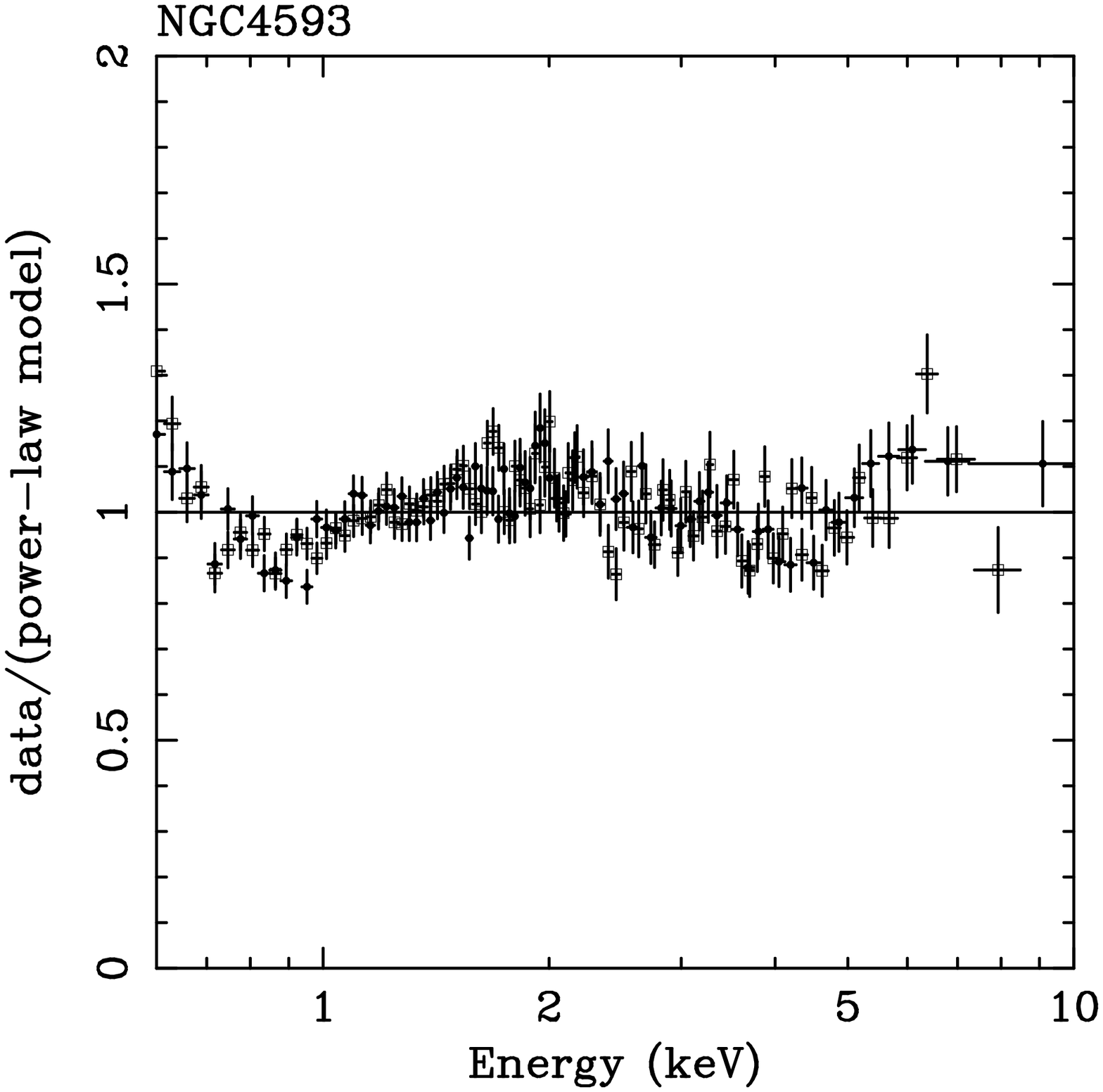,width=0.49\textwidth,height=0.23\textheight,angle=270}
\psfig{figure=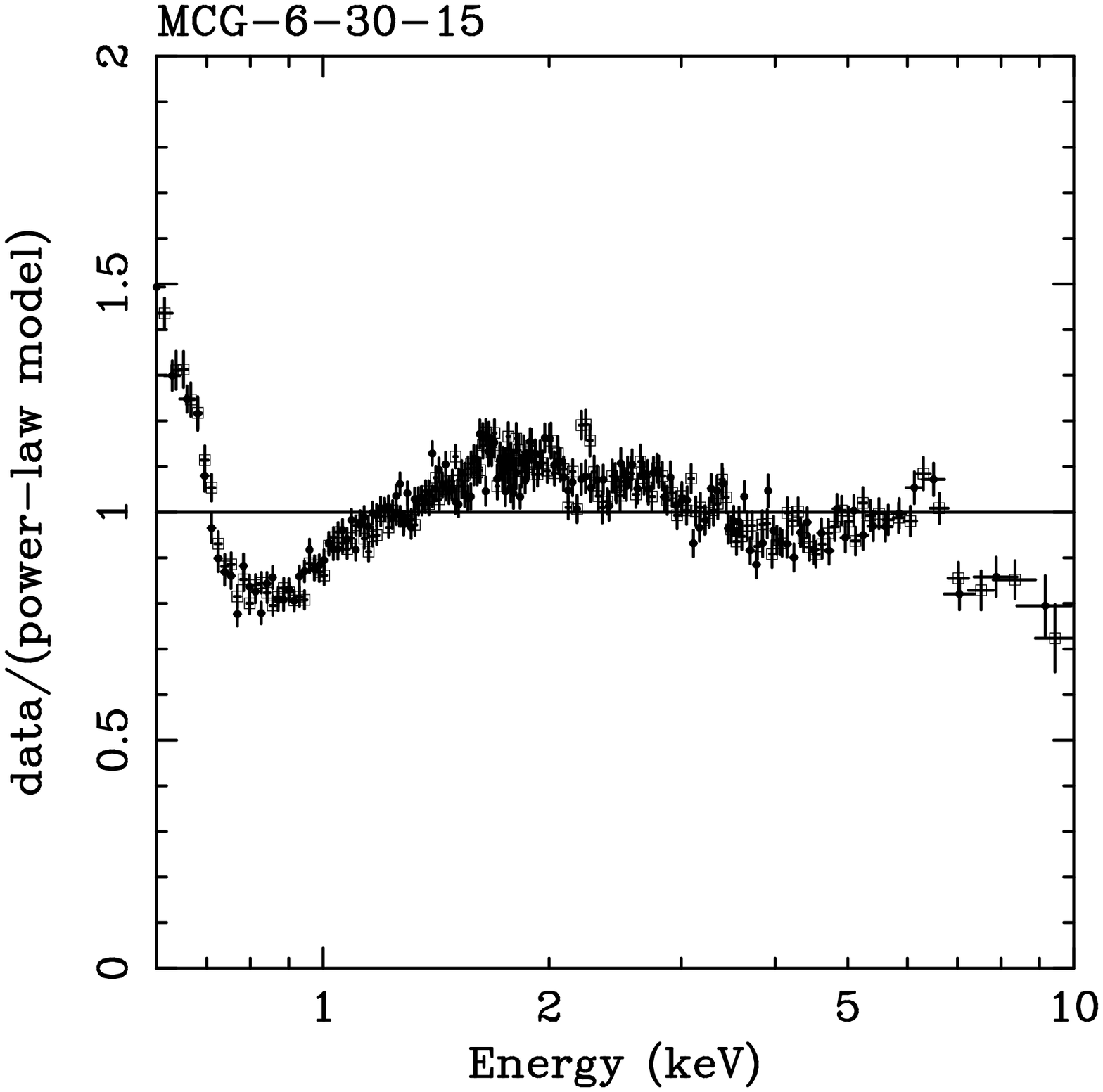,width=0.49\textwidth,height=0.23\textheight,angle=270}
}
\hbox{
\psfig{figure=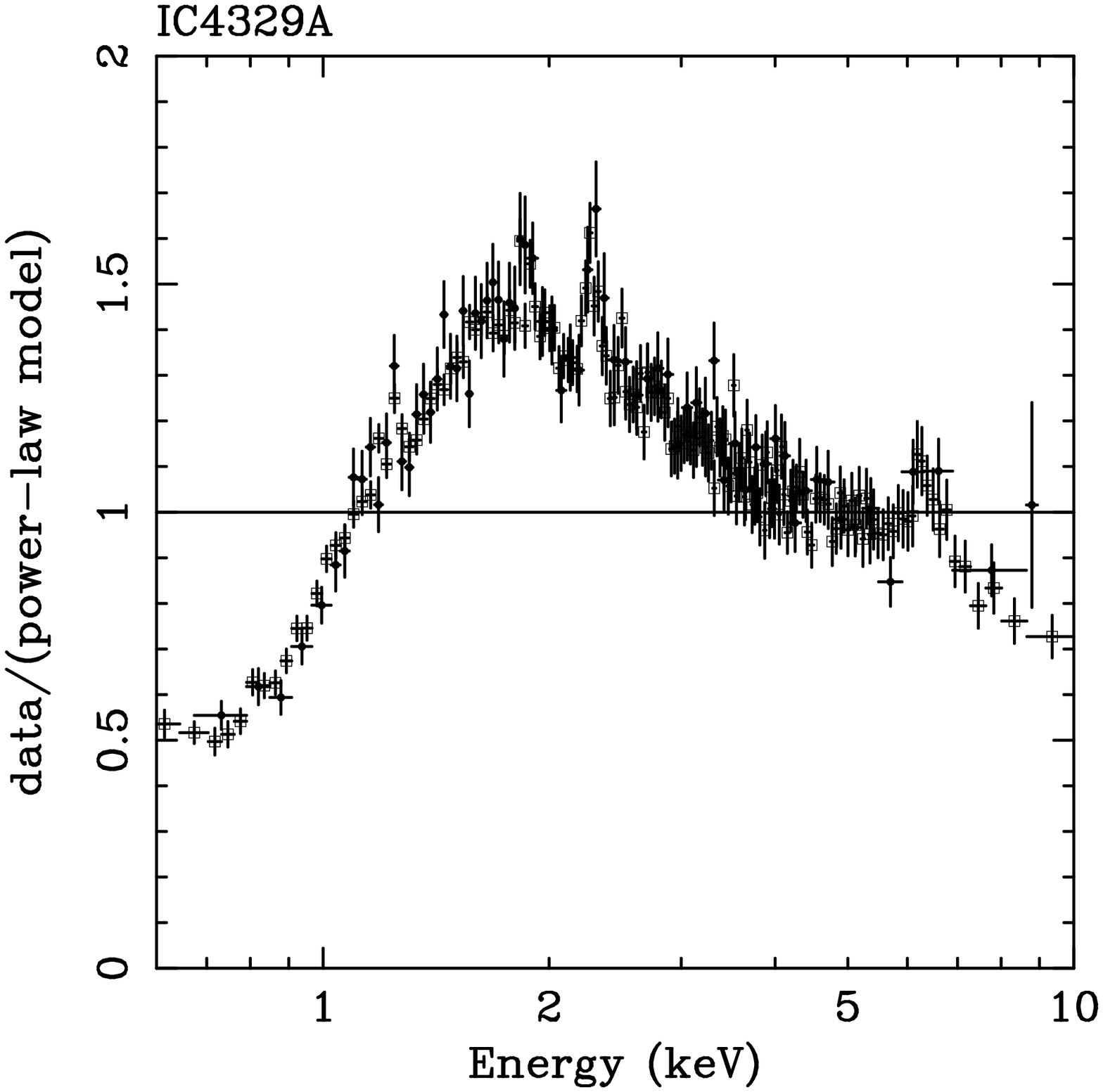,width=0.49\textwidth,height=0.23\textheight,angle=270}
\psfig{figure=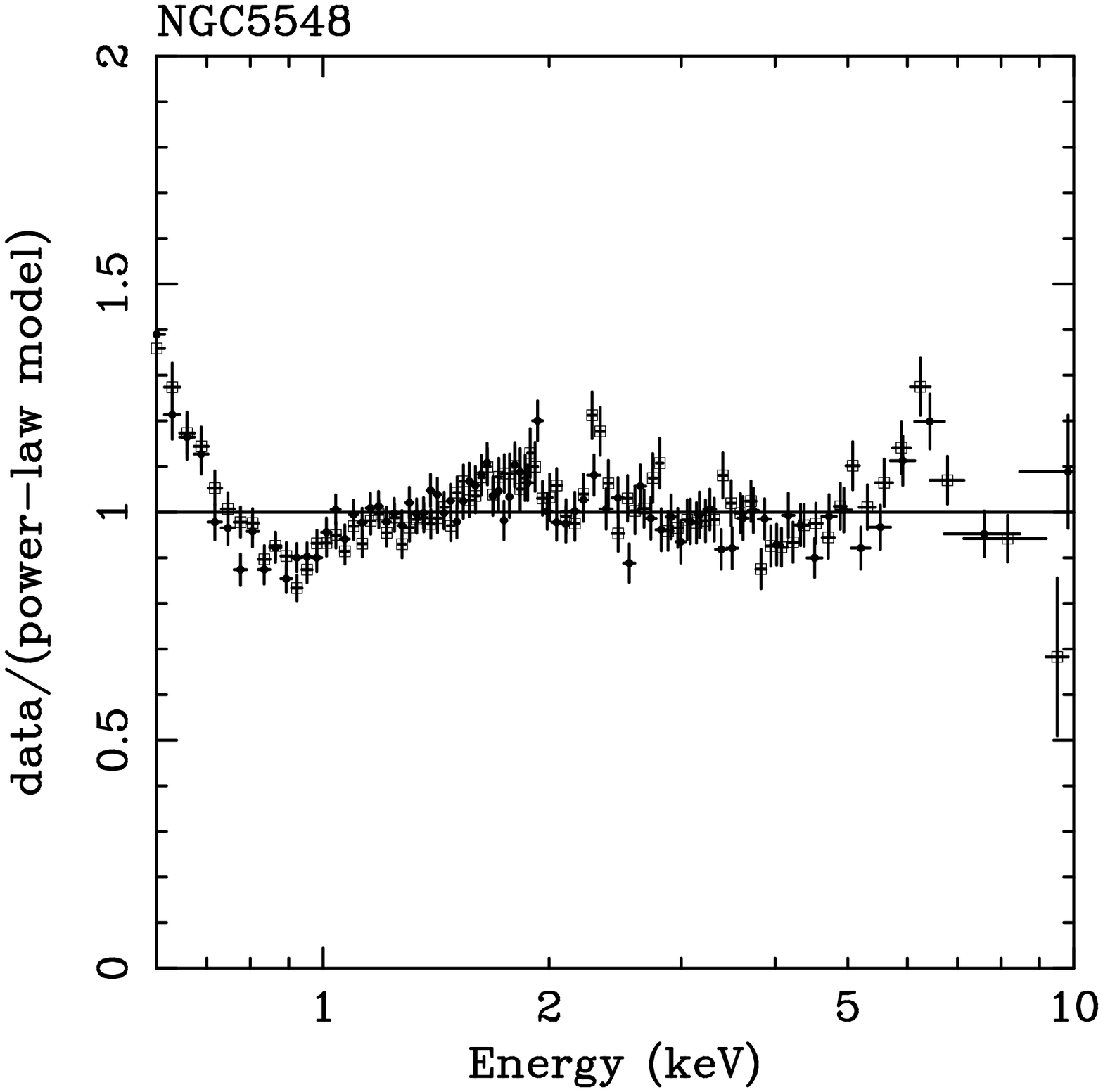,width=0.49\textwidth,height=0.23\textheight,angle=270}
}
\hbox{
\psfig{figure=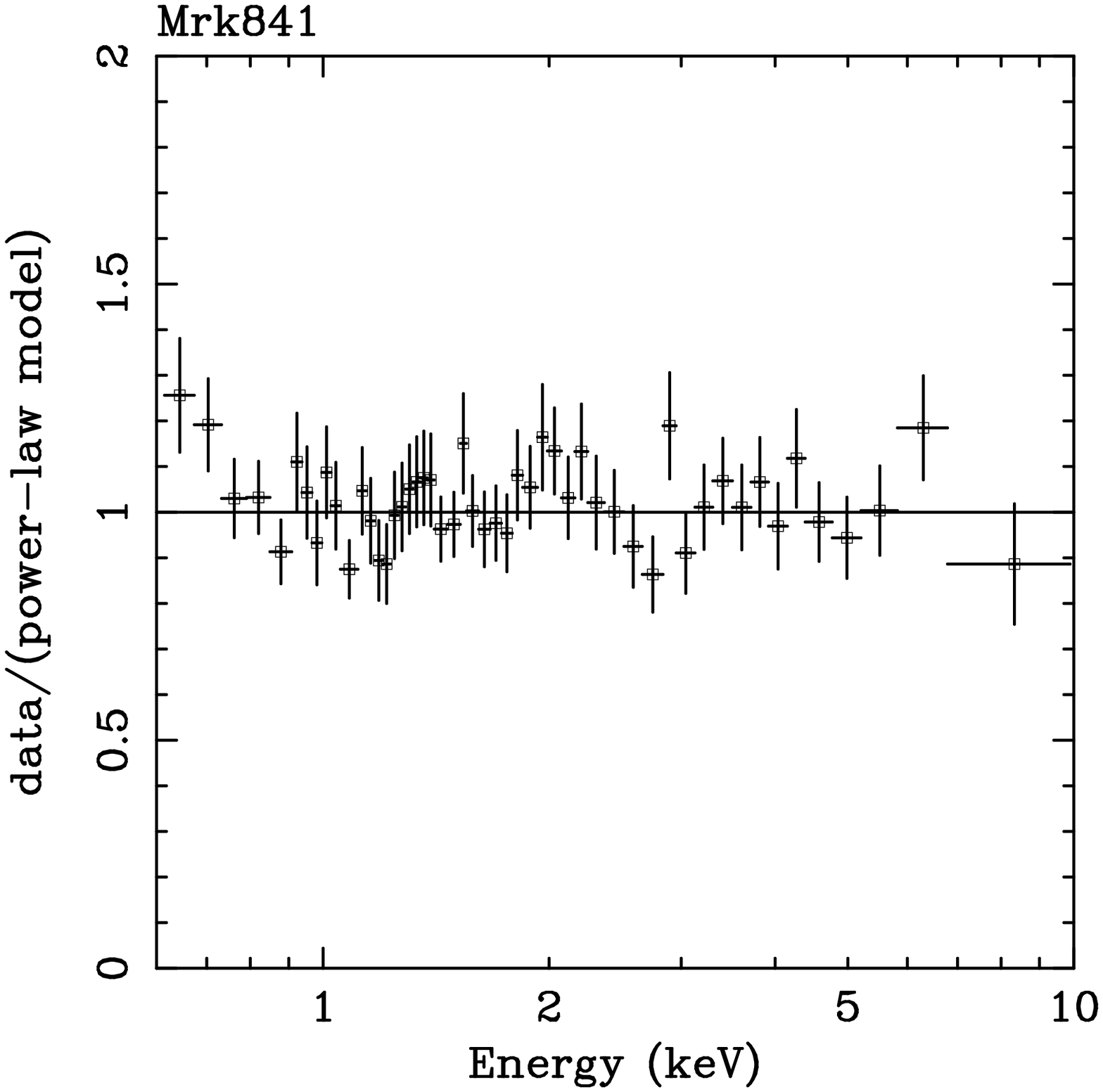,width=0.49\textwidth,height=0.23\textheight,angle=270}
\psfig{figure=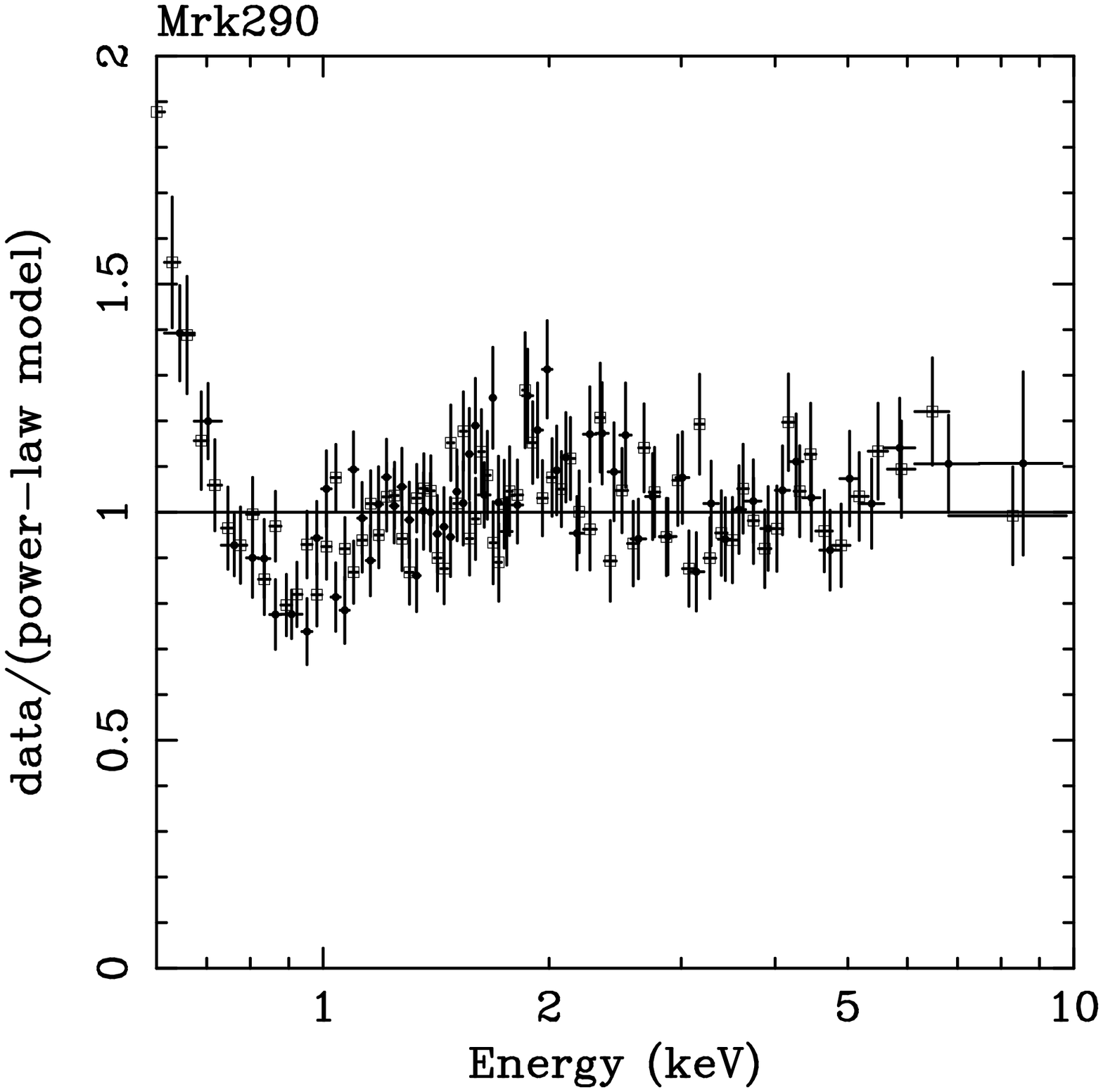,width=0.49\textwidth,height=0.23\textheight,angle=270}
}
\caption{cont.}
\end{figure*}

\addtocounter{figure}{-1}

\begin{figure*}
\hbox{
\psfig{figure=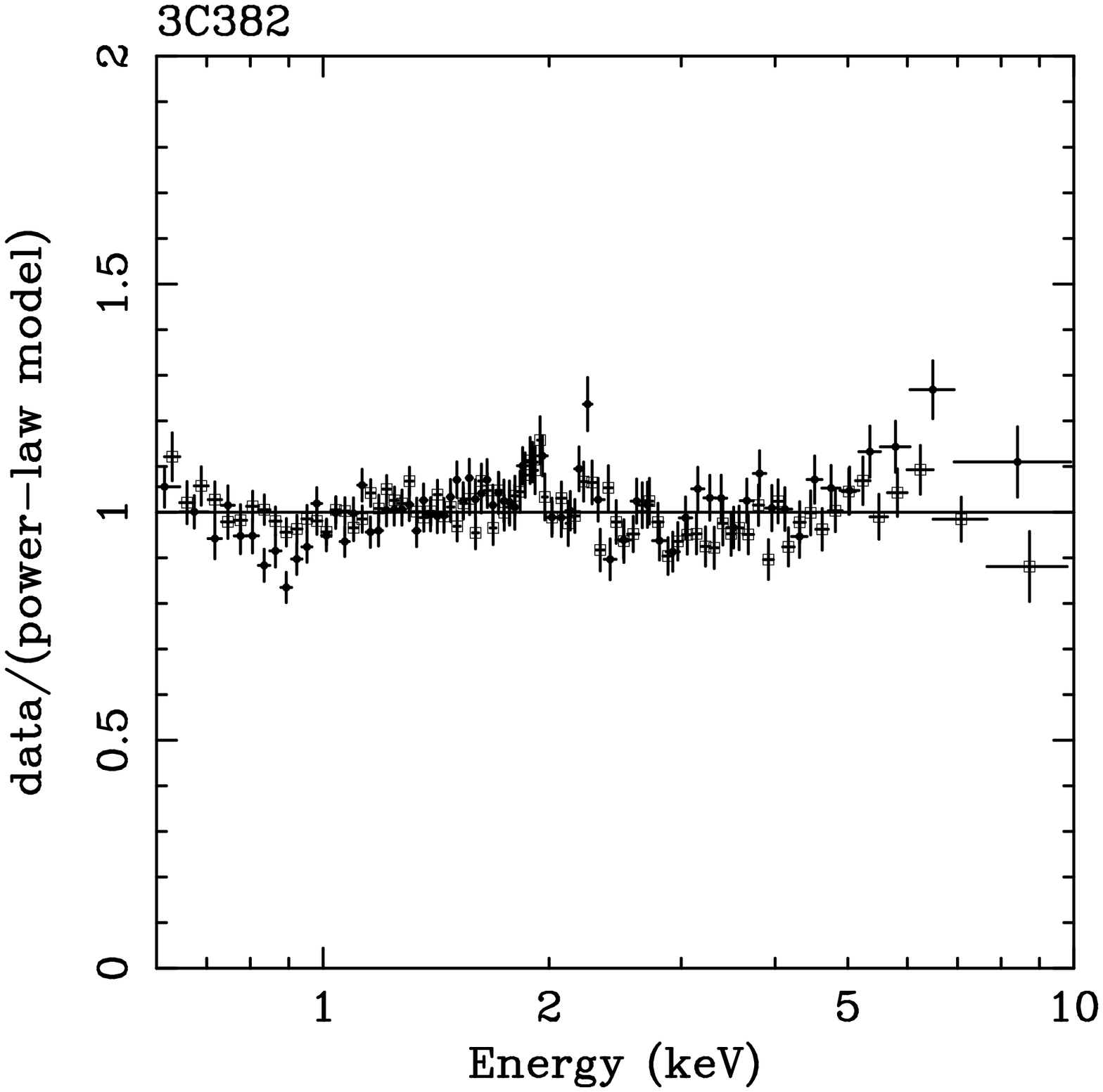,width=0.49\textwidth,height=0.23\textheight,angle=270}
\psfig{figure=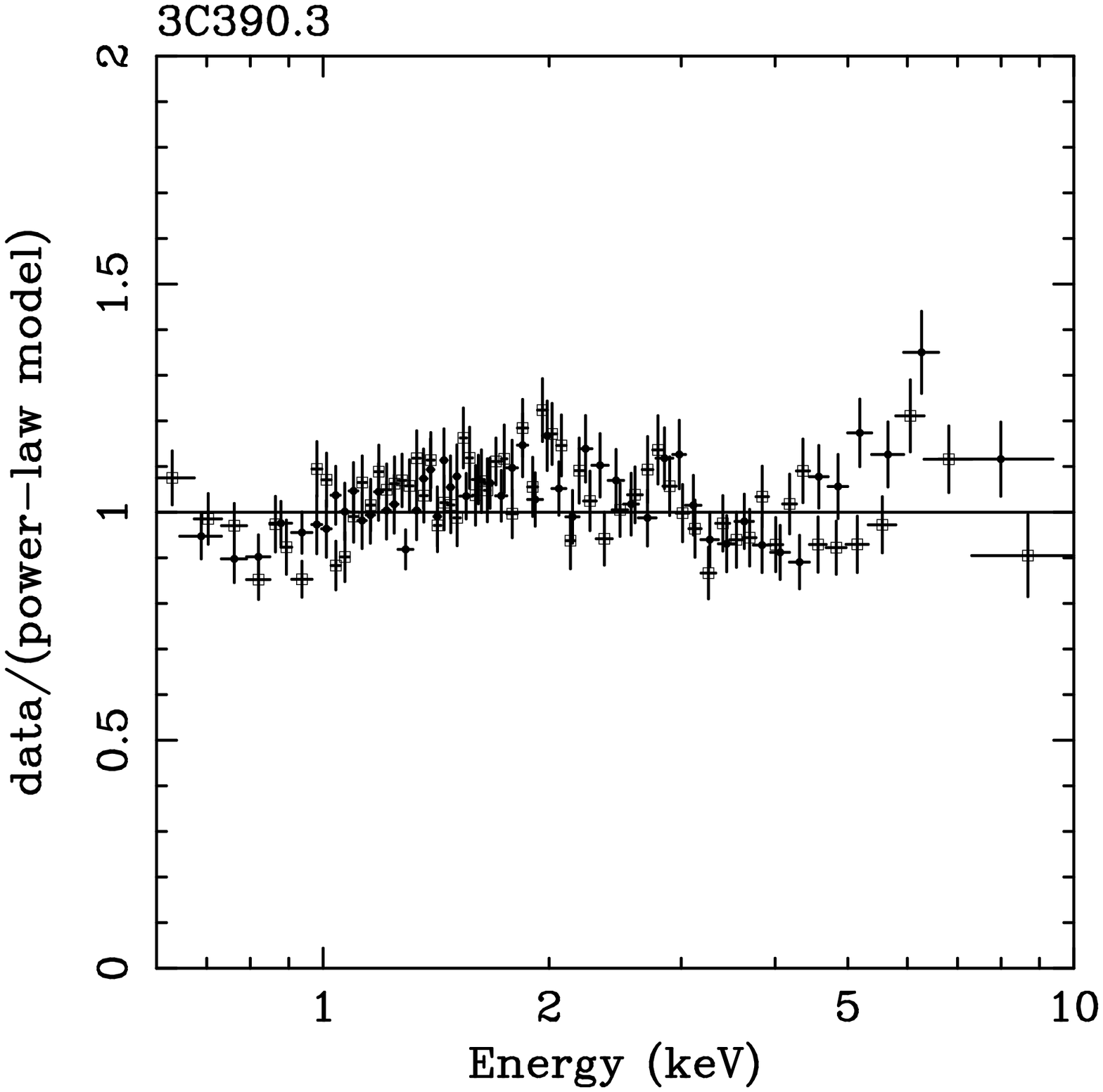,width=0.49\textwidth,height=0.23\textheight,angle=270}
}
\hbox{
\psfig{figure=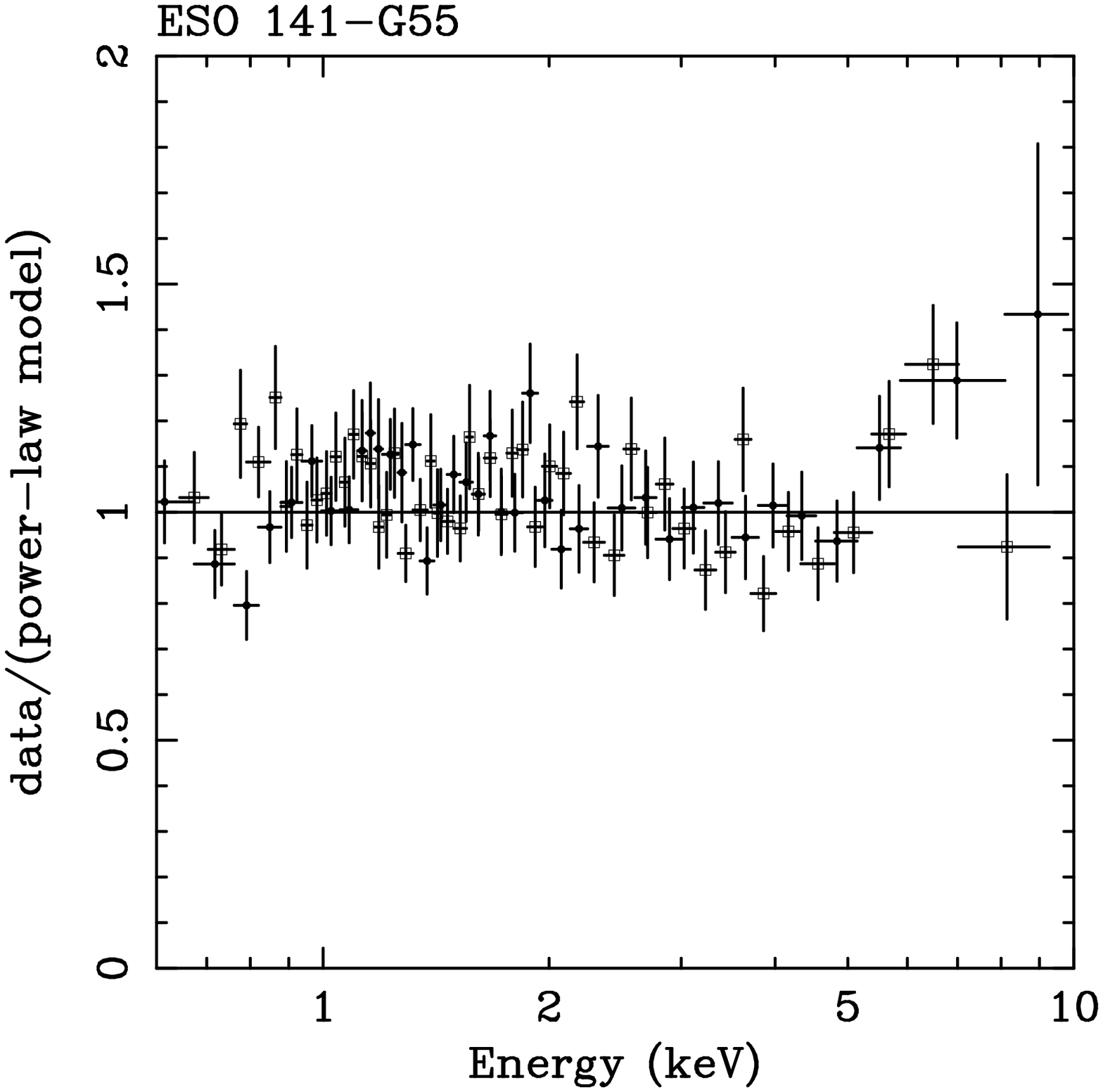,width=0.49\textwidth,height=0.23\textheight,angle=270}
\psfig{figure=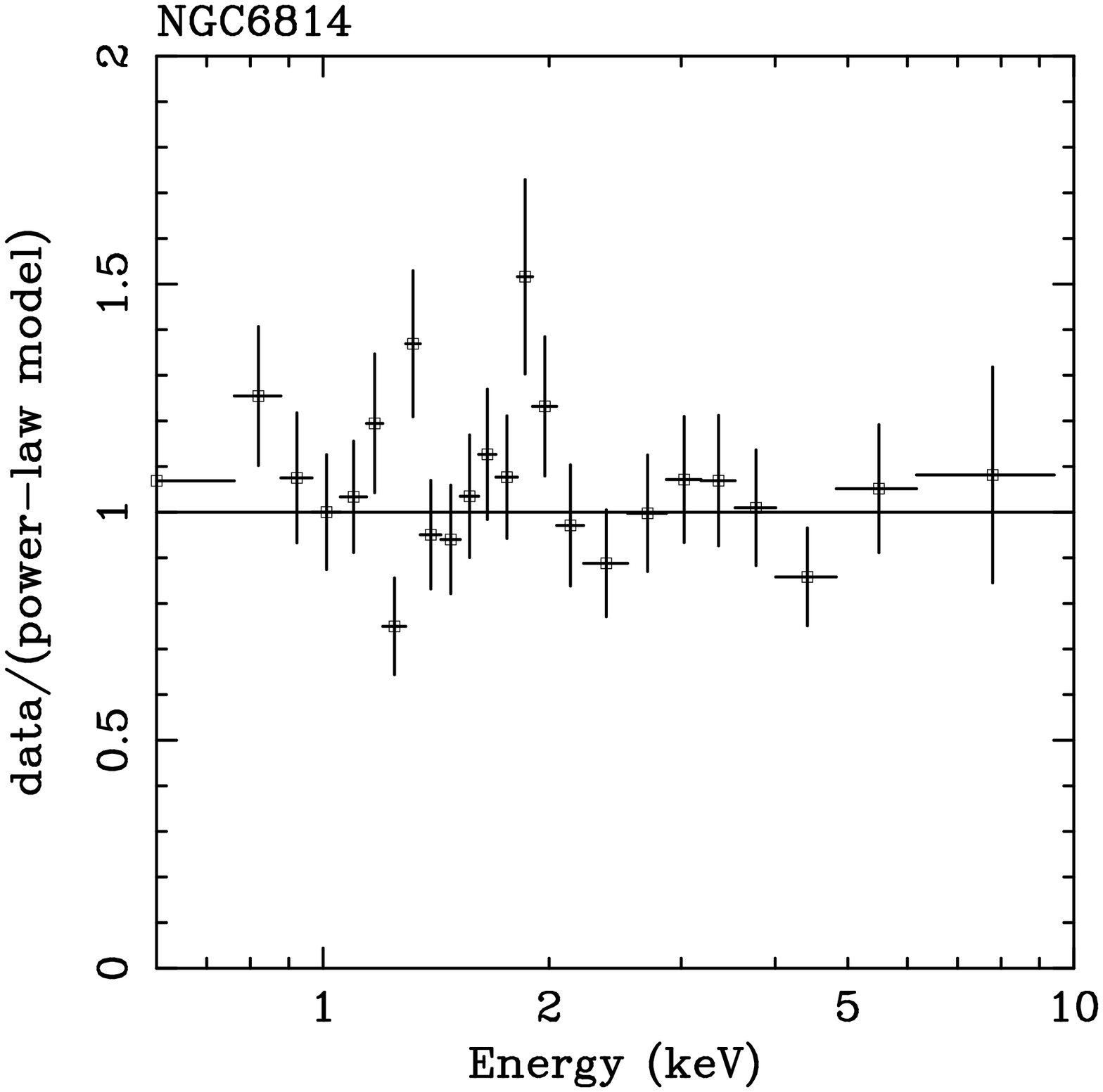,width=0.49\textwidth,height=0.23\textheight,angle=270}
}
\hbox{
\psfig{figure=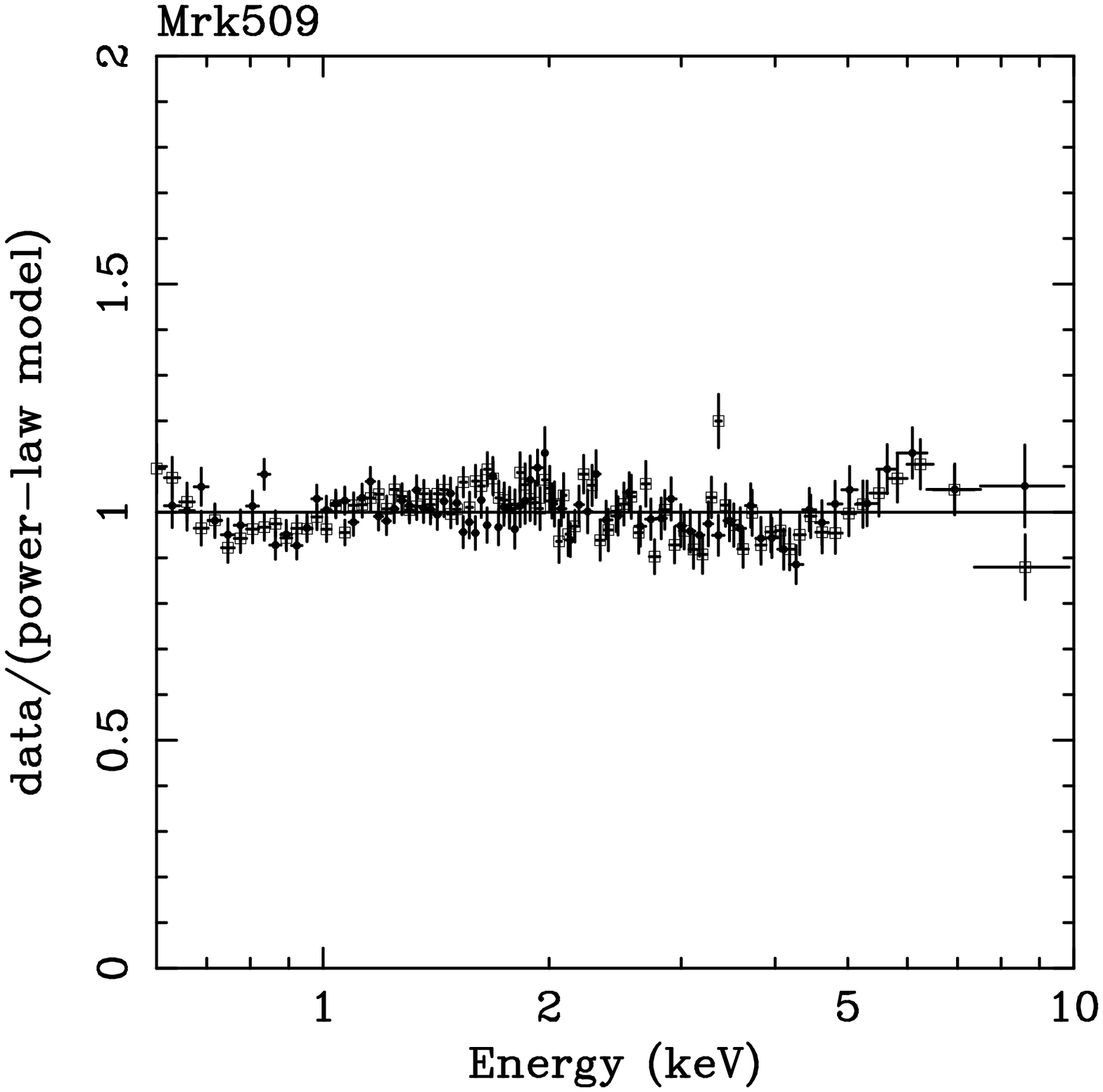,width=0.49\textwidth,height=0.23\textheight,angle=270}
\psfig{figure=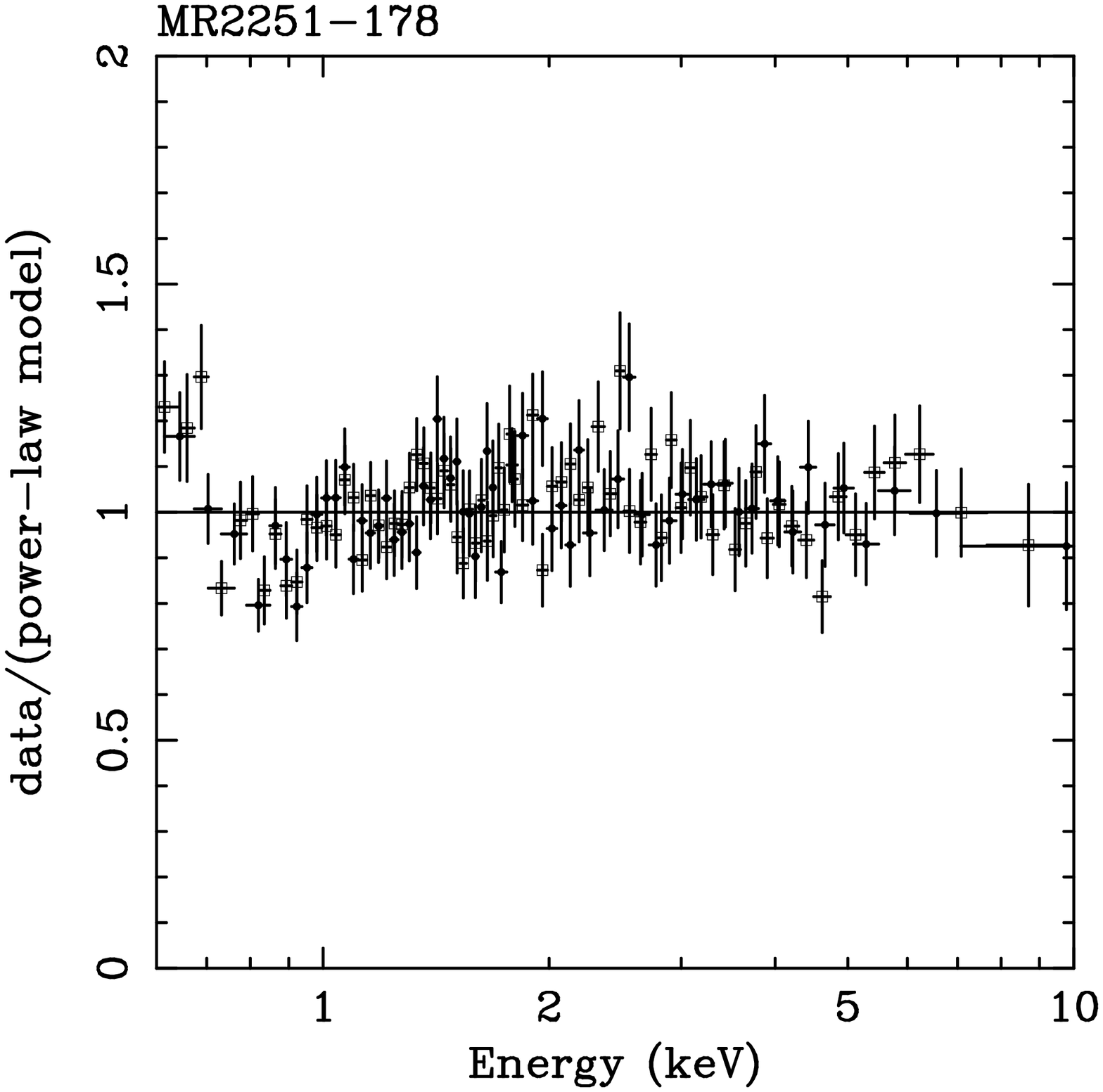,width=0.49\textwidth,height=0.23\textheight,angle=270}
}
\hbox{
\psfig{figure=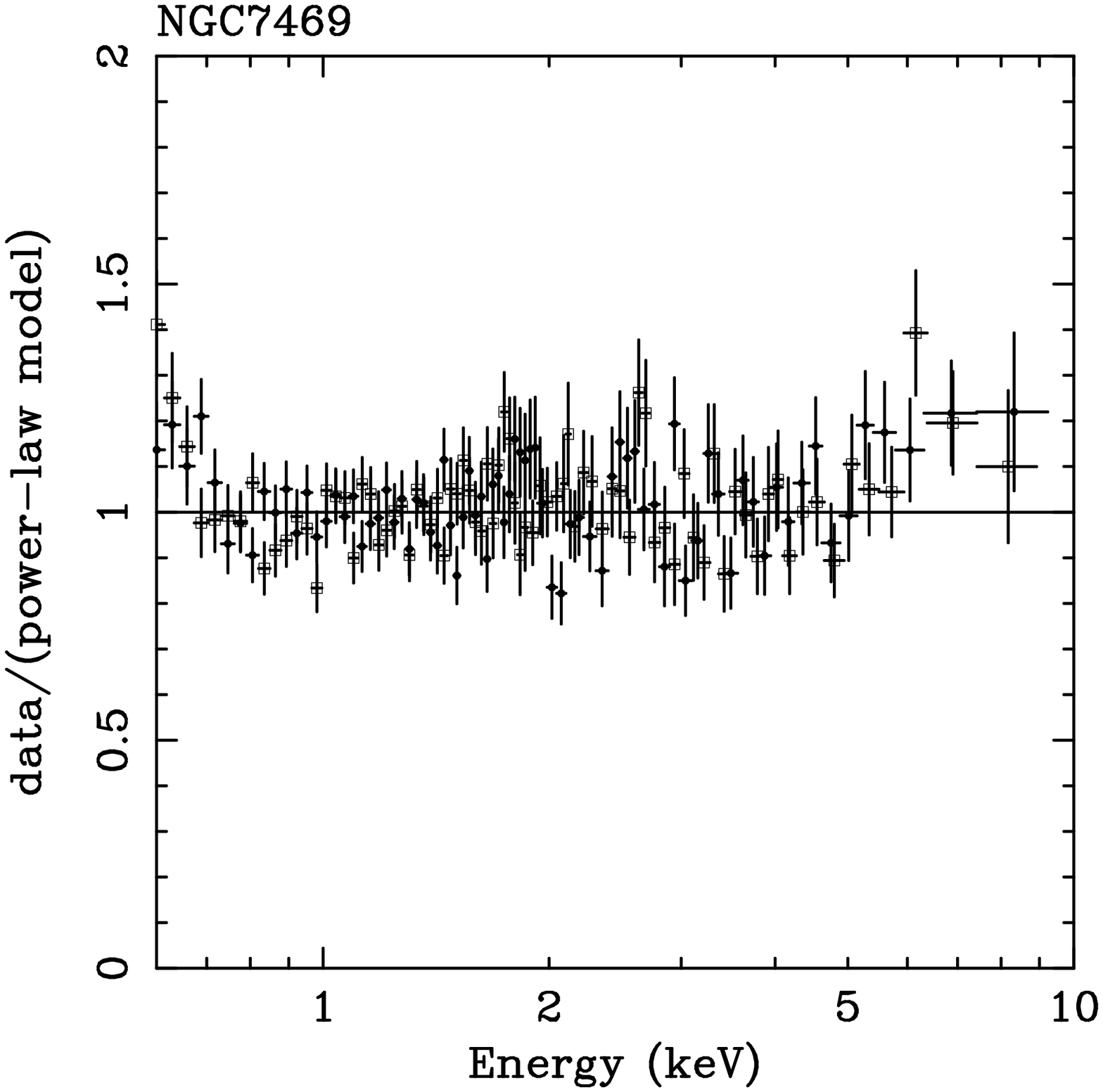,width=0.49\textwidth,height=0.23\textheight,angle=270}
\psfig{figure=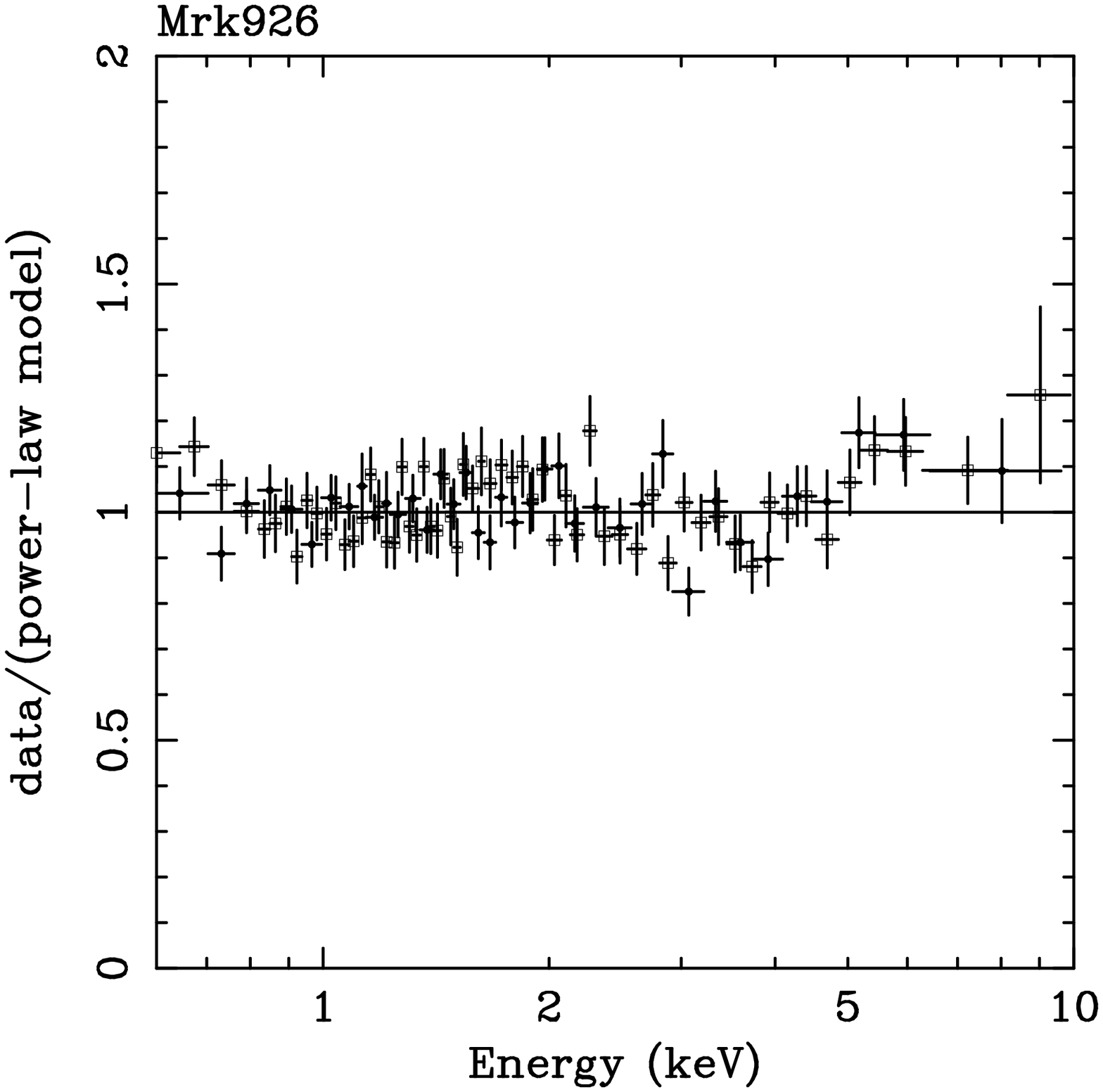,width=0.49\textwidth,height=0.23\textheight,angle=270}
}
\caption{cont.   Ratio of the data to the best fit power-law modified 
by cold Galactic absorption (model-A).  All 4 instruments were used
for spectral fitting but only SIS data are shown in order to maintain
clarity. Open squares are the SIS0 data whereas plain crosses are the
SIS1 data.  See text for a full discussion. }
\end{figure*}

\begin{table*}
\caption{Spectral fitting results for model-A (power-law modified by 
Galactic absorption).  Column 2 gives the average SIS0 count rate for
the good data.  Columns 3 and 4 show the 2--10\,keV flux and
luminosity, respectively, for these best fit power-law models ({\it
with no correction for the effects of absorption applied}).  Column 5
shows the best fit power-law photon index.  Column 6 shows the
goodness of fit and the number of degrees of freedom (dof).  All
errors are quoted at the 90 per cent confidence level for one
interesting parameter ($\Delta\chi^2=2.7$) apart from the error on the
average SIS0 count rate, which is quoted at the 1-$\sigma$ level for one
interesting parameter ($\Delta\chi^2=1.0$).}
\begin{center}
\begin{tabular}{lcccccc}\hline
source & SIS0 count & $F_{\rm 2-10}$ & $L_{\rm 2-10}$ & Photon index & $\chi^2$/dof\\
name & rate (cts s$^{-1}$) & ($10^{-12}\ergpcmsqps$) & ($10^{43}\ergps$) & $\Gamma$ & \\\hline
Mrk~335 & $0.677\pm 0.006$ & 9.2 & 2.5 & $2.17\pm 0.01$ & 899/741  \\
Fairall~9 & $1.001\pm 0.007$ & 19.0 & 18 & $1.93\pm 0.01$ & 1283/1149  \\
Mrk~1040 & $0.121\pm 0.003$ & 5.4 & 0.60 & $1.20\pm 0.03$ & 530/371  \\
3C~120 & $1.858\pm 0.007$ & 45.5 & 22 & $1.85\pm 0.01$ & 1894/1051  \\
NGC~2992 & $0.082\pm 0.002$ & 4.1 & 0.11 & $1.06\pm 0.04$ & 516/439  \\
NGC~3227 & $0.641\pm 0.006$ & 23.3 & 0.09 & $1.30\pm 0.01$ & 2294/1487  \\
NGC~3516 & $2.497\pm 0.010$ & 78 & 2.7 & $1.47\pm 0.01$ & 3692/1041  \\
NGC~3783 & $1.080\pm 0.009$ & 54 & 1.9 & $0.92\pm 0.1$ & 3894/1319  \\
NGC~4051 & $1.049\pm 0.006$ & 16 & 0.03 & $2.14\pm 0.01$ & 3641/1342  \\
3C~273 & $4.731\pm 0.022$ & 131 & 1530 & $1.59\pm 0.01$ & 1537/1476  \\
NGC~4593 & $1.660\pm 0.008$ & 35 & 1.2 & $1.84\pm 0.01$ & 1672/1391  \\
MCG$-6-30-15$ & $1.905\pm 0.004$ & 46 & 1.3 & $1.70\pm 0.01$ & 8804/2400  \\
IC~4329a & $2.896\pm 0.009$ & 78 & 8.6 & $1.30\pm 0.01$ & 16000/1781  \\
NGC~5548 & $2.501\pm 0.010$ & 43 & 5.4 & $1.79\pm 0.01$ & 2330/1633  \\
Mrk~841 & $0.359\pm 0.004$ & 10 & 5.7 & $1.63\pm 0.03$ & 334/331  \\
Mrk~290 & $0.321\pm 0.003$ & 9.1 & 3.6 & $1.61\pm0.02$ & 1168/1017  \\
3C~382 & $1.944\pm 0.008$ & 41 & 64 & $1.89\pm 0.01$ & 1722/1452  \\
3C~390.3 & $0.551\pm 0.004$ & 17 & 24 & $1.59\pm 0.01$ & 1475/1251  \\
ESO~141$-$G55 & $0.716\pm 0.009$ & 17 & 10 & $1.79\pm 0.02$ & 816/725  \\
NGC~6814 & $0.044\pm 0.001$ & 1.1 & 0.017 & $1.61\pm 0.07$ & 267/258  \\
Mrk~509 & $2.095\pm 0.008$ & 44 & 24 & $1.87\pm 0.01$ & 1676/1500  \\
MR~2251$-$178 & $1.817\pm 0.017$ & 45 & 93 & $1.65\pm 0.02$ & 871/846  \\
NGC~7469 & $1.453\pm 0.011$ & 29 & 3.6 & $2.02\pm 0.01$ & 1052/986   \\
Mrk~926 & $0.613\pm 0.005$ & 14 & 14 & $1.73\pm 0.02$ & 1106/1004 \\\hline
\end{tabular}
\end{center}
\end{table*}

\noindent Initially I fitted the time-average spectrum of each object 
with a model consisting of a power-law (with photon index $\Gamma$)
modified by Galactic absorption (as determined from {\sc H\,i} 21-cm
measurements and quoted in Table~1).  This will be denoted as {\it
model-A}.  Where available, data from all four {\it ASCA} instruments
were fitted simultaneously.  SIS data between 0.5--10\,keV and GIS
data between 0.8--10\,keV were used.  The data were rebinned so that
each spectral bin contained at least 20 counts.  Background subtracted
spectra were then fitted using the standard $\chi^2$ minimization
technique implimented in version 9.00 of the {\sc xspec} spectral
fitting package.  Table~2 gives the results of these fits.  Figure~1
shows the ratio of the SIS data with the best-fitting power-law spectrum.

It is clear from Table~2 and Figure~1 that model-A is a poor
description of the data for many of the objects.  Several objects show
low-energy excesses characteristic of the so-called soft-excess (with
clear examples being Mrk~335, NGC~4051 and Mrk~841), many display a
`notch' between $\sim 0.7$--2\,keV which is the signature of a
warm-absorber (e.g. NGC~3227, NGC~3516, NGC~3783, NGC~4593,
MCG$-6-30-15$, NGC~5548, Mrk~290 and MR~2251$-$178) and almost all
objects have an excess at energies characteristic of iron K$\alpha$
line emission.  The instrumental feature at 2.2\,keV (due to the
M-shell absorption edge of the gold coating the X-ray mirrors) can
also be seen in many of the objects.  The effect of this feature on
the statistics of the spectral fitting described here is negligible.

\subsection{Phenomenological model}

\begin{table*}
\caption{Absorption results from spectral fitting with model-B, as 
defined in Section 3.2 of the main text.  Column 6 gives the intrinsic
(i.e. corrected for all forms of absorption) 2--10\,keV luminosity.
Uncertainties on these luminosities are $\sim 5$ per cent and are
dominated by instrumental systematics.  Column 7 gives the goodness of
fit parameter.  Column 8 reports the improvement in $\chi^2$ upon the
addition of the oxygen absorption edges ($\Delta\chi^2>9\chi^2_\nu$
for 2 additional degrees of freedom corresponds to an improvement at
the 99 per cent confidence level) if only data above 0.6\,keV is
considered.  All errors and limits are quoted at the 90 per cent
confidence level for one interesting parameter ($\Delta\chi^2=2.7$). }
\begin{center}
\begin{tabular}{lccccccc}\hline
source & Photon Index & Intrinsic $N_{\rm H}$ & {\sc O\,vii} edge & {\sc O\,viii} edge & $L_{2-10}$ & $\chi^2$/dof & $\Delta\chi^2$ upon\\
name & $\Gamma$ & $(10^{20}\pcmsq)$ & depth, $\tau_{\rm O7}$ & depth, $\tau_{\rm O8}$ & ($10^{43}\ergps$) & &adding edges \\\hline
Mrk~335 & $2.07^{+0.07}_{-0.06}$ & $<0.9$ & $0.27\pm 0.11$ & $<0.09$ & 2.79 & 678/735 & 11\\
Fairall~9 & $1.99^{+0.02}_{-0.03}$ & $1.2^{+0.8}_{-0.7}$ & $<0.01$ & $<0.03$ & 18.0 & 1219/1143 & 0\\
Mrk~1040 & $1.69^{+0.09}_{-0.07}$ & $2.2^{+0.8}_{-0.7}$ & $0.41^{+0.50}_{-0.41}$ & $0.24^{+0.50}_{-0.24}$ & 0.566 & 352/361 & 2\\
3C~120 & $2.08^{+0.05}_{-0.03}$ & $8.2^{+1.0}_{-0.6}$ & $<0.01$ & $<0.05$ & 21.4 & 1106/1045 & 0\\
NGC~2992 & $1.25^{+0.09}_{-0.07}$ & $8.5^{+4.5}_{-4.0}$ & $<0.19$ & $0.21^{+0.44}_{-0.21}$ & 0.112 & 447/433 & 3\\
NGC~3227 & $1.57\pm 0.02$ & $9.5\pm 1.3$ & $0.53^{+0.11}_{-0.10}$ & $0.18^{+0.08}_{-0.07}$ & 0.089 & 1497/1481 & 60\\
NGC~3516 & $1.72^{+0.02}_{-0.01}$ & $0.2^{+0.6}_{-0.2}$ & $0.80\pm 0.05$ & $0.40^{+0.05}_{-0.04}$ & 2.64 & 1188/1035 & 1477\\
NGC~3783 & $1.43^{+0.02}_{-0.03}$ & $4.6\pm 0.9$ & $1.2\pm 0.1$ & $1.4\pm 0.1$ & 1.77 & 1562/1371 & 785\\
NGC~4051 & $2.04^{+0.03}_{-0.02}$ & $<0.6$ & $0.19\pm 0.06$ & $0.23^{+0.04}_{-0.05}$ & 0.0289 & 1479/1335 & 102\\
3C~273 & $1.66^{+0.01}_{-0.02}$ & $1.9^{+0.6}_{-0.5}$ & $0.01^{+0.05}_{-0.01}$ & $0.07^{+0.04}_{-0.03}$ & 1530 & 1482/1470 & 9\\
NGC~4593 & $1.97\pm 0.03$ & $2.0^{+0.5}_{-0.6}$ & $0.26\pm 0.04$ & $0.09^{+0.04}_{-0.03}$ & 1.23 & 1418/1385 & 123\\
MCG$-6-30-15$ & $1.92^{+0.02}_{-0.01}$ & $1.7^{+0.4}_{-0.3}$ & $0.64\pm 0.02$ & $0.19^{+0.02}_{-0.03}$ & 1.23 & 3020/2394 & 3336\\
IC~4329a & $1.85\pm 0.02$ & $30.8^{+0.09}_{-0.10}$ & $0.59\pm 0.07$ & $0.12\pm 0.05$ & 8.80 & 2201/1779 & 123\\
NGC~5548 & $1.88\pm 0.01$ & $<0.22$ & $0.25^{+0.04}_{-0.04}$ & $0.16\pm 0.03$ & 5.44 & 1727/1627 & 310\\
Mrk~841 & $1.64^{+0.08}_{-0.06}$ & $<10$ & $0.18^{+0.28}_{-0.18}$ & $0.03^{+0.20}_{-0.03}$ & 5.89 & 300/325 & 0\\
Mrk~290 & $1.77^{+0.03}_{-0.02}$ & $<0.60$ & $0.37\pm 0.09$ & $0.33^{+0.08}_{-0.07}$ & 3.76 & 937/1011 & 107\\
3C~382 & $2.02^{+0.06}_{-0.03}$ & $1.1^{+1.2}_{-0.8}$ & $0.16\pm 0.05$ & $0.10\pm 0.04$ & 65.3 & 1588/1446 & 41\\
3C~390.3 & $1.74^{+0.07}_{-0.04}$ & $5.8^{+1.5}_{-0.9}$ & $0.09^{+0.07}_{-0.08}$ & $0.07^{+0.07}_{-0.06}$ & 23.7 & 1303/1245 & 18\\
ESO~141-G55 & $1.90^{+0.03}_{-0.04}$ & $5.2^{+1.5}_{-1.2}$ & $<0.03$ & $<0.02$ & 9.89 & 759/719 & 0\\
NGC~6814 & $1.65^{+0.18}_{-0.09}$ & $<5.8$ & $<0.25$ & $0.05^{+0.33}_{-0.05}$ & 0.0178 & 261/252 & 0\\
Mrk~509 & $1.98\pm 0.02$ & $2.1\pm 0.6$ & $0.11^{+0.03}_{-0.04}$ & $0.04^{+0.04}_{-0.03}$ & 23.9 & 1549/1494 & 33\\
MR~2251-178 & $1.73^{+0.02}_{-0.03}$ & $<0.6$ & $0.32\pm 0.10$ & $0.15\pm 0.08$ & 93.2 & 764/840 & 83\\
NGC~7469 & $2.11^{+0.04}_{-0.03}$ & $<0.4$ & $0.17^{+0.08}_{-0.07}$ & $0.03^{+0.05}_{-0.03}$ & 3.69 & 963/980 & 13\\
Mrk~926 & $1.80^{+0.05}_{-0.04}$ & $<1.5$ & $0.06^{+0.08}_{-0.06}$ & $0.04^{+0.07}_{-0.04}$ & 13.9 & 1055/998 & 3\\\hline
\end{tabular}
\end{center}
\end{table*}

\begin{table*}
\caption{Iron emission line results from spectral fitting with model-B, as 
defined in Section 3.2 of the main text.  Given here are the best
fitting energies $E$, widths $\sigma$ (defined as the normal deviation
of the emission profile) and equivalent widths $W_{\rm Fe}$. For
ESO~141-G55 the line energy and width had to be fixed (as indicated by
the `f') in order to avoid a disasterous statistical degeneracy.  All
errors and limits are quoted at the 90 per cent confidence level for
one interesting parameter ($\Delta\chi^2=2.7$).}
\begin{center}
\begin{tabular}{lcccc}\hline
object & $E$ & $\sigma$ & $W_{\rm Fe}$ & $\Delta\chi^2$ upon allowing\\
name & (keV) & (keV) & (eV) & line to be broad\\\hline
Mrk~335 & $6.1\pm 0.4$ & $0.89^{+0.80}_{-0.34}$ & $560^{+600}_{-260}$ & 15\\
Fairall~9 & $6.37^{+0.11}_{-0.13}$ & $0.45^{+0.26}_{-0.17}$ & $380^{+150}_{-100}$ & 22\\
Mrk~1040 & $6.37^{+0.12}_{-0.10}$ & $0.25^{+0.15}_{-0.13}$ & $480^{+260}_{-130}$ & 5\\
3C~120 & $6.43^{+0.23}_{-0.24}$ & $1.5^{+0.6}_{-0.4}$ & $960^{+520}_{-270}$ & 119\\
NGC~2992 & $6.41\pm 0.02$ & $<0.06$ & $560^{+110}_{-130}$ & 0\\
NGC~3227 & $6.34\pm 0.07$ & $0.24^{+0.12}_{-0.08}$ & $210^{+80}_{-40}$ & 29\\
NGC~3516 & $6.41^{+0.05}_{-0.06}$ & $0.12^{+0.10}_{-0.06}$ & $120^{+30}_{-40}$ & 7\\
NGC~3783 & $6.42\pm 0.03$ & $0.04\pm 0.04$ & $130\pm 30$ & 1\\
NGC~4051 & $5.87^{+0.24}_{-0.23}$ & $0.82^{+0.28}_{-0.23}$ & $400^{+160}_{-130}$ &18\\
3C~273 & $6.78^{+0.27}_{-0.26}$ & $0.78^{+0.24}_{-0.20}$ & $200^{+70}_{-60}$ & 30\\
NGC~4593 & $6.43\pm 0.15$ & $0.66^{+0.22}_{-0.20}$ & $410\pm 120$ & 22\\
MCG$-6-30-15$ & $6.05\pm 0.08$ & $0.59^{+0.09}_{-0.08}$ & $220\pm 40$ & 133\\
IC~4329a & $6.39^{+0.08}_{-0.10}$ & $0.44^{+0.15}_{-0.13}$ & $270^{+100}_{-60}$ & 62\\
NGC~5548 & $6.37\pm 0.11$ & $0.49^{+0.20}_{-0.14}$ & $280^{+90}_{-70}$ & 39\\
Mrk~841 & $6.46^{+0.14}_{-0.30}$ & $0.15^{+0.42}_{-0.15}$ & $120^{+110}_{-70}$ & 1 \\
Mrk~290 & $6.27^{+0.15}_{-0.16}$ & $0.47^{+0.18}_{-0.14}$ & $420\pm 120$ & 41\\
3C~382 & $6.44^{+0.32}_{-0.44}$ & $1.8^{+1.0}_{-0.5}$ & $900^{+1042}_{-310}$ & 75\\
3C~390.3 & $6.46^{+0.17}_{-0.11}$ & $0.39\pm 0.18$ & $300^{+440}_{-90}$ & 17\\
ESO~141-G55 & $6.4^{\rm f}$ & $0.1^{\rm f}$ & $140\pm 70$ & --\\
NGC~6814 & $6.46^{+0.38}_{-0.37}$ & $<0.9$ & $430^{+800}_{-300}$ & 0\\
Mrk~509 & $6.64^{+0.24}_{-0.21}$ & $1.27^{+0.34}_{-0.31}$ & $680^{+230}_{-190}$ & 69\\
MR~2251-178 & $6.1^{+0.4}_{-0.3}$ & $0.53^{+0.52}_{-0.26}$ & $190^{+140}_{-90}$ & 8\\
NGC~7469 & $6.56^{+0.29}_{-0.26}$ & $1.19^{+0.56}_{-0.36}$ & $1000^{+400}_{-280}$ & 46\\
Mrk~926 & $6.27^{+0.29}_{-0.25}$ & $0.84^{+0.42}_{-0.45}$ & $540^{+300}_{-220}$ & 47\\\hline
\end{tabular}
\end{center}
\end{table*}

\noindent A more complex model is required in order to explore the 
various spectral features displayed by these data.  Therefore, to the
spectrum of each object was fitted a spectral model consisting of the
following components:

(i) a power-law representing the primary continuum (photon index
$\Gamma$).

(ii) Gaussian emission line representing iron K$\alpha$ emission
(energy $E$, width $\sigma$ and equivalent width $W_{\rm Fe}$ in the
rest-frame of the source).

(iii) two absorption edges with rest-frame threshold energies of
0.74\,keV and 0.87\,keV representing {\sc O\,vii} and {\sc O\,viii}
K-shell absorption (with maximum optical depths $\tau_{\rm O7}$ and
$\tau_{\rm O8}$) respectively.  These two edges provide, to a good
approximation, a description of the effects of the warm absorber over
the {\it ASCA} band.  A comparison of this two-edge model to a more
complete photoionization model will be performed in Section 5.

(iv) intrinsic absorption by a column density $N_{\rm H}$ of neutral
matter in the rest-frame of the source.  This absorbing material is
assumed to have cosmic abundances.

(v) Galactic absorption (at $z=0$) fixed at the level determined by
the {\sc H\,i} 21-cm measurements reported in Table~1.

This will be denoted as {\it model-B}.  The model was fitted
simultaneously to the background subtracted spectra from each of the
four {\it ASCA} instruments for each object of the sample.  Free
parameters in the fit were $\Gamma$, $E$, $\sigma$, $W_{\rm Fe}$,
$\tau_{\rm O7}$, $\tau_{\rm O8}$ and $N_{\rm H}$.  The normalizations
of the model for each of the four instruments were also left as free
parameters.  This allows for the $\sim 10$--20 per cent uncertainties
between the normalizations of the different instruments.

Aspects of this model require justification.  First, although we
include the fluorescent iron-line emission, the associated reflection
continuum is not included.  In the `standard' case of reflection by a
cold slab of material covering a solid angle of $2\pi$ as seen by the
X-ray source, we expect a negligible effect in the {\it ASCA} band
[e.g. see Monte-Carlo simulations of George \& Fabian (1991) and
Reynolds, Fabian \& Inoue (1995).]
Objects with a stronger reflection continuum due to, for example,
time-delayed reflection from the molecular torus after the central
source has faded (e.g. IC~4329a, Cappi et al. 1996; NGC~2992, Weaver
et al. 1996) might be expected to display an unmodelled hard tail once
they have been fitted with model-B.  Secondly, the energies of the two
absorption edges have been fixed at the physical rest-frame energies
of the K-shell absorption edges of {\sc O\,vii} and {\sc O\,viii}
(0.74\,keV and 0.87\,keV respectively) as opposed to allowing the
threshold energies to vary as free parameters in the fit.  This method
is the most robust to the various unmodelled spectral complexities
such as recombination line/continuum emission, resonance absorption
lines and other absorption edges.  These complexities 
would require many additional
degrees of freedom to model (such as the covering fraction of the warm
material, its density, the chemical abundances 
and the velocity structure) and would lead to an
over modelling of the data for all but the very highest quality {\it
ASCA} data.  In fitting a simple two-edge model, these complexities
could lead to false shifts in the threshold edge energies of the edges
(for example, see the discussion of MCG$-6-30-15$ in Section 5.2).
An independent reason for fixing the edge energies is to allow us to
compare objects that show warm absorption and those that do not show
any evidence for warm absorption (i.e. only have upper limits for the
edge depths).

After fitting this model to the sample, three objects (Mrk~335,
NGC~4051 and Mrk~841) display residues clearly demonstrating the need
for a soft-excess component.  Thus, for these objects an additional
black-body component (also subject to the same cold and ionized
absorption as the power-law continuum) is added to the model in order
to take account of the soft excess.  A more systematic study of the
presence of soft excesses in the rest of the sample will be performed
in the Section 4.4.

Table~3 reports the results for the absorption features of spectral
fitting with model-B.  Table~4 reports the iron-line results.  Model-B
(with the addition of soft excesses in the case of Mrk~335, NGC~4051
and Mrk~841) provides a good description of the data for most objects.

\section{Results}

\subsection{The primary X-ray continuum}

\begin{figure}
\hspace{-1.5cm}
\hbox{
\psfig{figure=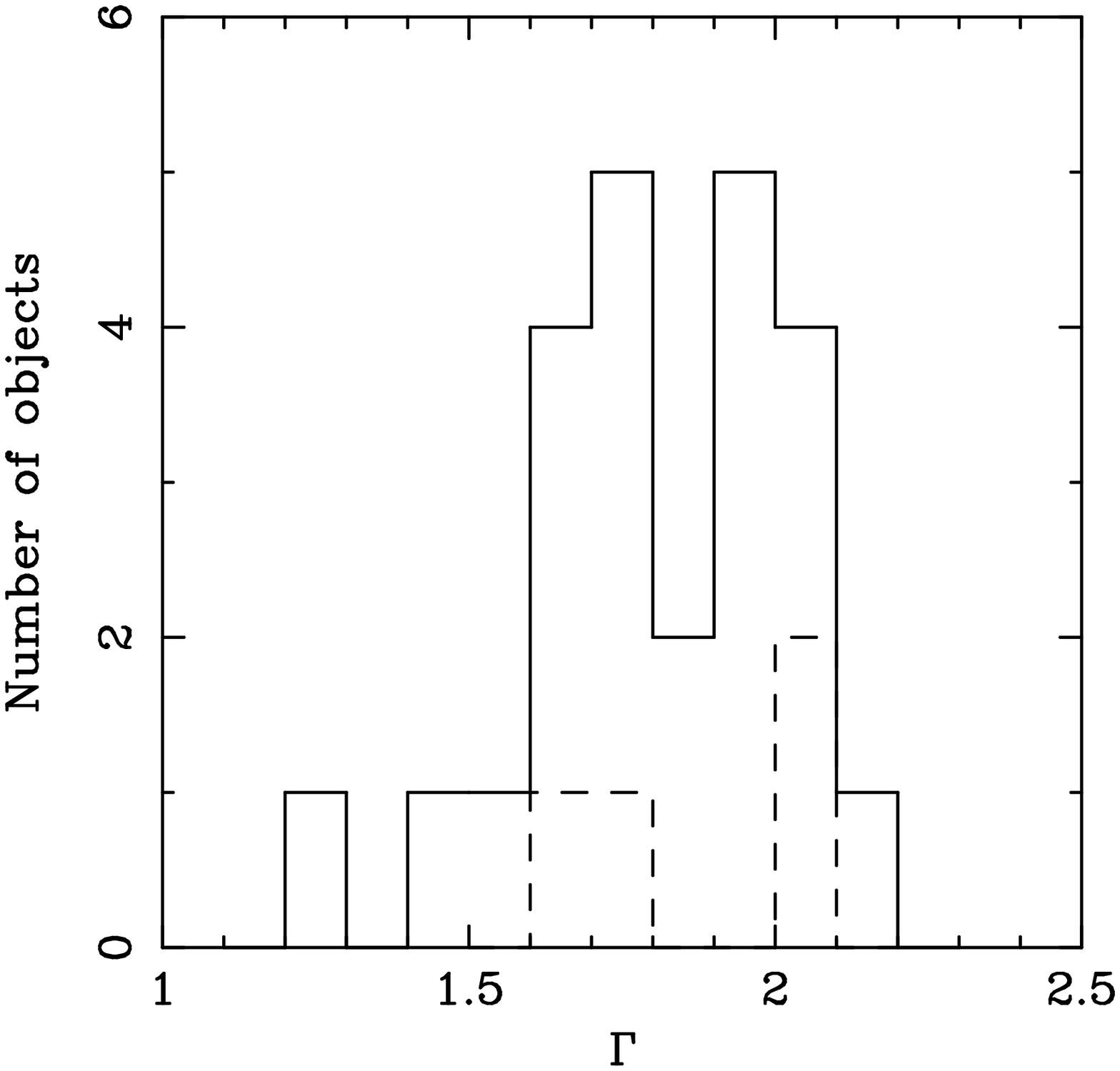,width=0.65\textwidth,angle=270}
}
\caption{Distribution of photon indices derived from model-B, 
as defined in Section 3.2 of the main text.  The solid line 
shows the distribution of the whole sample whereas the dashed lines
shows the contribution to this from the four radio-loud objects.}
\end{figure}

\begin{figure}
\hspace{-1.5cm}
\hbox{
\psfig{figure=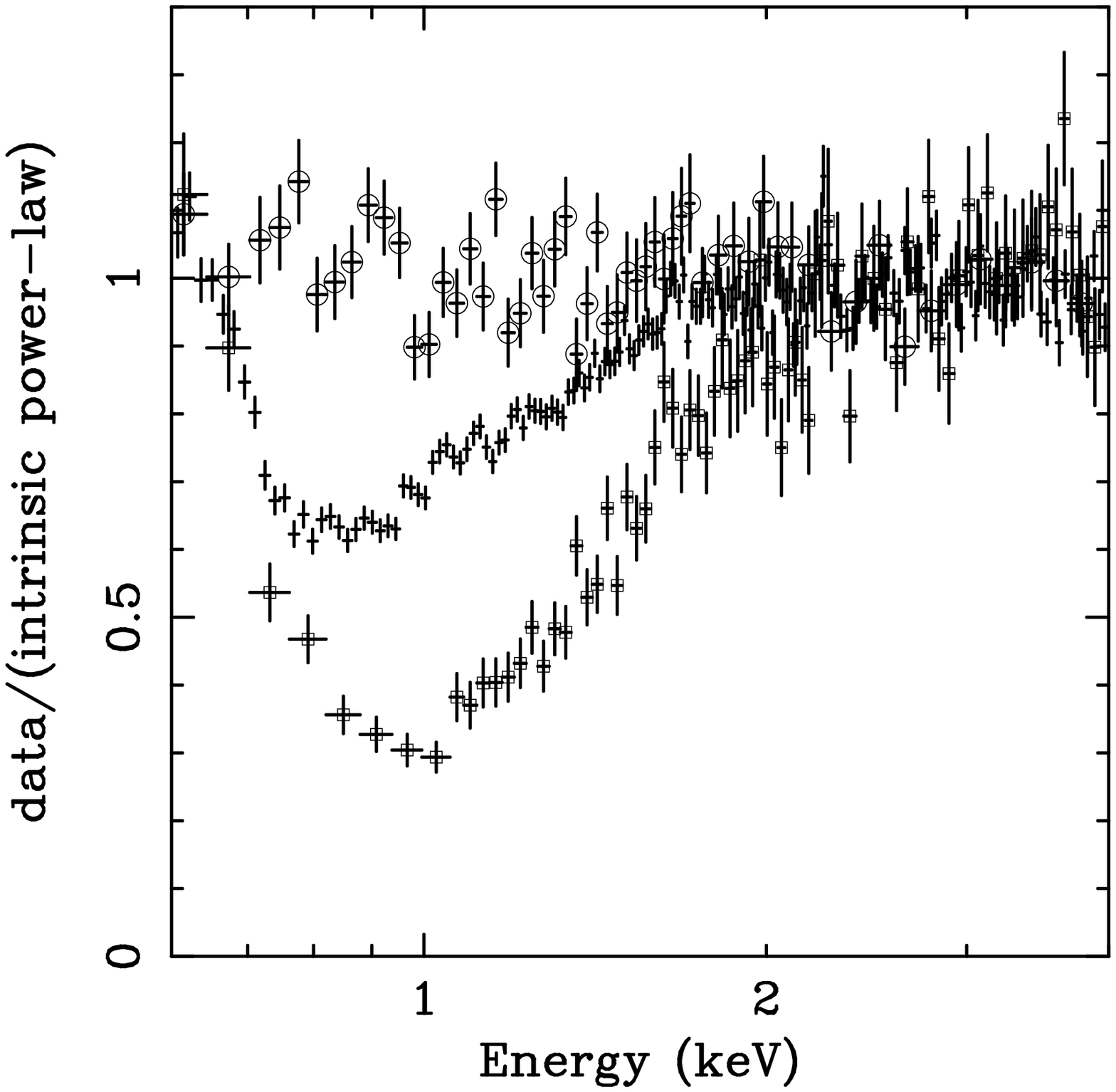,width=0.65\textwidth,angle=270}
}
\caption{Ratio of the data for three objects to the intrinsic power-law 
spectrum (modified by neutral Galactic absorption) inferred from spectral 
fitting with model-B.  The objects are Fairall~9 (top dataset, large 
circles), MCG$-6-30-15$ (middle dataset, plain crosses) and NGC~3783 
(bottom dataset, small squares).  The deep ionized oxygen edges can be 
clearly seen in both MCG$-6-30-15$ and NGC~3783.  By contrast, Fairall-9 
shows no deviation from the intrinsic spectrum (implying no soft X-ray 
reprocessing).  No correction for redshift has been applied to the
observed energies.}
\end{figure}

\noindent The photon index of the primary X-ray continuum is a crucial
observational constraint on our understanding of high-energy radiation
mechanisms associated with the inner accretion flow.  If we assume
that model-B correctly accounts for all X-ray reprocessing phenomena
relevant in the {\it ASCA} band, the best fitting value of $\Gamma$
should represent the photon index of the primary continuum.  Figure~2
shows the distribution of derived photon indices for this sample.  The
photon index distribution for the sample as a whole has mean
$\langle\Gamma\rangle=1.81$ and dispersion $\sigma_{\Gamma}=0.21$.
The distributions for the radio-loud and radio-quiet sub-samples have
$\langle\Gamma\rangle=1.88$ (dispersion $\sigma_{\Gamma}=0.18$) and
$\langle\Gamma\rangle=1.80$ (dispersion $\sigma_{\Gamma}=0.21$)
respectively.  Thus the photon indices of the radio-loud and
radio-quiet sub-samples do not appear to be significantly different,
although the number of radio-loud objects is small.

\subsection{Ionized absorption}

\subsubsection{Ubiquity of the warm absorber}

An important result of the spectral fitting reported in Table~3 is the
frequency with which statistically-significant ionized absorption is
detected (as indicated by statistically-significant oxygen absorption
edges).  12 out of the 24 objects show a significant improvement in
the goodness of fit (according to the F-test using a 99 per cent
confidence level; Bevington 1969) when the two oxygen edges are
included.  Such a harsh statistical criterion was used to protect
against any small calibration uncertainties.  These objects are listed
in Table~6 (see Section 5 for a full discussion of Table~6).  This
result is robust to neglecting all data below 0.6\,keV.  The result is
also unaffected if we include the possibility of a black body soft
excess in the spectral fitting (see Section 4.4).  The effect of the
ionized absorption is demonstrated in Fig.~3.  Here we show the ratio
of the data to the inferred underlying power-law continuum (corrected
for neutral absorption) for Fairall-9 (which displays no significant
warm absorption), MCG$-6-30-15$ (which shows strong evidence for
ionized absorption) and NGC~3783 (which shows the deepest ionized
absorption edges of all the sources in the present sample).  Note the
sharp onset of absorption at $\sim 0.7\keV$ and the gradual recovery
over the 1--2\,keV range.

This sample has sufficient sources to search for general trends.
Figure~4 shows the warm oxygen edge depths as a function of luminosity
for the sample.  Whilst there is no clear correlation of $\tau_{\rm
O7}$ or $\tau_{\rm O8}$ with luminosity, there does appear to be a
luminosity-dependent upper envelope to these distributions such that
there are no high-luminosity systems displaying deep edges.  To
formally address this, the full sample of 24 objects was divided into
a high luminosity half and a low luminosity half (on the basis of the
2--10\,keV luminosities derived from model-B fits and reported in
Table~3).  The distribution of $\tau_{\rm O7}$ and $\tau_{\rm O8}$ in
each of these subsamples were compared using a KS test.  The high-$L$
and low-$L$ distributions of $\tau_{\rm O7}$ are found to be
inconsistent at the 95 per cent level (i.e. there is only a 5 per cent
probability that the two distributions were drawn from the same parent
distribution).  A much stronger result is obtained for the high-$L$
and low-$L$ distributions of $\tau_{\rm O8}$: these are found to be
inconsistent at the 99.5 per cent level (i.e. only a 0.5 per cent
probability that the two distributions were drawn from the same parent
distribution).

Care must be exercised in the physical interpretation of this result
since the four radio-loud objects are amongst the most luminous of the
sample.  Performing the same statistical analysis on the radio-quiet
subsample only, no significant relationship between the ionized edge
depths and luminosity is found.  Thus, the underlying dependence is
unclear: the apparent luminosity relationship of the previous
paragraph may be a manifestation of a true dependence of the ionized
absorption on radio-loudness.  The present dataset is too small to
disentangle the effects of any true luminosity dependence from any
dependence based on radio properties.  A detailed spectral examination
of low-luminosity radio-loud objects or high luminosity radio quiet
objects would be necessary to settle this issue.  

\subsubsection{Ionized absorption and the optical/UV properties}

\begin{figure*}
\hspace{-1.5cm}
\hbox{
\psfig{figure=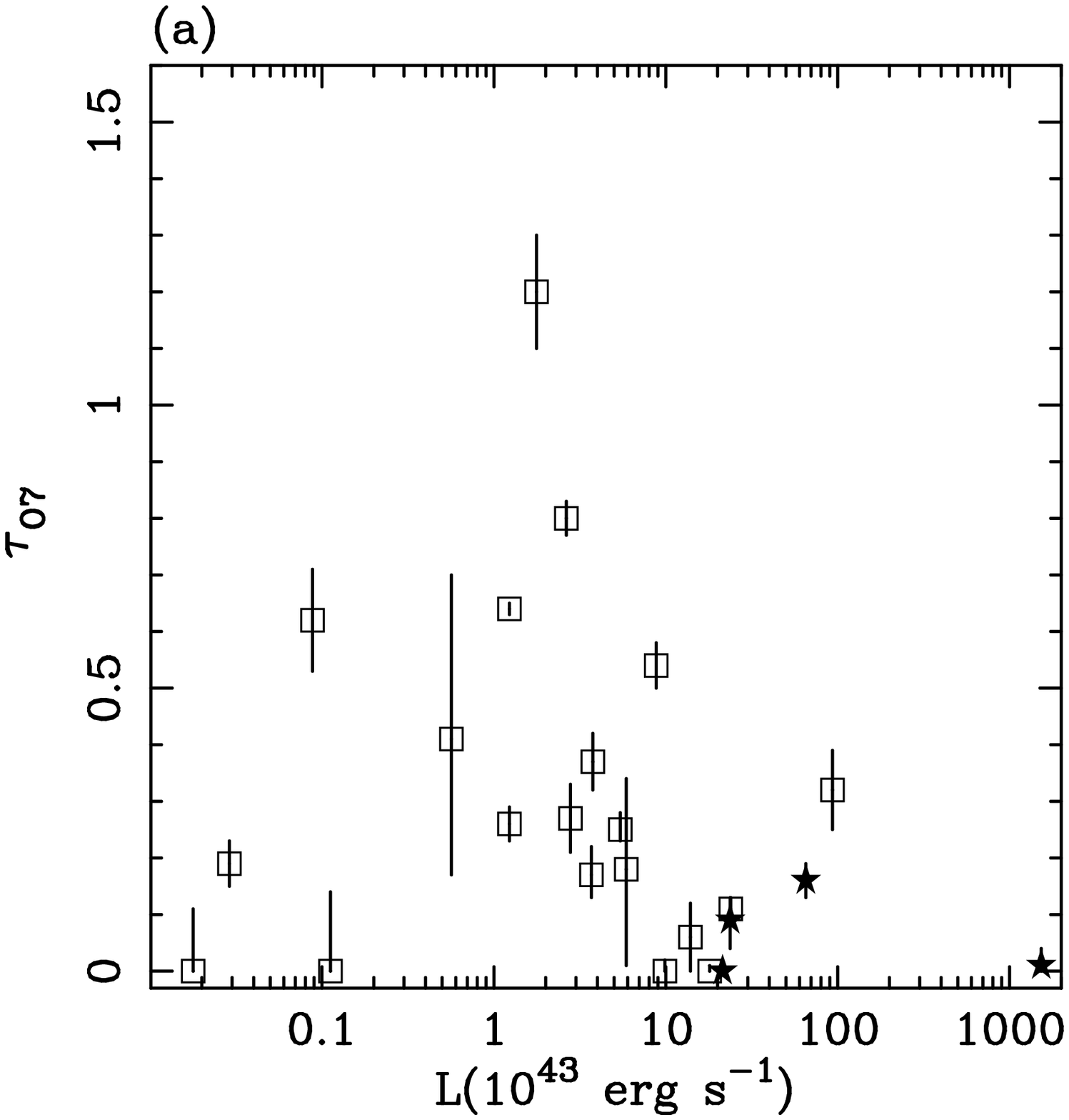,width=0.65\textwidth,angle=270}
\hspace{-3cm}
\psfig{figure=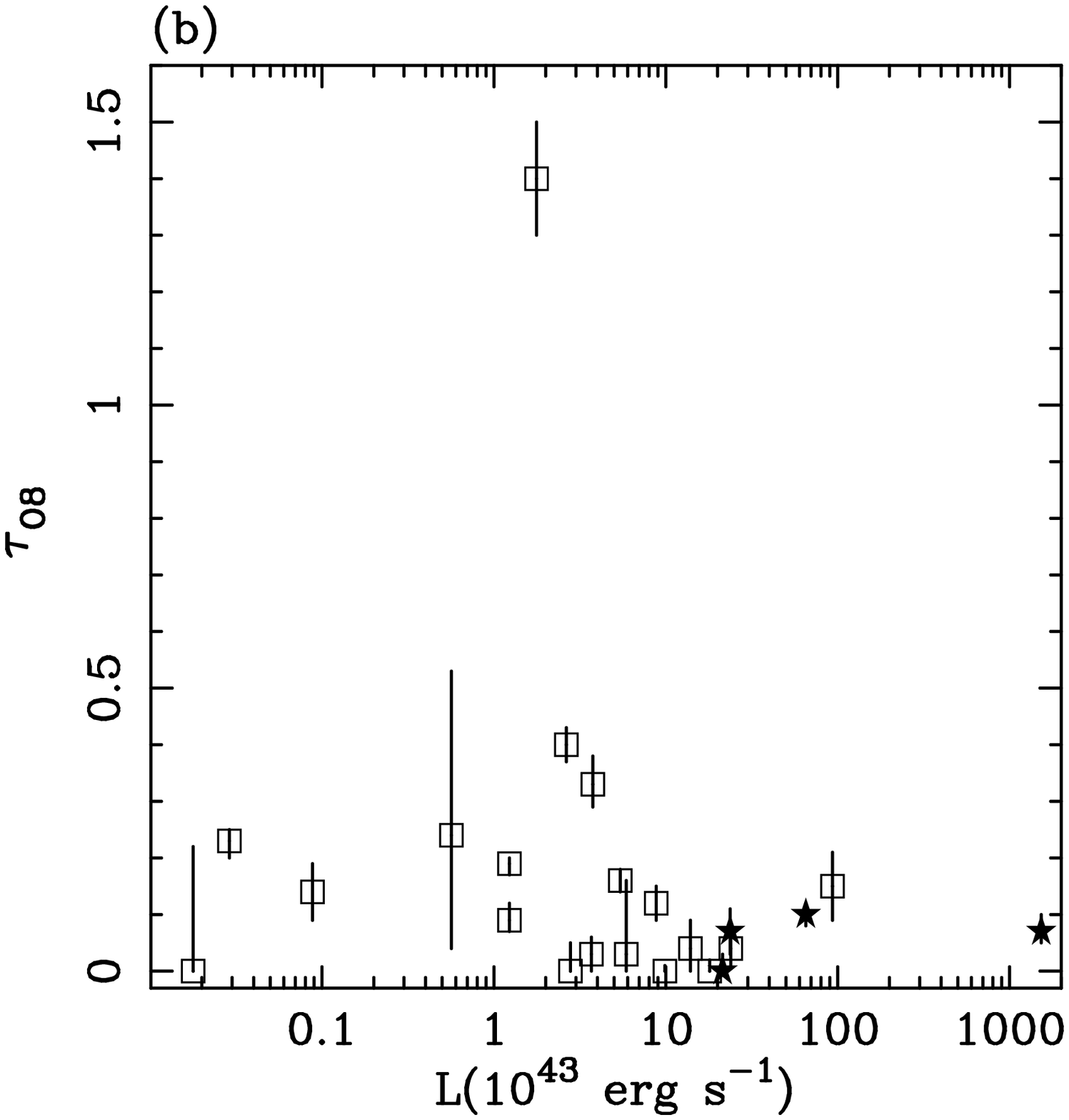,width=0.65\textwidth,angle=270}
}
\caption{Maximum oxygen edge depth ($\tau_{\rm O7}$ and $\tau_{\rm O8}$) 
as a function of (unabsorbed) 2--10\,keV 
X-ray luminosity, $L$.  Squares represent data points for radio-quiet objects 
whereas stars show the radio-loud data points.   Panel (a) reports 
the threshold optical depth of the {\sc O\,vii} edge as a function
of $L$.  Whilst there 
is no formal correlation present, there is an dearth of high 
luminosity systems also showing a large {\sc O\,vii} edge.   This is 
discussed in the text.   Panel (b) shows the threshold optical depth 
of the {\sc O\,viii} edge as a function of $L$.   A similar dependence
is seen. Errors are displayed at the 1-$\sigma$ level for one
interesting parameter.}
\end{figure*}

\begin{figure*}
\hspace{-1.5cm}
\hbox{
\psfig{figure=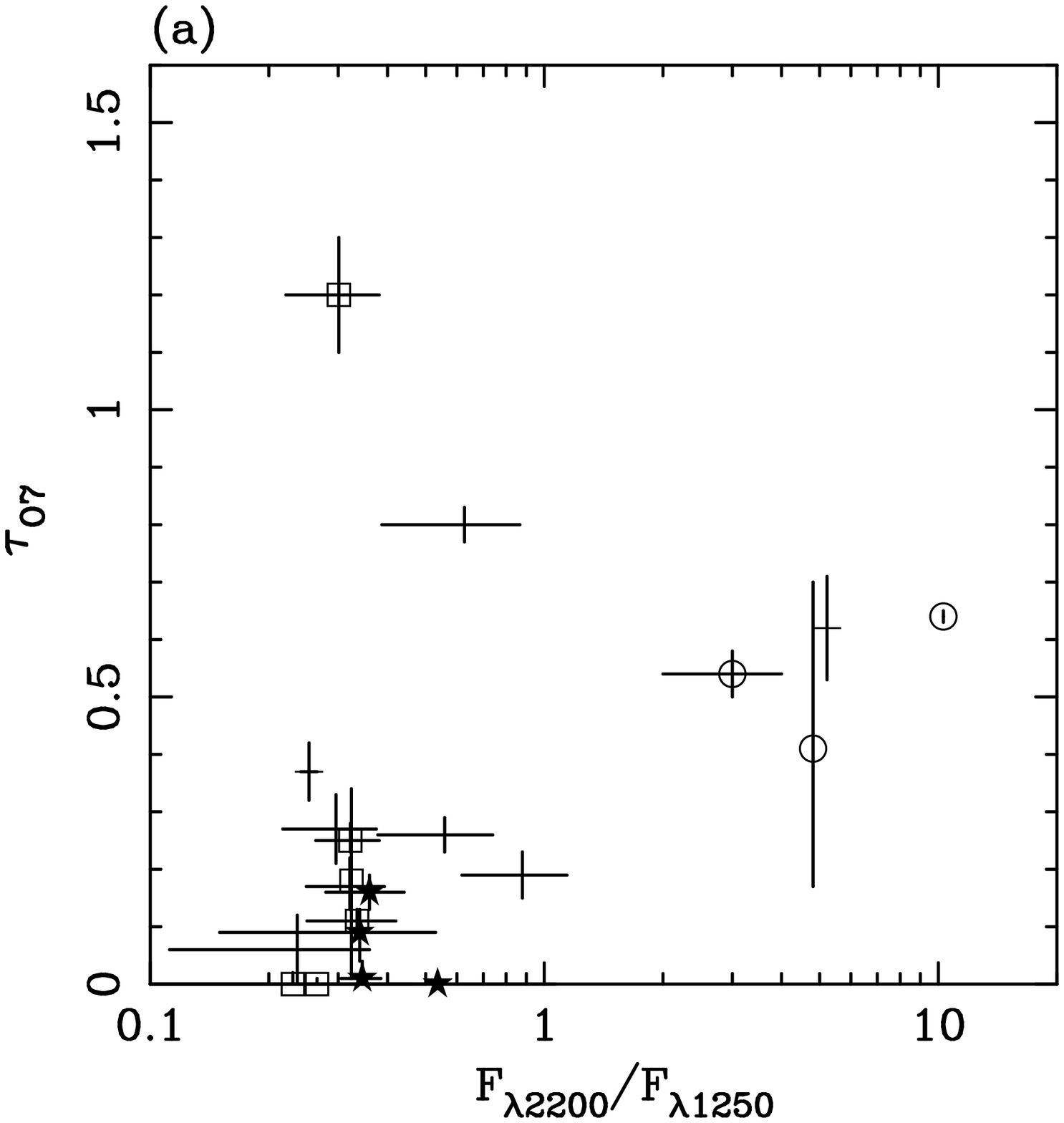,width=0.65\textwidth,angle=270}
\hspace{-3cm}
\psfig{figure=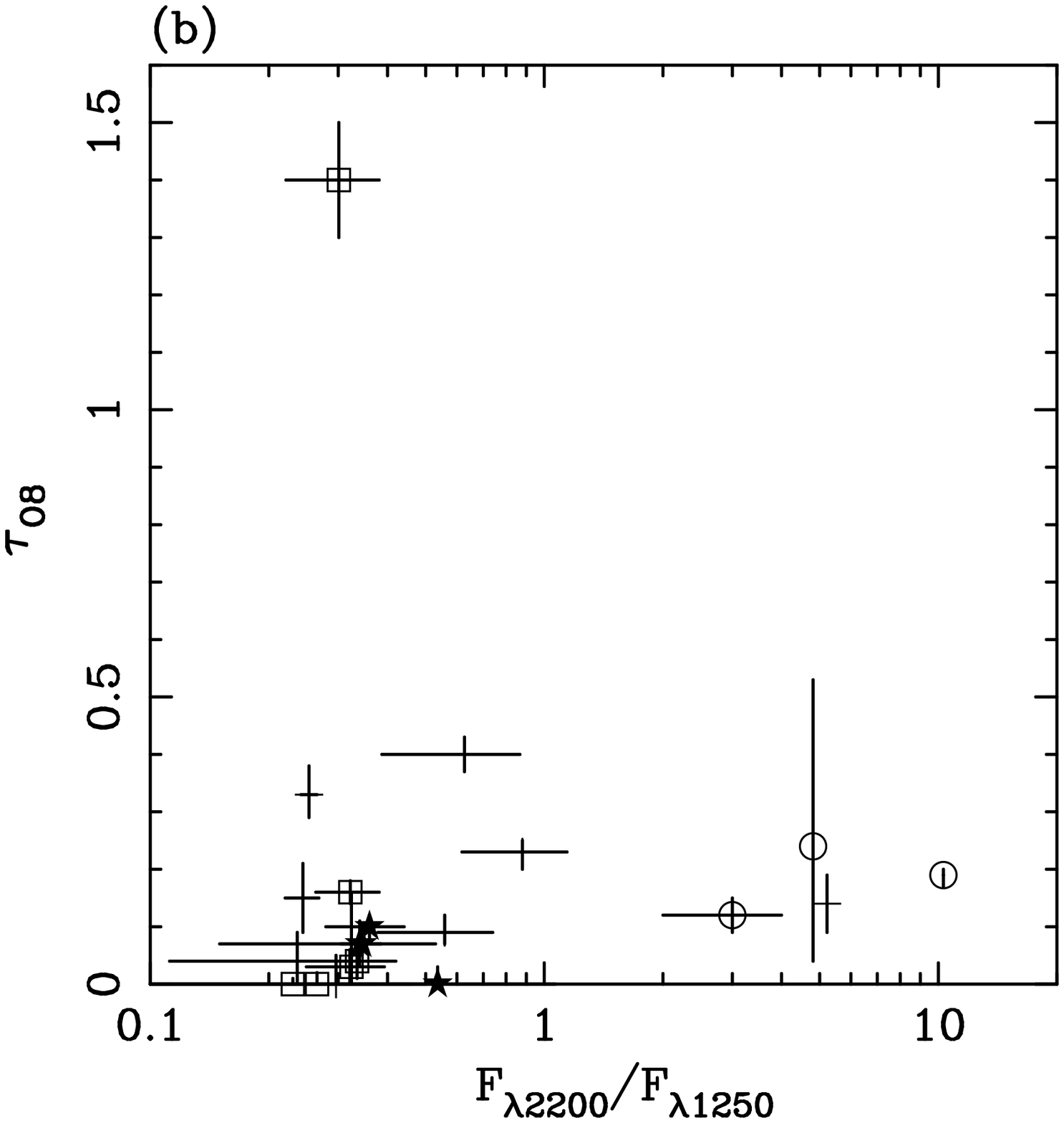,width=0.65\textwidth,angle=270}
}
\caption{Maximum oxygen edge depth ($\tau_{\rm O7}$ and $\tau_{\rm O8}$) 
as function of a reddening indicator, $X$, given by the ratio of the
continuum flux at 2200\AA\ and 1250\AA.  Squares show those objects
classifed by Ward et al. (1987) to be `bare' (unreddened) whereas
circles denote those objects classified as heavily reddened.  Note the
agreement between Ward et al. and the reddening indicator used here.
Stars show the data for the radio-loud objects of the sample.  Trends
are discussed in the text.  Errors are shown at the 1-$\sigma$ level
for one interesting parameter.}
\end{figure*}

\noindent Recent evidence has emerged linking the ionized absorption 
seen in the X-ray spectrum of AGN with optical/UV characteristics of
the objects.  Firstly, it has been suggested that the UV absorption
lines seen in the broad line profiles of many AGN arise from the same
ionized material as do the warm absorption features (Mathur et
al. 1994; Mathur 1994; Mathur, Elvis \& Wilkes 1995).  Given this
identification, the velocity structure of the absorbing material can
be accurately measured from the UV absorption lines.  Mathur, Elvis \&
Wilkes (1995) examine the Seyfert 1 galaxy NGC~5548 and deduce that
the absorber forms an outflow with a velocity of $1200\pm 260\kmps$.

There is also strong evidence that at least some warm absorbers are
intrinsically dusty leading to a connection between optical/UV
reddening and ionized absorption.  The two strongest cases are
MCG$-6-30-15$ (Reynolds \& Fabian 1995) and IRAS~13349$+$2438 (Brandt,
Fabian \& Pounds 1996).  Both of these objects display significant
optical reddening ($E(B-V)\approxgt 0.3$) which, assuming a Galactic
dust-to-gas ratio implies a line-of-sight gas column density in excess
of $2\times 10^{21}\pcmsq$ (Burstein \& Heiles 1978).  However, both
of these objects have tight X-ray limits on the amount of neutral gas
which are inconsistent with the reddening value by an order of
magnitude or more (i.e. the X-ray limits are $N_{\rm H}\approxlt
2\times 10^{20}\pcmsq$).  A resolution of this apparent discrepancy,
which avoids postulating ad-hoc geometries or exceptionally high
dust-to-gas ratios, is to place the dust responsible for the reddening
in the warm ionized material inferred from the oxygen edges seen in
the X-ray spectrum of these two objects.

Dust can survive in warm ionized plasma given that two conditions are
satisfied.  Firstly, the {\it gas} must be below the temperature at
which sputtering will destroy the dust ($T\sim 10^6\K$).  Since the warm gas
is photoionization dominated, the required high ionization states can
easily be achieved without exceeding this temperature (see Section 5
for photoionization models).  Turning this argument around, we can
cite observations of highly-ionized dusty gas as supprting evidence
for the hypothesis that this gas is photoionization dominated.
Secondly, the {\it dust} must be below its sublimation temperature
($T\sim 1000$--2000\,K depending on the exact dust grain type under
consideration).  It is important to note that the grain temperature
can differ greatly from the gas temperature due to the tenuous nature
of the gas.  In fact, to a good approximation, the grains near the
central engine of an AGN will be in thermal equilibrium with the
primary radiation field.  Assuming that each grain radiates as a black
body, the temperature, $T$, of the grains can be estimated by equating
the radiative flux incident on the grain from the primary source with
the black body radiation emitted by the grain.  This gives
\begin{equation}
T=\left(\frac{L_{\rm B}}{16\pi\sigma R^2}\right)^{1/4},
\end{equation}
where $\sigma$ is the Stefan-Boltzmann constant and $R$ is the
distance from an (isotropic) source with bolometric luminosity $L_{\rm
B}$.  If $L_{\rm B}=10^{44}L_{44}\ergps$ and $R=R_{\rm pc}\pc$, this
evaluates to give
\begin{equation}
T=250\,L_{44}^{1/4}\,R_{\rm pc}^{-1/2}\K
\end{equation}
Thus, for typical Seyfert luminosities ($L_{44}\sim 1$), dust can
survive in the warm gas provided it is more than $\sim 10^{17}\cm$
from the primary source.

Motivated by these studies, it is interesting to search for a
relationship between the optical/UV reddening and ionized absorption.
A difficulty in such a study is to determine the reddening in a
uniform manner across the sample.  Determining reddening via Balmer
decrements suffers from major uncertainties in the value of the
intrinsic Balmer decrement of the various emission line regions.
Radiative transfer effects can severely modify the Balmer decrements
predicted from the simple case-B recombination (Kwan \& Krolik 1981)
values.  Furthermore, determining reddening by comparing two wildly
different wavebands (for example, UV and hard X-ray bands) is a poor
technique for such a sample due to the lack of simultaneous
multiwavelength observations of most sources and the variable nature
of these sources.  Here, I use a reddening indicator, $X$, defined as
the ratio of the continuum flux at 2200\AA, $F_{\lambda 2200}$, to
that at 1250\AA, $F_{\lambda 1250}$,
\begin{equation}
X=\frac{F_{\lambda 2200}}{F_{\lambda 1250}}
\end{equation}
as determined by the International Ultraviolet Explorer ({\it IUE})
and reported in the AGN compilation of Courvoisier \& Paltani (1992).
Only quasi-simultaneous (i.e. same day) measurements of these two
fluxes were considered in order to minimise the effects of source
variability.  For objects with more than one such quasi-simultaneous
measurement, the mean and standard deviation of the various values of
$X$ were computed.  For a given primary continuum shape, reddening
will tend to increase $X$.

Figure~5 shows $\tau_{\rm O7}$ and $\tau_{\rm O8}$ as a function of
the reddening indicator, $X$.  To check that $X$ does indeed
distinguish reddened objects from unreddened objects, the
classifications of Ward et al. (1987) have been used.  It can be seen
from Fig.~5 that those objects classified as reddened by Ward et
al. (shown as open circles) have $X\approxgt 3$ whereas those
classified as unreddened (shown as open squares) have $X\approxlt0.4$.
Thus, $X$ appears to be an excellent indicator of the reddening.

Trends are seen in the $\tau_{\rm O7}$-X plot (Fig.~5a).  In
particular, three classes of objects can be distinguished: unreddened
objects with small {\sc O\,vii} edges, reddened objects with
appreciable {\sc O\,vii} edges and two outlyers (NGC~3516 and
NGC~3783) which have a small reddening and a very large {\sc O\,vii}
edge.  To formally assess this trend, the sample was divided into an
unreddened subsample (with $X<1$) and a reddened subsample ($X>1$).
The distributions of $\tau_{\rm O7}$ for these two subsamples are
inconsistent (according to a KS test) at more than the 99 per cent
level.  This result is independent of whether the radio-loud objects
are included or not.  No strong trends are seen in the $\tau_{\rm
O8}$-X plot (Fig.~5b) except that all reddened objects have non-zero
{\sc O\,viii} edge depths.

The implications of these results for AGN models are discussed in
Section 8.

\subsection{Iron-line emission}

All objects in the current sample show a statistically significant
improvement in the goodness of fit upon the addition of a Gaussian
emission line at energies characteristic of iron K$\alpha$ line
emission.  The resulting line properties for each object are shown in
Table~4.  20 out of the 24 objects show a significantly broadened line
(i.e. if the line is initially fixed to be narrow and the width is
then introduced as a free parameter, 20 out of 24 objects demonstrate
a significant improvement in the goodness of fit according to the
F-test).  In most cases this can be interpreted as fluorescent iron
K$\alpha$ emission from the inner regions of an accretion disk: this
hypothesis has been dramatically strengthened by the accurate line
profile obtained with a long {\it ASCA} observation of MCG$-6-30-15$
(Tanaka et al. 1995; Fabian et al. 1995) which matches the model
prediction.

Given that the broad iron emission originates from the central most
regions of the accretion flow, comparing any such emission from
radio-loud and radio-quiet nuclei should give important clues as to
the fundamental differences between these two classes of objects.
Although the number of radio-loud objects in the present sample is
small (only 4), some interesting trends are suggested.  For the
radio-quiet objects alone, the distribution of line properties has
$\langle E\rangle=6.33\keV$ (dispersion $\sigma_E=0.18\keV$),
$\langle\sigma\rangle=0.50\keV$ (dispersion $\sigma_\sigma=0.37\keV$)
and $\langle W_{\rm Fe}\rangle=380\eV$ (dispersion $\sigma_W=220\eV$).
The corresponding quantities for the four radio-loud objects are
$\langle E\rangle=6.53\keV$ (dispersion $\sigma_E=0.15\keV$),
$\langle\sigma\rangle=1.12\keV$ (dispersion $\sigma=0.56\keV$) and
$\langle W_{\rm Fe}\rangle=590\eV$ (dispersion $\sigma_W=340\eV$).
Whilst the number of objects are small there is a clear trend for
radio-loud objects to have features which are broader than those in
radio-quiet objects.  There may also be a suggestive trend of higher
centroid energies and larger equivalent widths in the radio-louds.
This maybe related to the presence of a relativistic jet in the radio
loud objects (3C~120 and 3C~273 show superluminal motion of knots
within the radio jet; Cohen et al. 1977).

\subsection{Soft excesses}

\begin{table}
\caption{Best-fitting soft excess parameters for objects showing a 
statistically sugnificant improvement (at the 99 per cent level) upon 
addition of a black body component.  Column 2 gives the best-fitting 
black body temperature with errors shown at the 90 per cent level 
for one interesting parameter.  Column 3 reports the total (bolometric) 
luminosity of this besting fitting black body and Column 4 shows the 
ratio of this bolometric luminosity to the 2--10\,keV X-ray luminosity 
(from Table~3).} 
\begin{center}
\begin{tabular}{lccc}\hline
source & blackbody $T$ & blackbody $L_{\rm B}$ & $L_{\rm B}/L_{2-10}$ \\
name & (keV) & ($10^{43}\ergps$) & \\\hline
Mrk~335 & $0.14^{+0.02}_{-0.01}$ & 1.3 & 0.47 \\
NGC~3516 & $0.08\pm 0.01$ & 1.2 & 0.45 \\
NGC~4051 & $0.12\pm 0.01$ & 0.023 & 0.80 \\
NGC~5548 & $0.24^{+0.03}_{-0.02}$ & 0.13 & 0.024 \\
Mrk~841 & $0.11^{+0.07}_{-0.04}$ & 1.3 & 0.22\\\hline
\end{tabular}
\end{center}
\end{table}

\noindent Soft excesses are clearly seen in the data for Mrk~335, NGC~4051 and
Mrk~841, as mentioned above.  A thorough investigation of soft
excesses in the current sample is hampered by the following issue:
soft excesses are expected to be most noticeable below $\sim 0.6\keV$.
However, this is the energy below which SIS calibration uncertainties
exist.  In principle, the most powerful technique for examining soft
excesses would be simultaneous fitting of {\it ROSAT} PSPC and SIS
data.  Unfortunately, such joint spectral fitting suffers from the
problems of non-simultaneous observations and cross calibration
problems.

For these reasons a detailed study of soft excesses was not performed.
However, for completeness, a black-body component (subject to neutral
and ionized absorption) was added to model-B and the data were
refitted.  To avoid being biased by calibration uncertainties below
0.6\,keV, all data below this energy were discarded.  Each source fell
into one of three catagories:

(A) {\it No significant soft excess (17 objects) : }For most sources
the addition of the extra black body component led to no significant
statistical improvement in the quality of the fit (according to the
F-test using a 99 per cent confidence threshold; Bevington 1969).
Such a harsh significance level was set to protect against any small
calibration uncertainties affecting this result.  Furthermore, the
addition of the black-body component did not affect the best-fitting
values of the other model-B parameters.  Sources in this catagory are
Fairall~9, Mrk~1040, 3C~120, NGC~2992, 3C~273, NGC~4593,
MCG$-6-30-15$, IC~4329a, Mrk~290, 3C~382, 3C~390.3, ESO~141-G55,
NGC~6814, Mrk~509, MR~2251$-$178, NGC~7469 and Mrk~926.

(B) {\it Statistically significant soft excess (5 objects) : }For most
of the remaining sources, a small soft excess was inferred by fitting
of the black body component.  Sources in this catagory are Mrk~335,
NGC~3516, NGC~4051, NGC~5548 and Mrk~841.  These soft excesses signal
a gentle change in spectral index below 1\,keV and do not affect the
other fit parameters (in particular $\Gamma$, $\tau_{\rm O7}$ and
$\tau_{\rm O8}$) significantly.  Details of the black-body fits are
reported in Table~5.  Note that the total luminosity of the black body
component from these fits is always less than the 2--10\,keV luminosity.
This is in agreement with expectations based on reprocessing models for
soft excess components.

(C) {\it A dramatically different model is inferred by inclusion of a
black-body component (2 objects) : }For NGC~3227 and NGC~3783, a {\it
qualitatively} different spectral model results from the addition of
the black body component.  In both cases a strong and very soft excess
produces a large change of spectral slope such as to mimic the effect
of one of the oxygen edges.  Including the slightly less well
calibrated soft end of the SIS range (0.4--0.6\,keV) strongly argues
against these soft excess models: the spectral models strongly diverge
(factor of 5--10) from the data over this small energy range by an
amount far in excess of any calibration uncertainty.  Thus, we reject
these soft excess models in favour of pure warm absorber models.

\section{One-zone photoionization models}

\begin{table*}
\caption{Results from spectral fitting with a one-zone photoionization
model (model-C).  Column 2 gives the best fitting photon index.
Column 3 gives the best fitting column density of intrinsic neutral
absorbing material (placed at the reshift of the source).  Columns 4
and 5 give the best-fitting column density and ionization parameter
$\xi$ of the warm photoionized plasma.  Column 6 reports the goodness
of fit parameter.  Errors and limits are shown at the 90 per cent
confidence level for one interesting parameter ($\Delta\chi^2=2.7$).}
\begin{center}
\begin{tabular}{lcccccl}\hline
source & Photon Index & Intrinsic $N_{\rm H}$ & $N_{\rm W}$ & $\xi$ & $\chi^2$/dof \\ 
name & $\Gamma$ & ($10^{20}\pcmsq$) & ($10^{20}\pcmsq$) & (erg\,cm\,s$^{-1}$) &  \\\hline
NGC~3227 & $1.49\pm 0.02$ & $<3$ & $36^{+5}_{-8}$ & $17^{+8}_{-4}$ & 1390/1436  \\
NGC~3516 & $1.86^{+0.05}_{-0.02}$ & $<1.7$ & $100^{+7}_{-5}$ & $30\pm 1$ & 1251/1035  \\
NGC~3783 & $1.66\pm 0.01$ & $4.7^{+1.9}_{-1.6}$ & $204\pm 5$ & $45^{+3}_{-2}$ & 1503/1371  \\
NGC~4051 & $1.89^{+0.04}_{-0.02}$ & $<0.7$ & $11^{+7}_{-4}$ & $10^{+10}_{-4}$ & 1538/1336 \\
NGC~4593 & $1.93^{+0.05}_{-0.03}$ & $1.0^{+1.0}_{-0.8}$ & $26^{+10}_{-5}$ & $23^{+7}_{-5}$ & 1416/1385 \\
MCG$-6-30-15$ & $1.97\pm 0.01$ & $0.5^{+0.5}_{-0.4}$ & $57^{+3}_{-2}$ & $21\pm 1$ & 2865/2394  \\
IC~4329a & $1.76\pm 0.02$ & $26\pm 1$ & $25^{+5}_{-2}$ & $9^{+1}_{-3}$ & 2232/1779  \\
NGC~5548 & $1.89^{+0.02}_{-0.01}$ & $<0.1$ & $51^{+4}_{-5}$ & $35\pm 2$ & 1792/1627  \\
Mrk~290 & $1.80^{+0.10}_{-0.07}$ & $1.1^{+2.4}_{-1.1}$ & $89^{+3}_{-2}$ & $45^{+9}_{-10}$ & 958/1011 \\
3C~382 & $2.00^{+0.12}_{-0.02}$ & $1.3^{+2.2}_{-1.0}$ & $32^{+12}_{-8}$ & $34^{+6}_{-7}$ & 1612/1446  \\
Mrk~509 & $1.90^{+0.06}_{-0.03}$ & $1.4^{+1.3}_{-0.7}$ & $26^{+14}_{-10}$ & $48^{+11}_{-16}$ & 1559/1494  \\
MR~2251-178 & $1.70^{+0.05}_{-0.03}$ & $<1.3$ & $50^{+20}_{-10}$ & $43^{+12}_{-8}$ & 777/840  \\\hline
\end{tabular}
\end{center}
\end{table*}

\noindent Photoionization is believed to dominate the physics of 
the warm absorbing material.  This belief results from the very high
and variable ionization state.  Quantitative evidence for this comes
from the observed anti-correlation between the {\sc O\,viii}
absorption edge depth and the primary ionizing flux observed during
the long observation of MCG$-6-30-15$ (Otani et al. 1996).  Many
authors have addressed the issue of warm absorption by fitting
spectral models constructed with a standard photoionization code such
as {\sc cloudy} (Ferland 1991).  Such codes solve the equations of
thermal and ionization equilibrium for the situation in which ionizing
radiation (with a given incident flux and spectral form) passes
through a slab of Thomson-thin matter.  Most authors consider only
one-zone models in the sense that the absorbing material is taken to
be a slab with uniform properties throughout.  These models are then
characterised by the column density of the ionized plasma, $N_{\rm
W}$, and the ionization parameter, $\xi$, defined by
\begin{equation}
\xi:=\frac{L}{nR^2}
\end{equation}
where $n$ is the number density of the warm plasma and $R$ is the
distance of the slab of plasma from an ionizing source of radiation
with isotropic ionizing luminosity $L$.

A caution is in order.  Spectral variability of MCG$-6-30-15$ strongly
argues for a multi-zone absorber (see Section 6.2.1 and Otani et
al. 1996).  The simultaneous UV/X-ray studies of NGC~3516 by Kriss et
al. (1996a,b) also imply the existence of a complex stratified ionized
absorber.  Thus, one-zone photoionization models can only be regarded as
a useful parameterisation of the spectral data and the physical
quantities resulting from spectral fitting of such models cannot be
regarded as measurements of true physical quantities.  For this
reason, I have not phrased the present discussion of ionized absorbers
in terms of $N_{\rm W}$ and $\xi$.  Instead, I have concentrated on
the phenomenological two-edge parameterisation of Sections 3 and 4.

For completeness (and to provide a useful comparison with other work),
those objects displaying statistically significant {\sc O\,vii} and/or
{\sc O\,viii} absorption edges were examined using {\sc cloudy} models
of the type described above.  The model is the same as model-B except
that the warm absorber is modeled using the photoionization model
rather than the two-edge parameterisation.  This is denoted as
model-C.  In detail, {\sc cloudy} was used to construct a
three-parameter grid of models with $\Gamma$, $N_{\rm W}$ and $\xi$
taken to be the variables.  The absorber was modelled as a
geometrically-thin slab at a distance of $10^{16}\cm$ from a point
source of radiation with ionizing luminosity $10^{43}\ergps$ (with a
power-law spectrum extending from 13.6\,eV to 40\,keV with photon
index $\Gamma$).  Models are then computed for various values of the
absorber electron number density, $n$, column density, $N_{\rm W}$,
and photon index, $\Gamma$.  Since the state of such photoionized gas
is dominated by absorption of EUV and soft X-ray photons, this
simplistic input spectrum will suffice for the current application in
which we are interested in gross features of the soft X-ray opacity.
The ionization parameter $\xi$ was then computed from equation (4).
For a very large range of parameter space, the effects of differing
$n$ are completely described by the ionization parameter $\xi$.  The
results of these fits are shown in Table~6.

It is interesting to compare the detailed fits of the simple two-edge
model (model-B) with those of the one-zone photoionization model
(model-C).  Several interesting points can arise from such a
comparison.  First, the assertion that the two-edge model provides a
good description of the warm absorber can be tested.  Secondly,
spectral structure beyond the two-edge model can be addressed.
Thirdly, failures of the one-zone models can be probed.  Here, the
warm absorber fits for two objects (NGC~3783 and MCG$-6-30-15$) are
examined in detail.

\subsection{NGC~3783}

\begin{figure*}
\hspace{-1.5cm}
\hbox{
\psfig{figure=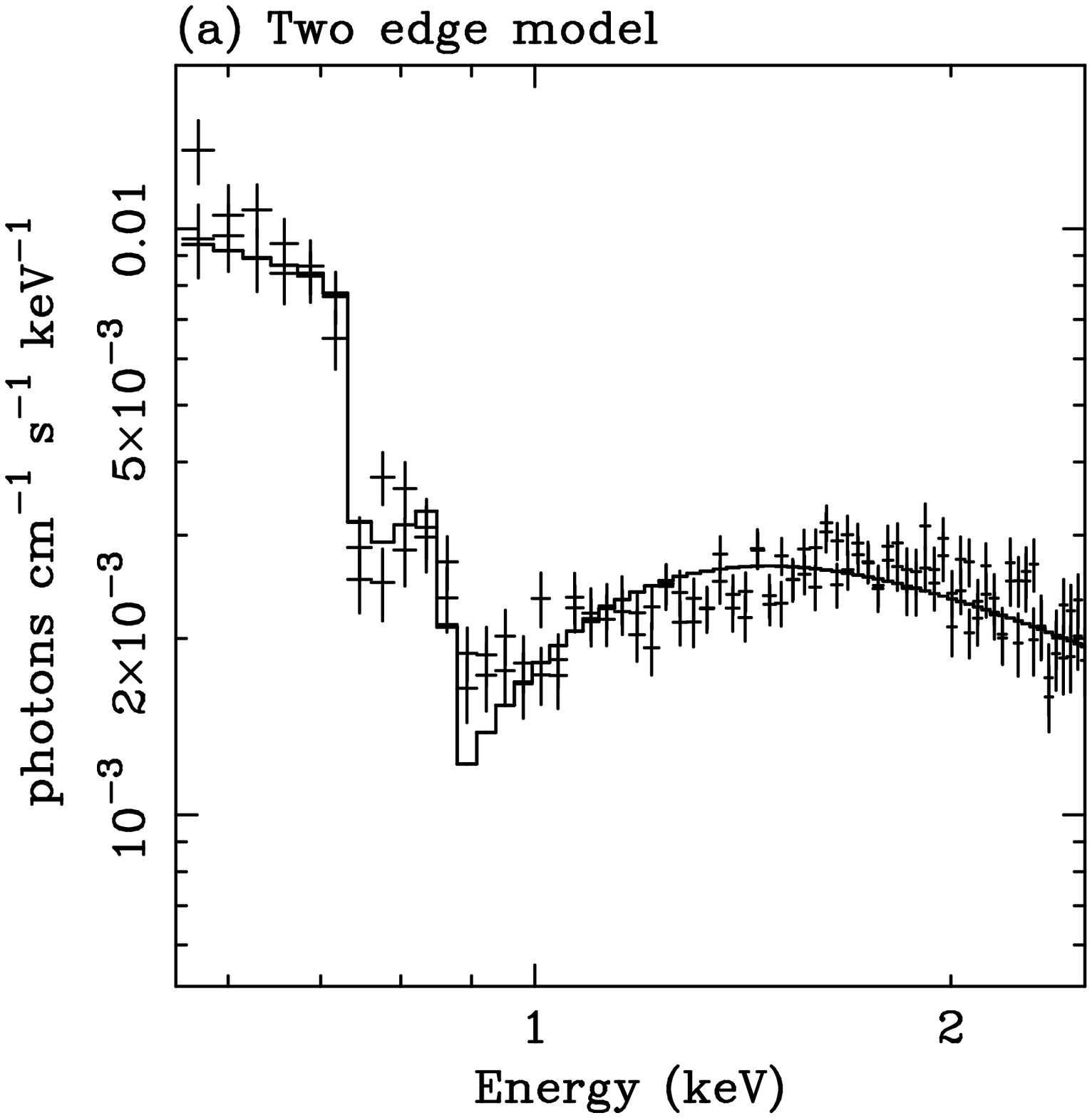,width=0.65\textwidth,angle=270}
\hspace{-3cm}
\psfig{figure=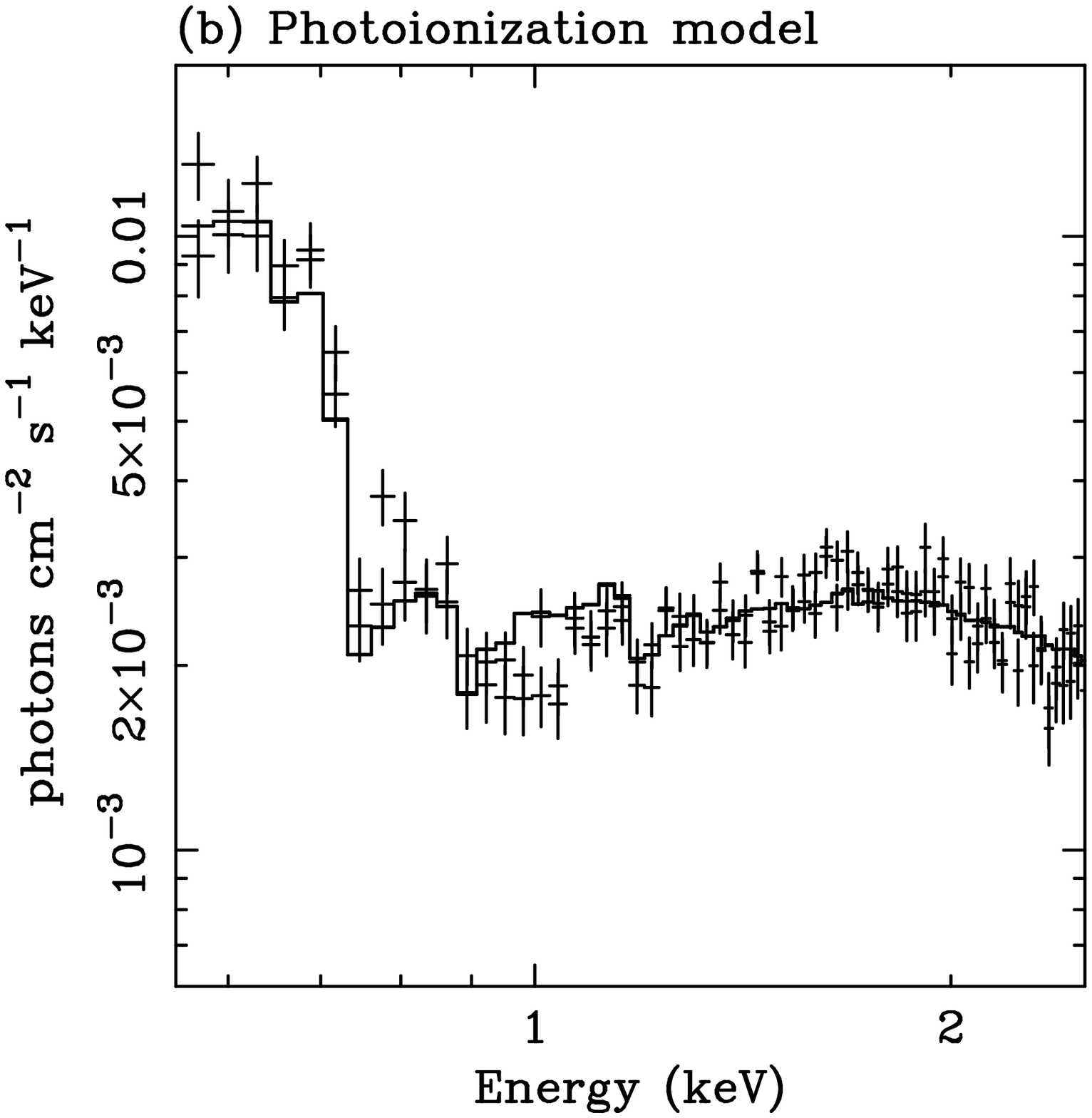,width=0.65\textwidth,angle=270}
}
\hspace{-1.5cm}
\hbox{
\psfig{figure=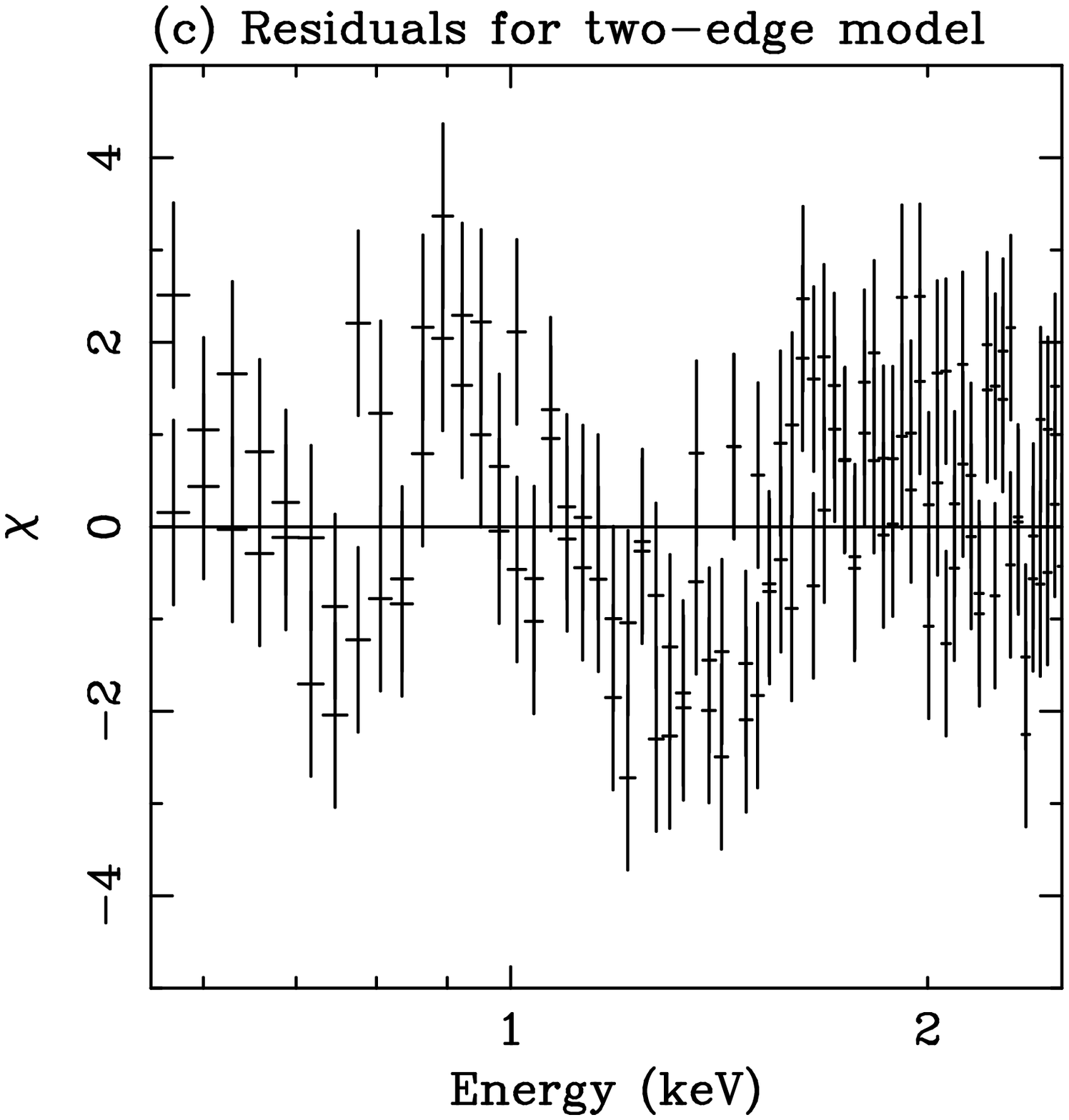,width=0.65\textwidth,angle=270}
\hspace{-3cm}
\psfig{figure=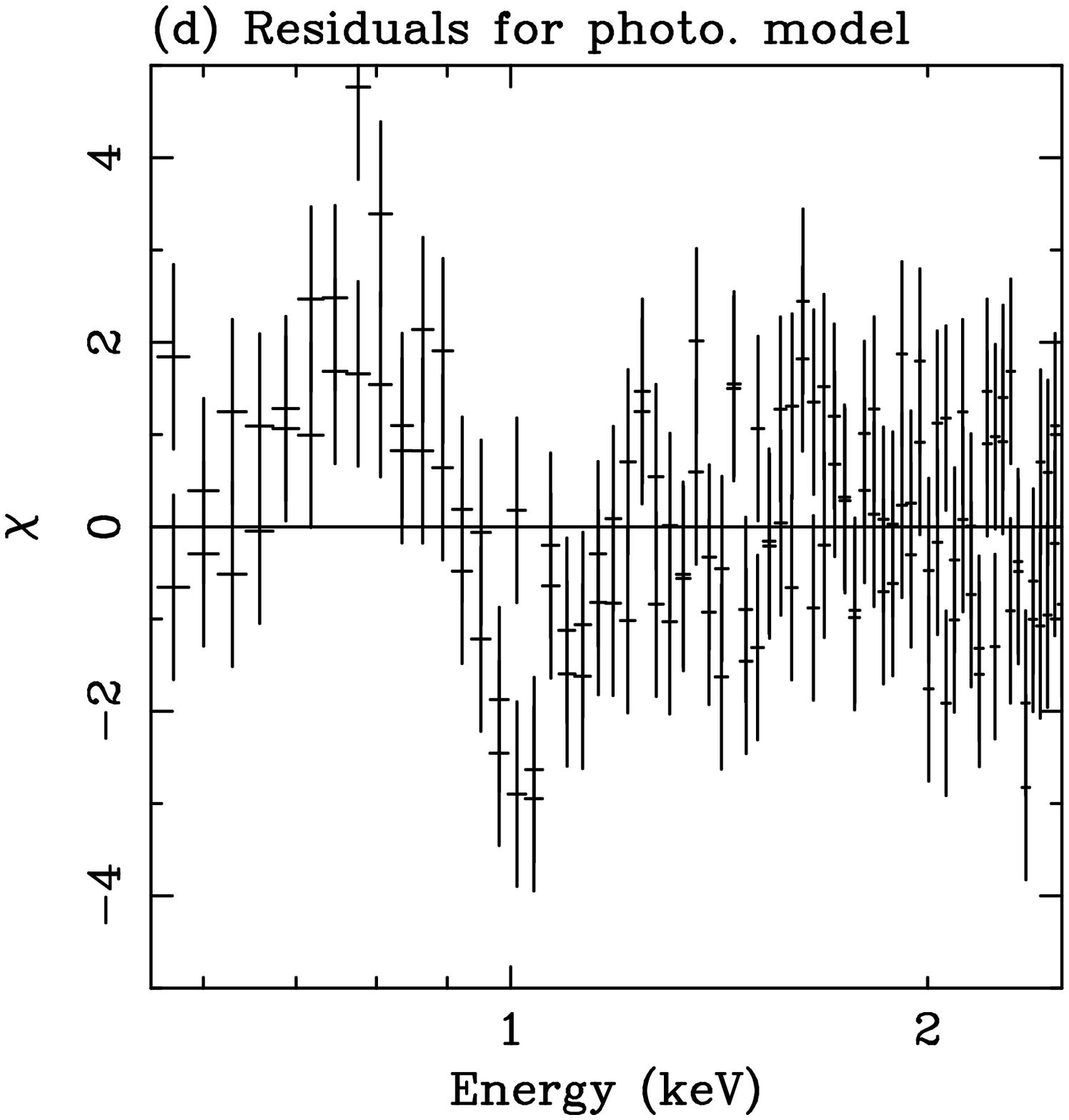,width=0.65\textwidth,angle=270}
}
\caption{The warm absorber in NGC~3783.  Panel (a) shows the best fit 
two-edge model (model-B) and the unfolded SIS data.  Panel (b) shows
the best fit photoionization model (model-C) and the unfolded data.
Panels (c) and (d) show the deviations of the data from the two-edge
and photoionization models, respectively.}
\end{figure*}

\noindent A two-edge fit shows NCG~3783 to have the deepest ionized
oxygen edges of the current sample.  These edges were first found and
extensively investigated with the {\it ROSAT} PSPC (Turner et
al. 1993a).  Also, George, Turner \& Netzer (1995) report the detection
of {\sc O\,vii} and {\sc O\,viii} recombination lines (at 0.57\,keV
and 0.65\,keV in the source rest frame) using {\it ASCA}.  Thus, this
object provides a good test case with which to probe details of the
ionized absorption.

The two-edge parametrization of the warm absorber in NGC~3783 results
in systematic residuals.  In particular, there is a small excess of
counts at $\sim 1\keV$ and a broad deficit of counts between
1--2\,keV.  Given the extreme nature of this ionized absorber
(inferred from the oxygen edge fits), these may be due to unmodelled
absorption edges of other species and/or emission lines.  Figure~6
addresses this issue.  Whilst the photoionization model confirms that
the dominant effect is indeed modelled by two ionized oxygen edges, it
reveals the presence of an additional absorption edge at $\sim
1.2\keV$ identified as the K-shell edge of Ne\,{\sc ix}.  The onset of
this extra-edge, and the consequent diminishing of the {\sc O\,viii}
edge explains some, but not all, of the residuals seen in the simple
two-edge fit.

\subsection{MCG$-6-30-15$}

\begin{figure*}
\hspace{-1.5cm}
\hbox{
\psfig{figure=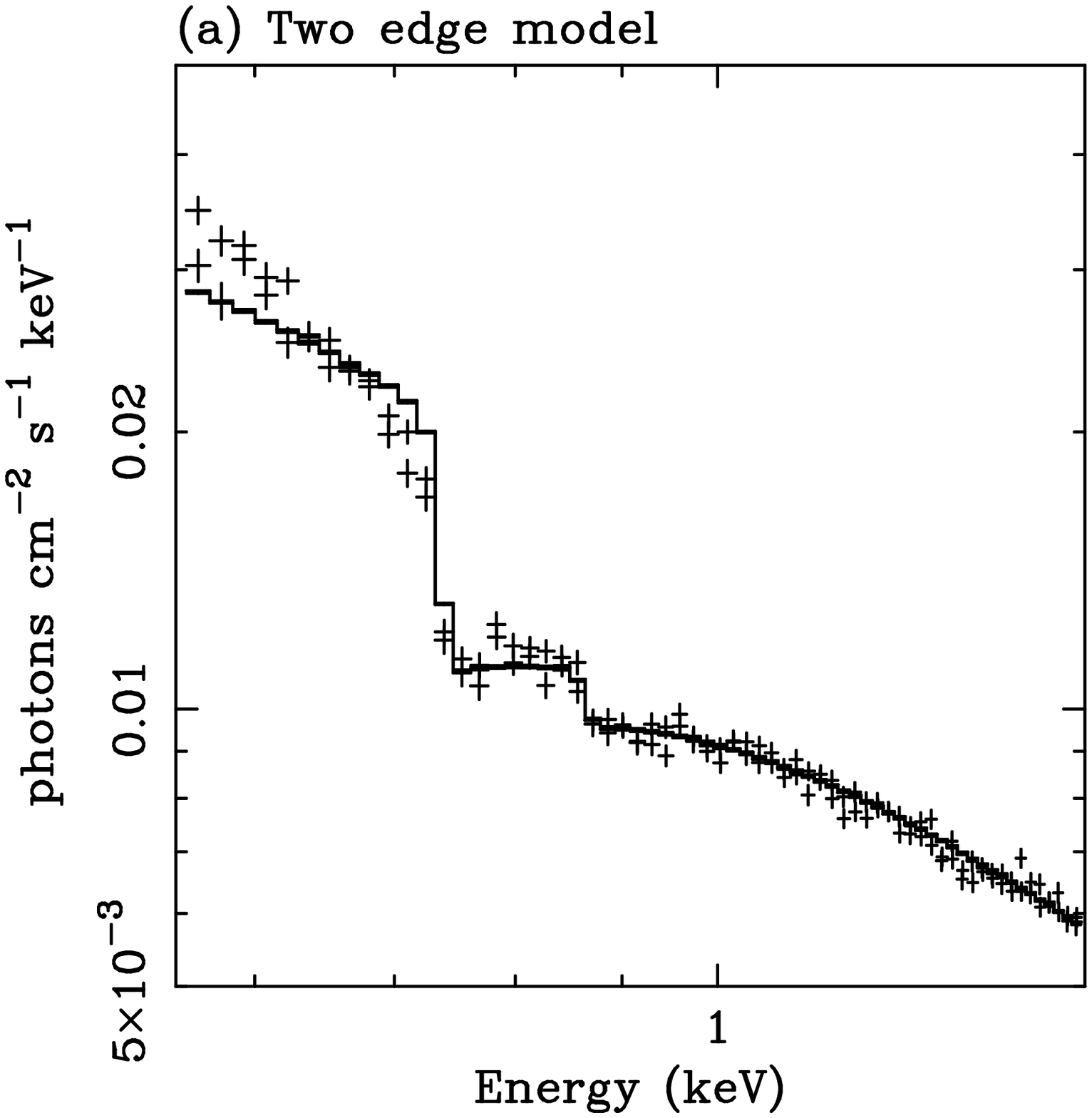,width=0.65\textwidth,angle=270}
\hspace{-3cm}
\psfig{figure=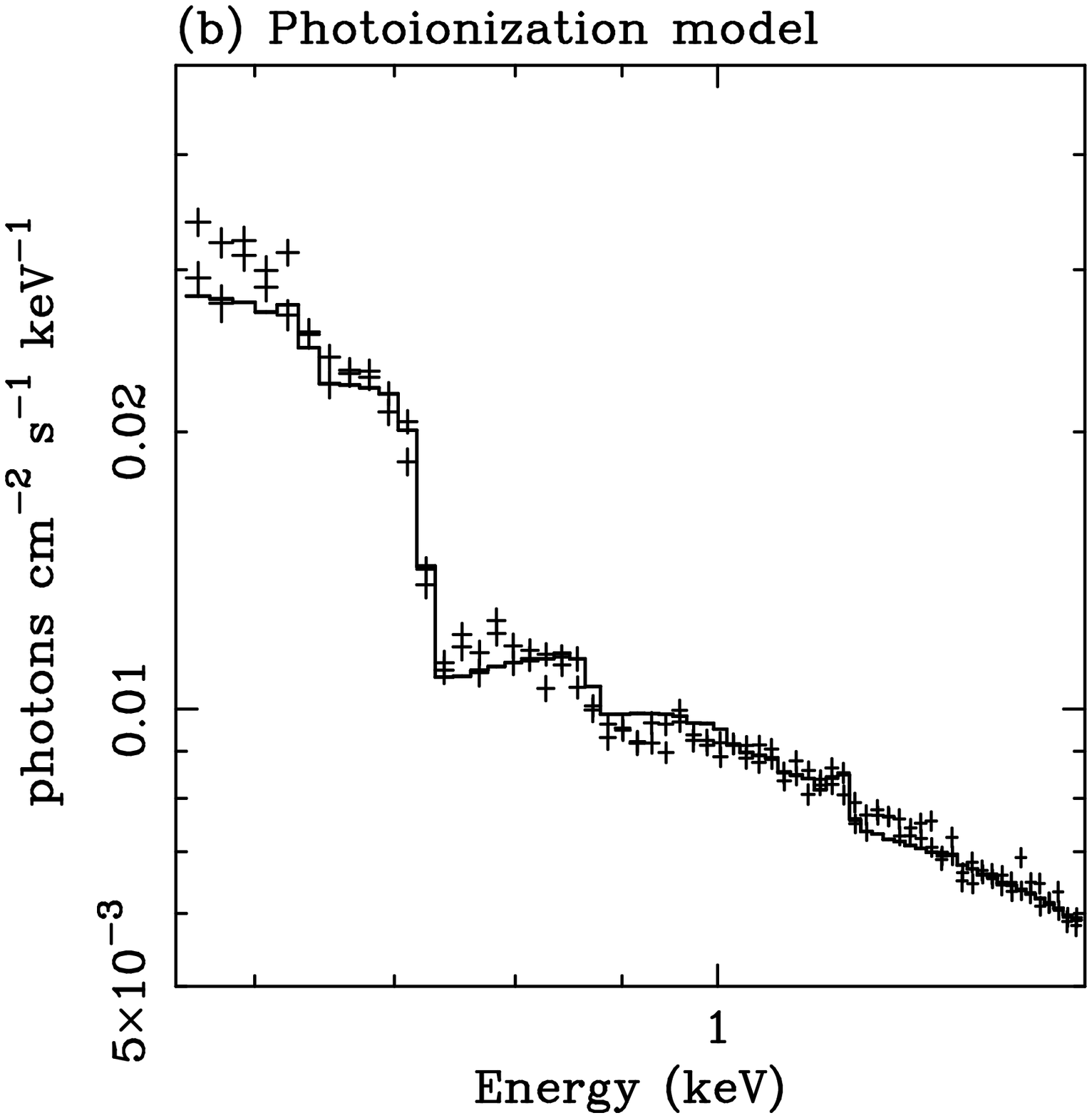,width=0.65\textwidth,angle=270}
}
\hspace{-1.5cm}
\hbox{
\psfig{figure=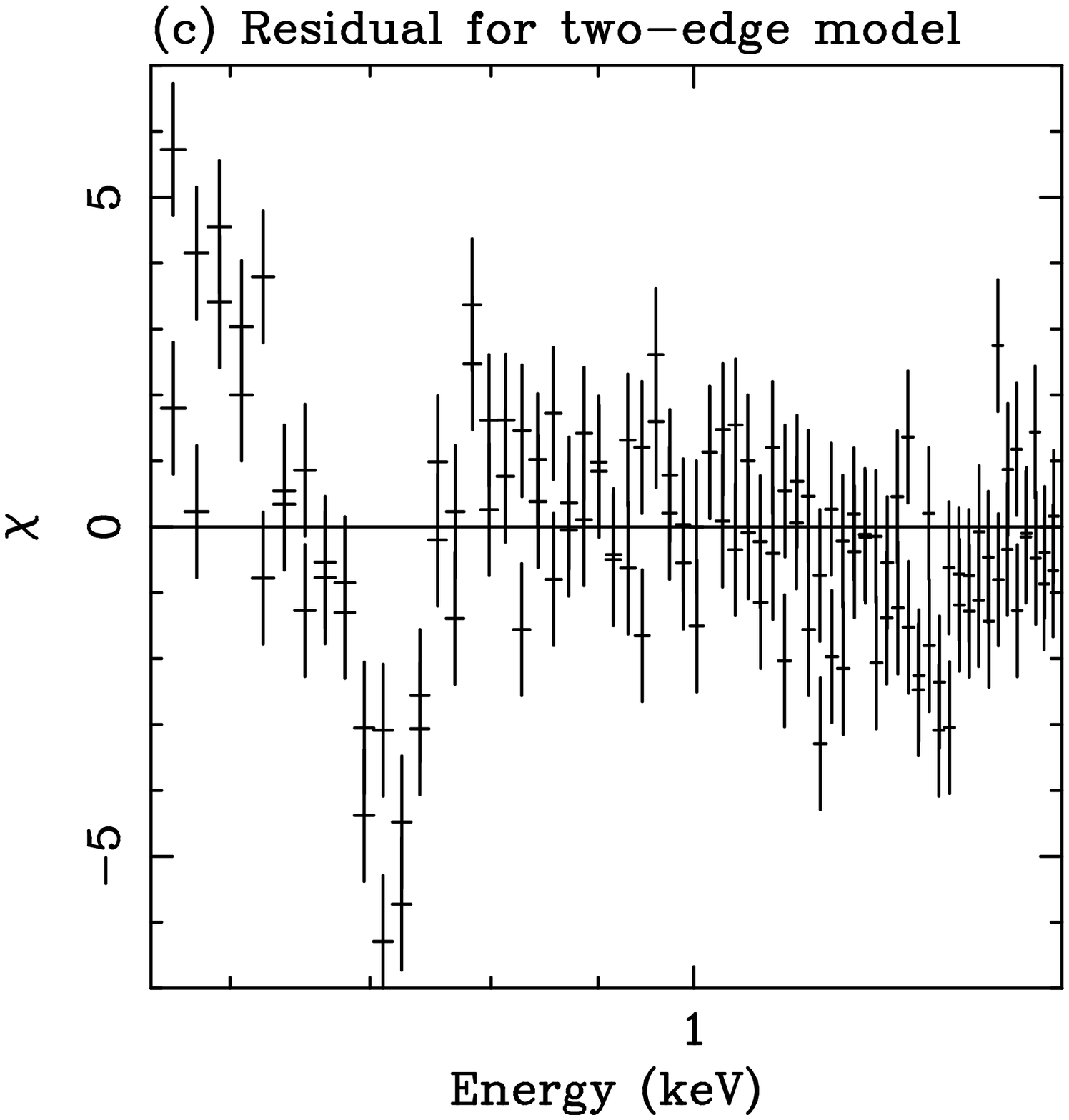,width=0.65\textwidth,angle=270}
\hspace{-3cm}
\psfig{figure=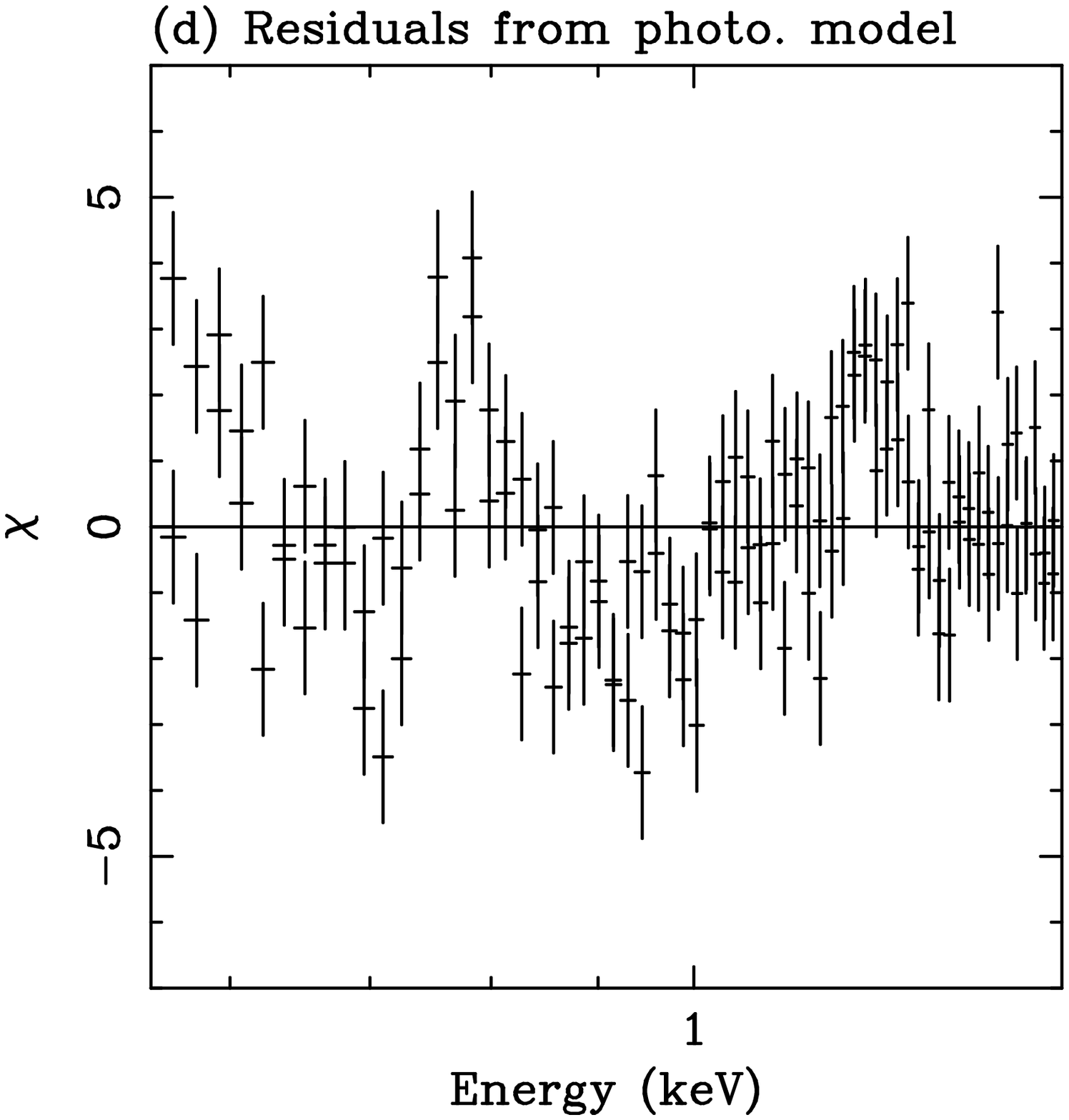,width=0.65\textwidth,angle=270}
}
\caption{The warm absorber in MCG$-6-30-15$.  Panel (a) shows the 
best fitting simple two-edge fit (model-B) and the unfolded SIS
spectrum.  Panel (b) shows the best fitting photoionization model
(model-C) and the unfolded SIS data. Panels (c) and (d) illustrate the
deviations from the two-edge and photoionization model, respectively.}
\end{figure*}

\noindent The extremely long observation of this bright source leads to very
good quality data at the energies characteristic of the warm absorber.
Complexities beyond the oxygen edge model (model-B) are likely to be
detectable in data of this quality.  Indeed, this object is
(statistically speaking) described the worst by the simple two-edge
model.  Figures~7a,c shows the best fitting two-edge model (model-B) and
the unfolded SIS data for this object.  The poor quality of the fit
largely results from a mismatch between the model and the data at
$\sim 0.7\keV$ (just below the energies of the {\sc O\,vii} absorption
edge threshold).  Fitting the data with a one-zone photoionization
model (Figs.~7b,d) produces a much better fit ($\Delta\chi^2=155$) due to
the emergence of a small {\sc O\,vi} edge at 0.68\,keV.  However, as
with NGC~3783, the photoionization calculation confirms that the
two-edge model is a good approximation to the effects of the ionized
absorption.  Note that the existence of this {\sc O\,vi} edge may
falsely imply a small redshift of the {\sc O\,vii} edge when a
two-edge model is fitted leaving the edge threshold energies as free
parameters.

The one-zone photoionization model also has failures: it predicts an
unobserved Ne\,{\sc ix} K-shell edge at $\sim 1.2\keV$ and fails to
reproduce an excess of emission at $\sim 0.6\keV$ (most likely
recombination line emission from highly ionized oxygen species).  This
failure is most likely related to the one-zone nature of the model.
To see this, note that (for a given ionizing spectrum) the depths of
the {\sc O\,vii} and {\sc O\,viii} edges uniquely determine the
ionization parameter and column density of any one-zone warm absorber
model that is fitted to the spectrum.  The predicted depths of all
other absorption edges are therefore fixed once the spectrum has fitted
the oxygen edges.  If the absorber is multi-zoned, the state of the
absorber is no longer uniquely determined by the depths of the oxygen
edges.  Hence, multi-zone absorbers can produce combinations of
absorption edges that are not possible from one-zone absorbers.

\subsection[Summary of comparison]
{Summary of comparision between photoionization model and
phenomenological (two-edge model) model}

To summarize this Section, it has been explicitly shown that the main
observable effects of the warm absorber in the {\it ASCA} band are
K-shell absorption edges of {\sc O\,vii} and {\sc O\,viii} at
0.74\,keV and 0.87\,keV respectively.  In addition, K-shell absorption
edges of {\sc O\,vi} and Ne\,{\sc ix}, at 0.68\,keV and 1.2\,keV,
respectively, can also be observed in high quality {\it ASCA} data.
For comparison to other work, those objects displaying statistically
significant oxygen edges have been fitted with a one-zone
photoionization model based on the code {\sc cloudy} (Ferland 1991).
Although these models provide a useful parameterisation of the data,
the belief that multi-zone absorption is important means that the
physical parameters derived from such models must be treated with
caution.  For this reason, the current paper concentrates on the
phenomenological, two-edge description of the warm absorber.

\section{Spectral variability}

\begin{figure*}
\hbox{
\psfig{figure=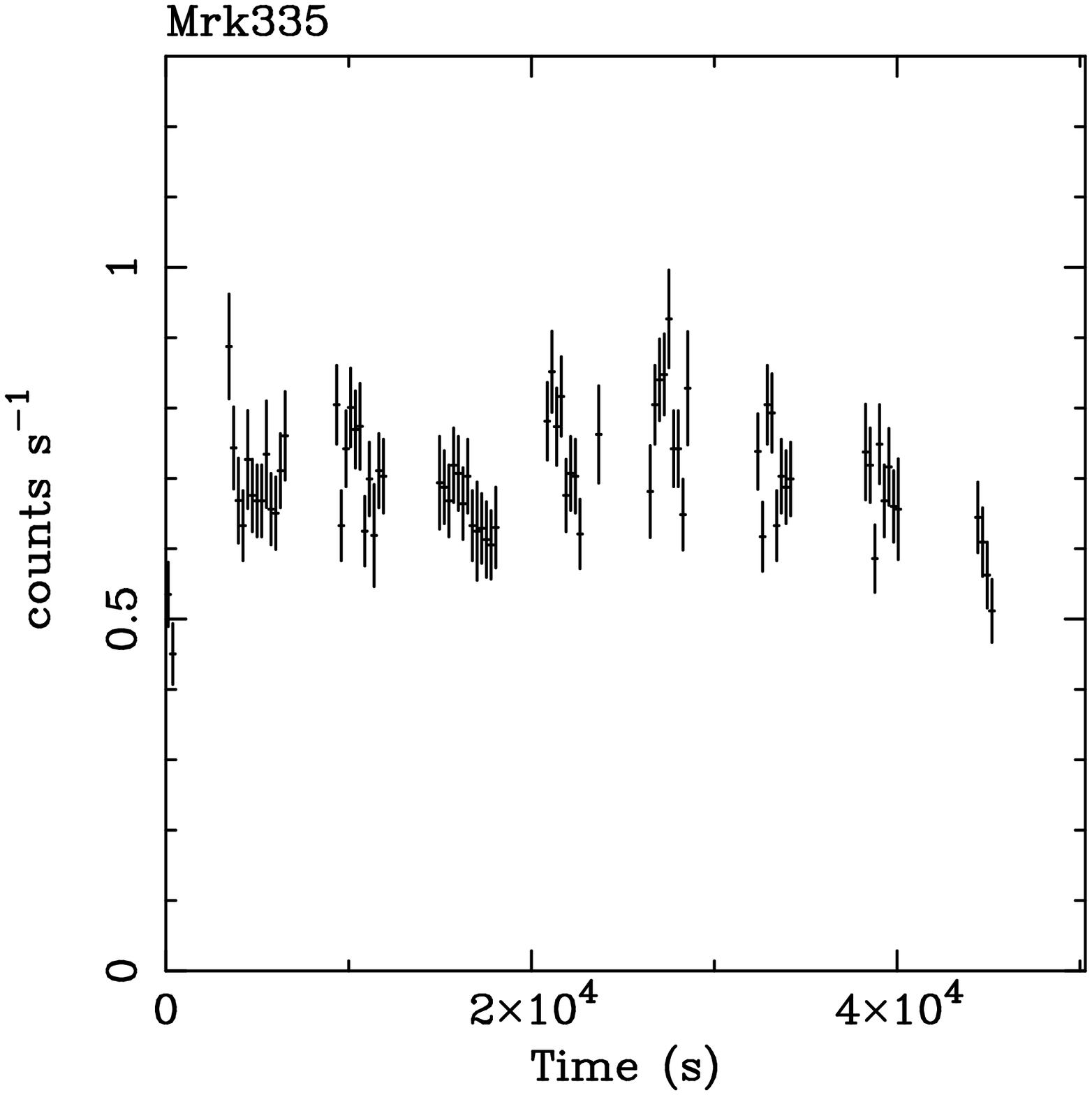,width=0.49\textwidth,height=0.23\textheight,angle=270}
\psfig{figure=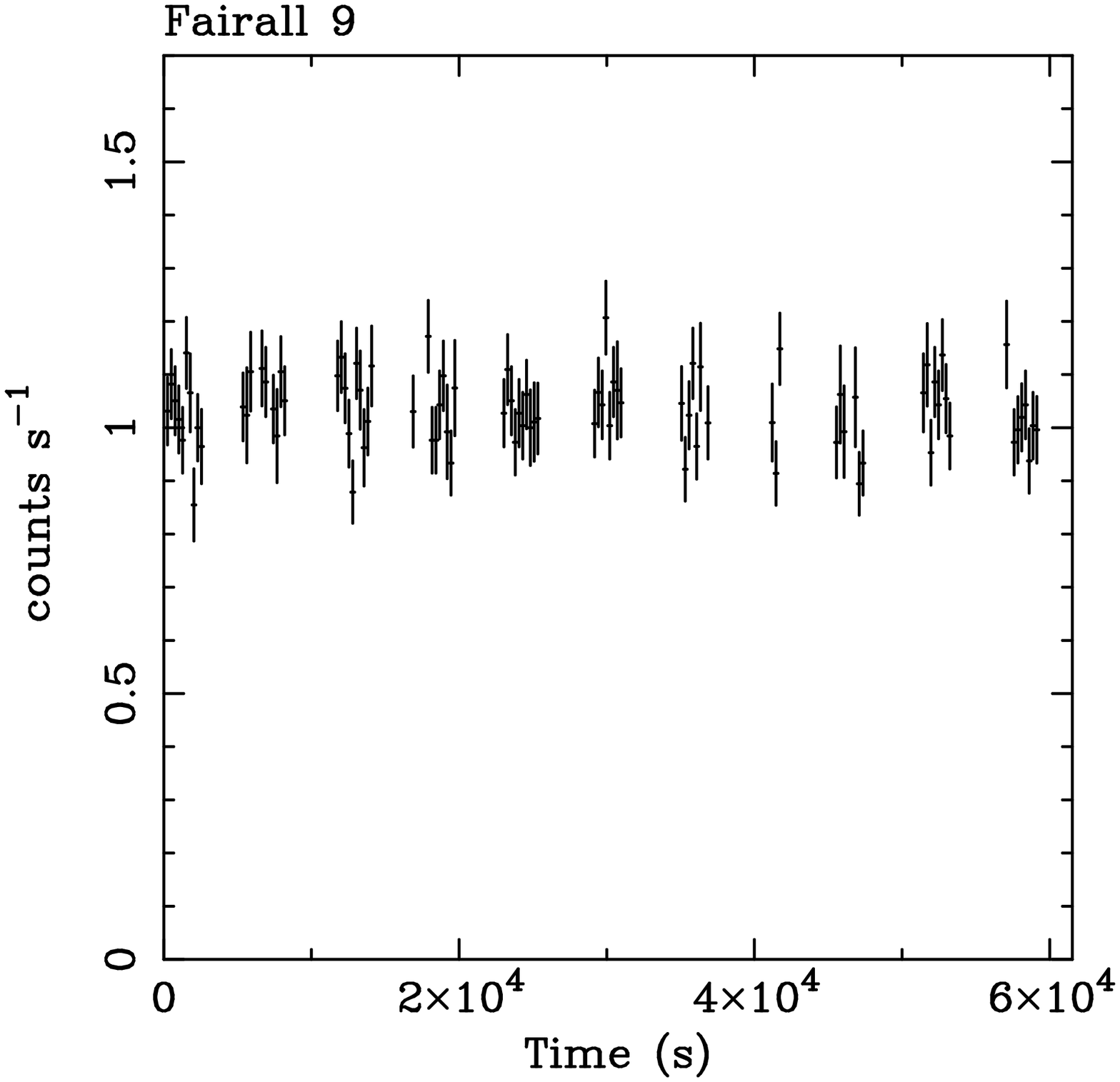,width=0.49\textwidth,height=0.23\textheight,angle=270}
}
\hbox{
\psfig{figure=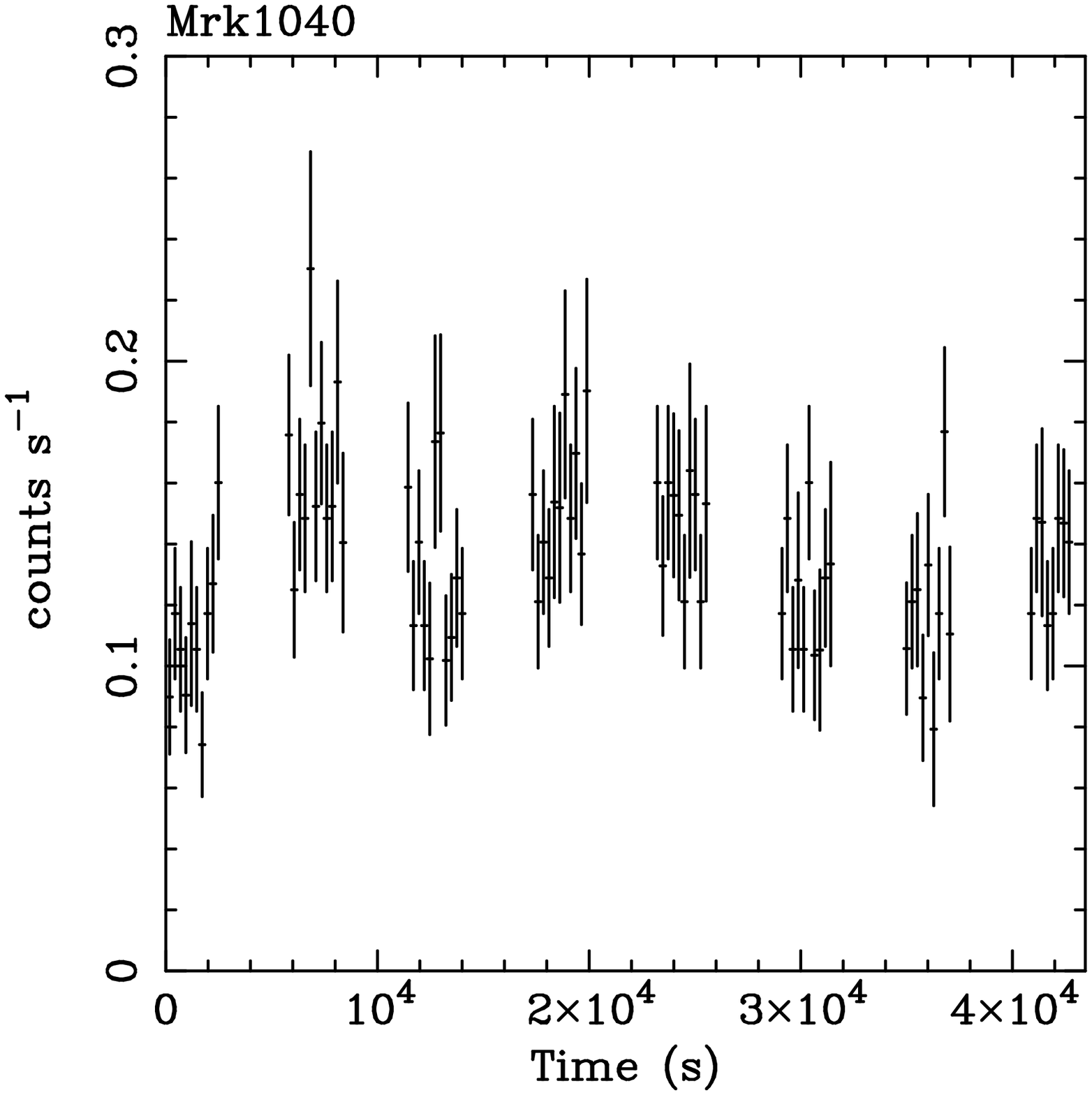,width=0.49\textwidth,height=0.23\textheight,angle=270}
\psfig{figure=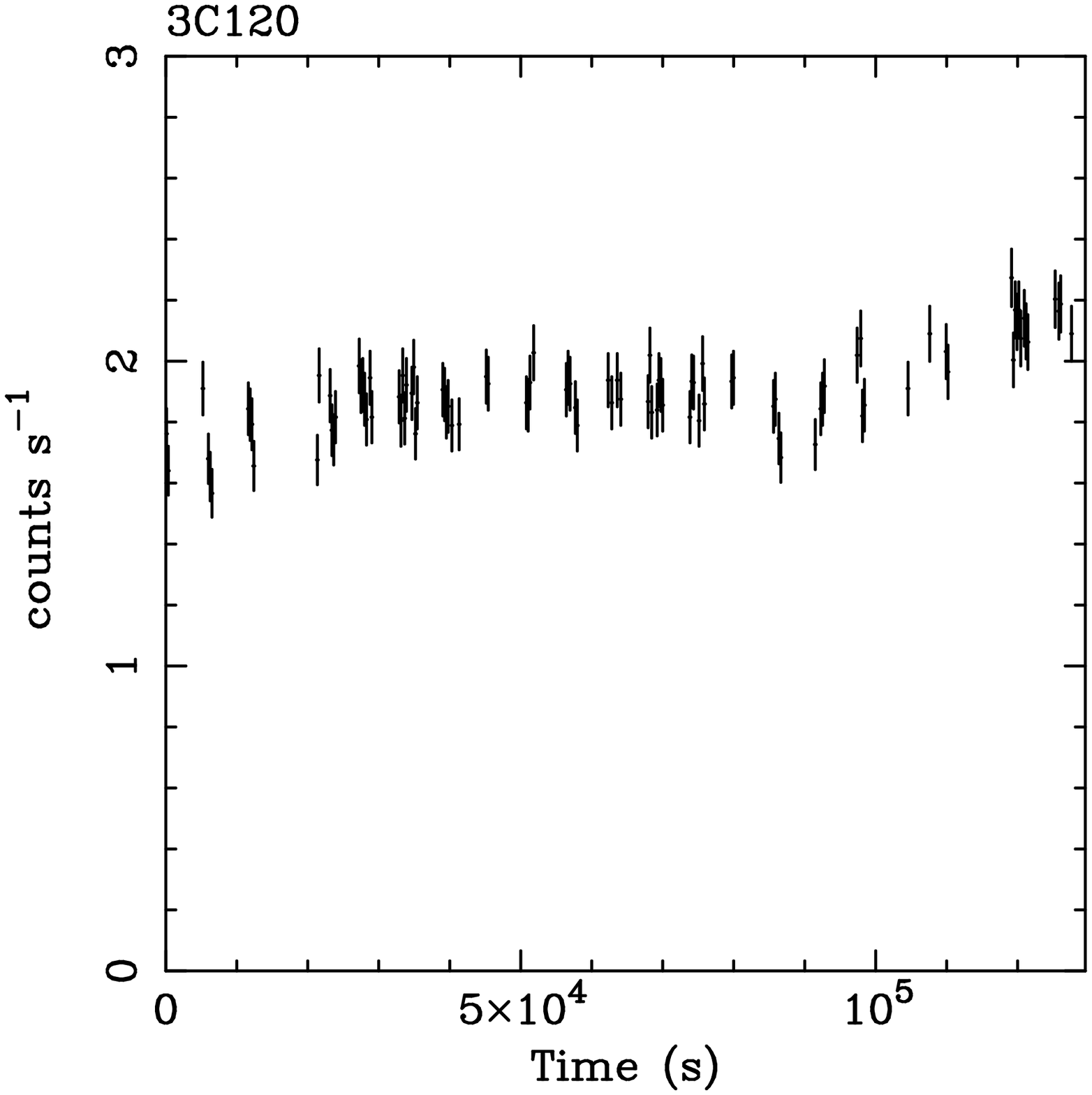,width=0.49\textwidth,height=0.23\textheight,angle=270}
}
\hbox{
\psfig{figure=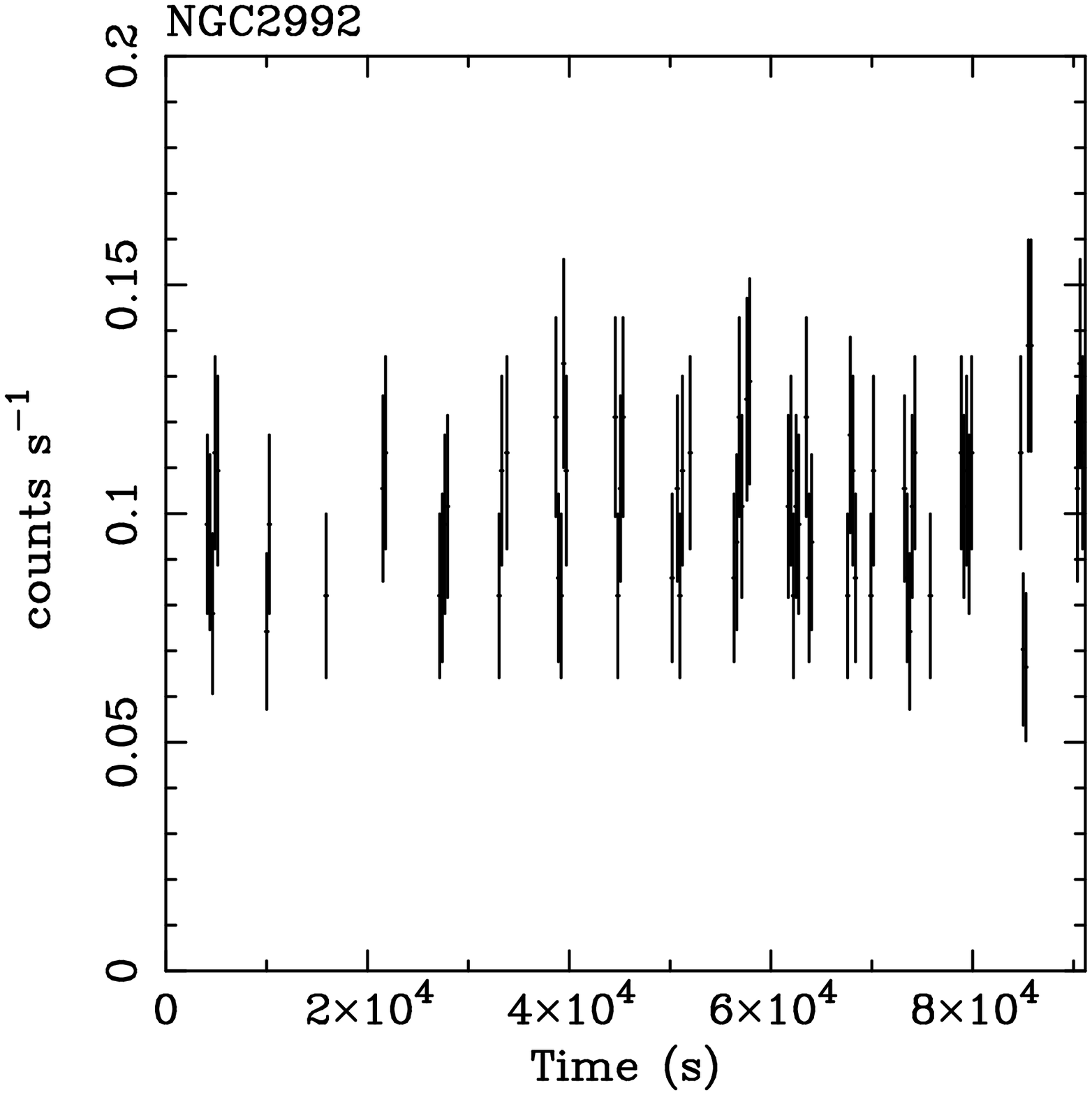,width=0.49\textwidth,height=0.23\textheight,angle=270}
\psfig{figure=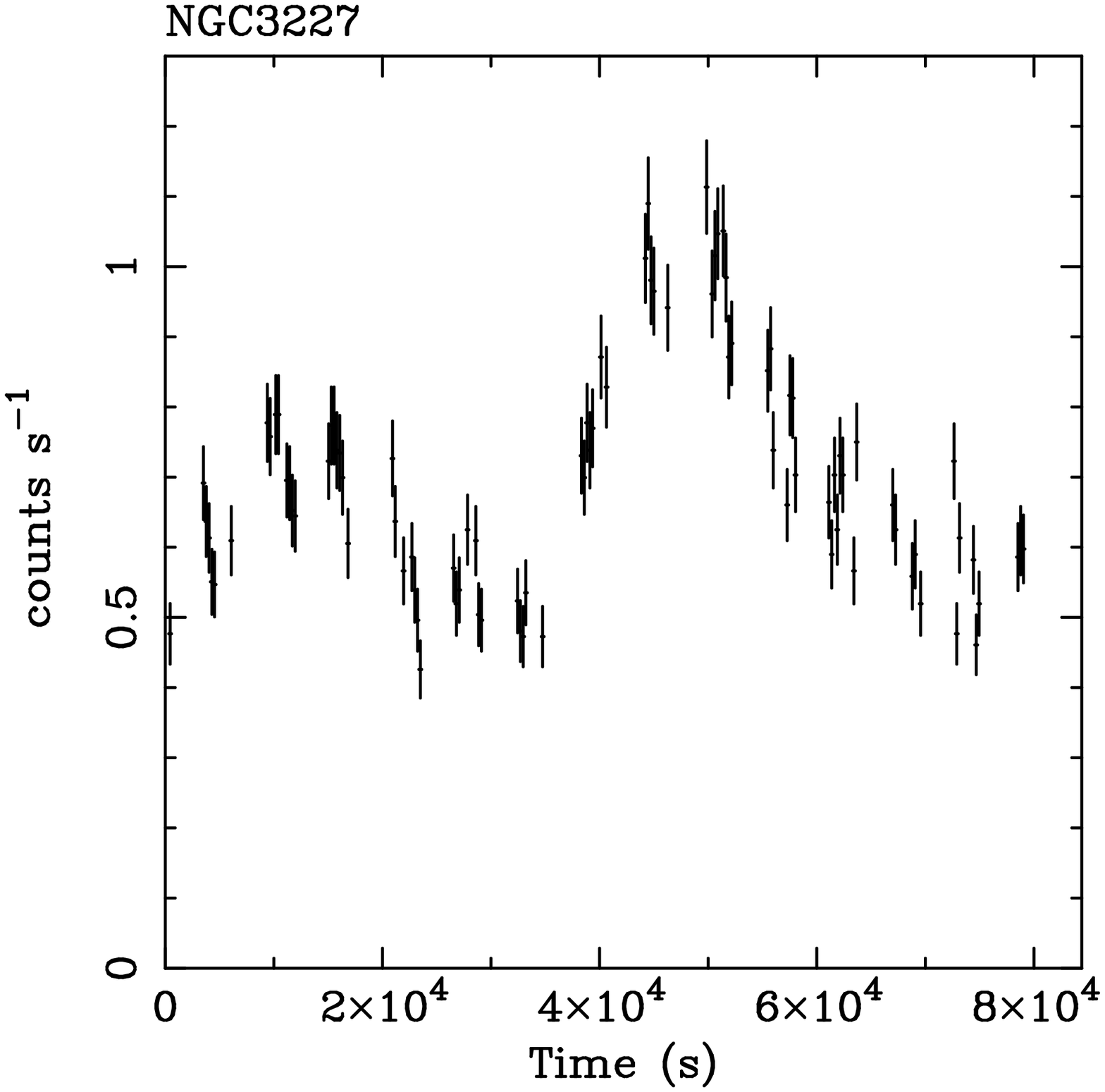,width=0.49\textwidth,height=0.23\textheight,angle=270}
}
\hbox{
\psfig{figure=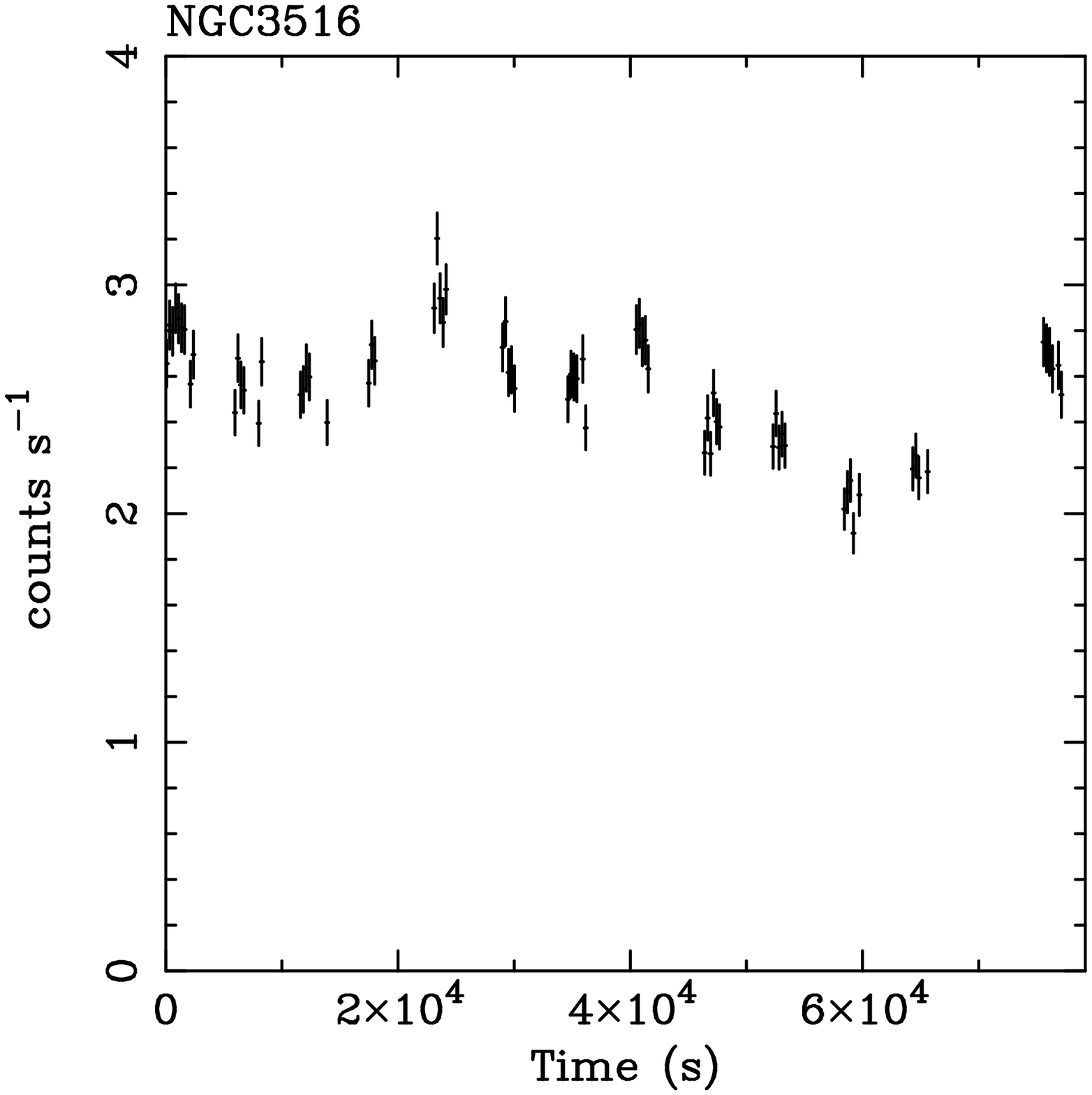,width=0.49\textwidth,height=0.23\textheight,angle=270}
\psfig{figure=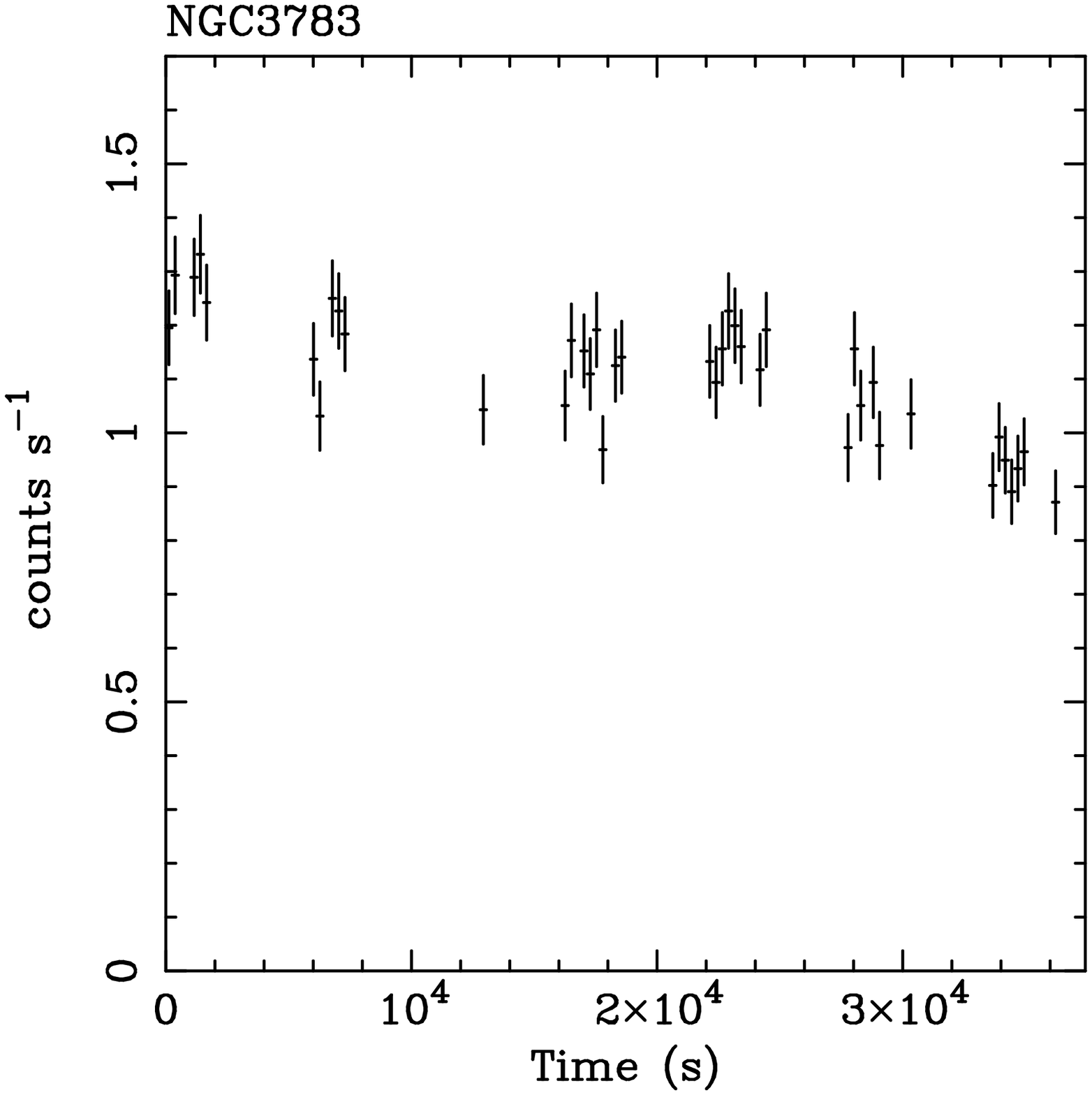,width=0.49\textwidth,height=0.23\textheight,angle=270}
}
\caption{}
\end{figure*}

\addtocounter{figure}{-1}

\begin{figure*}
\hbox{
\psfig{figure=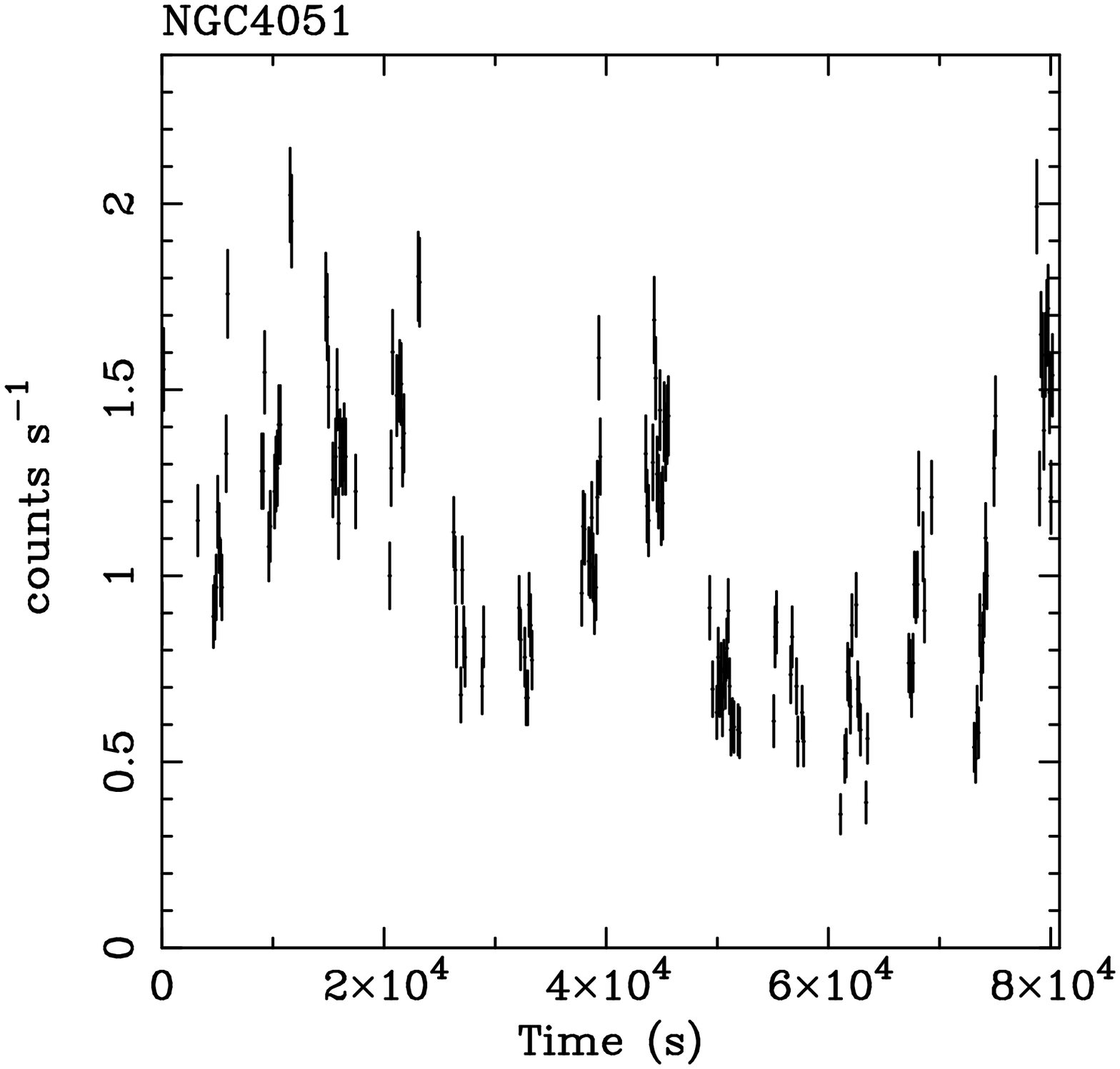,width=0.49\textwidth,height=0.23\textheight,angle=270}
\psfig{figure=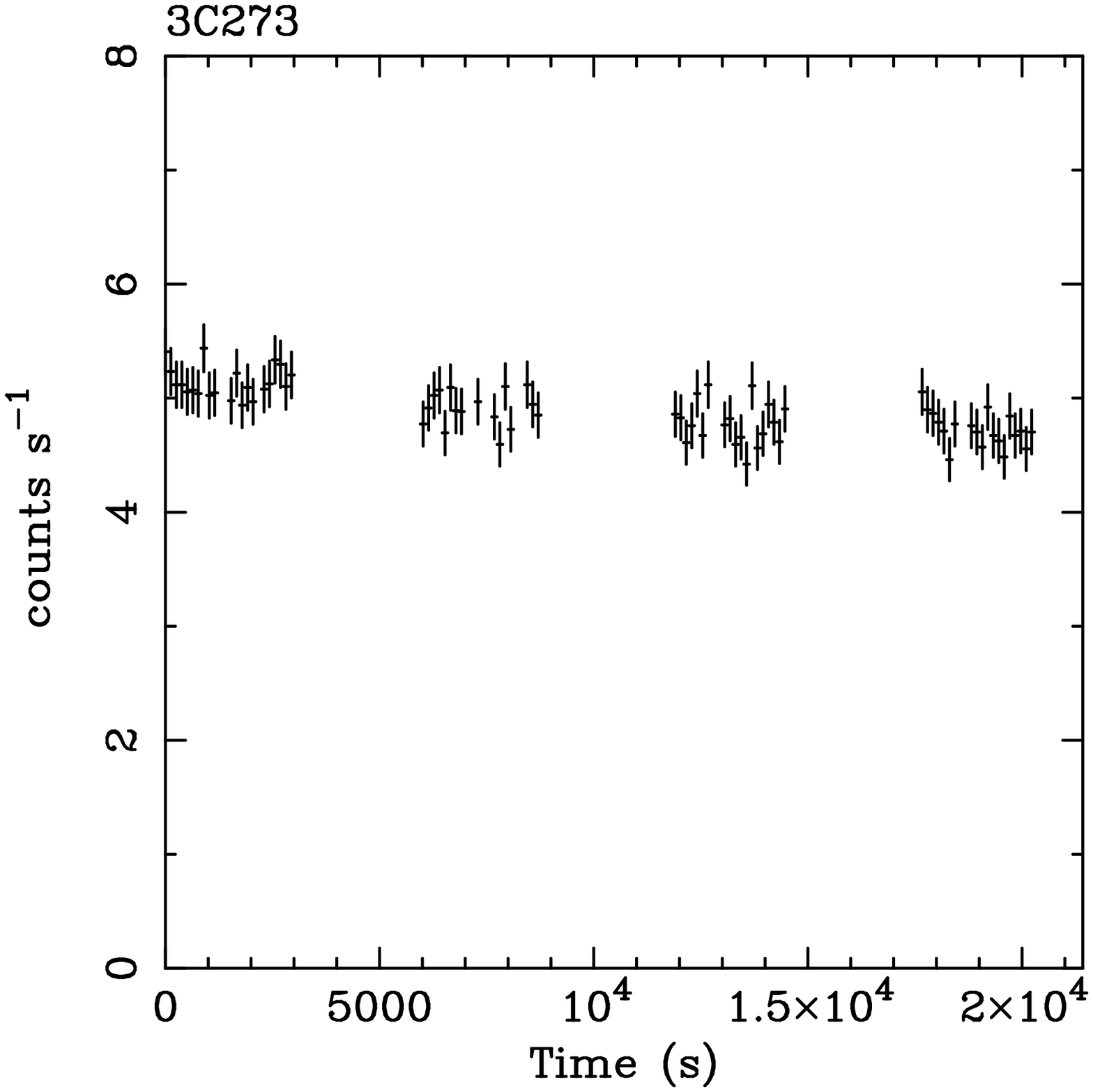,width=0.49\textwidth,height=0.23\textheight,angle=270}
}
\hbox{
\psfig{figure=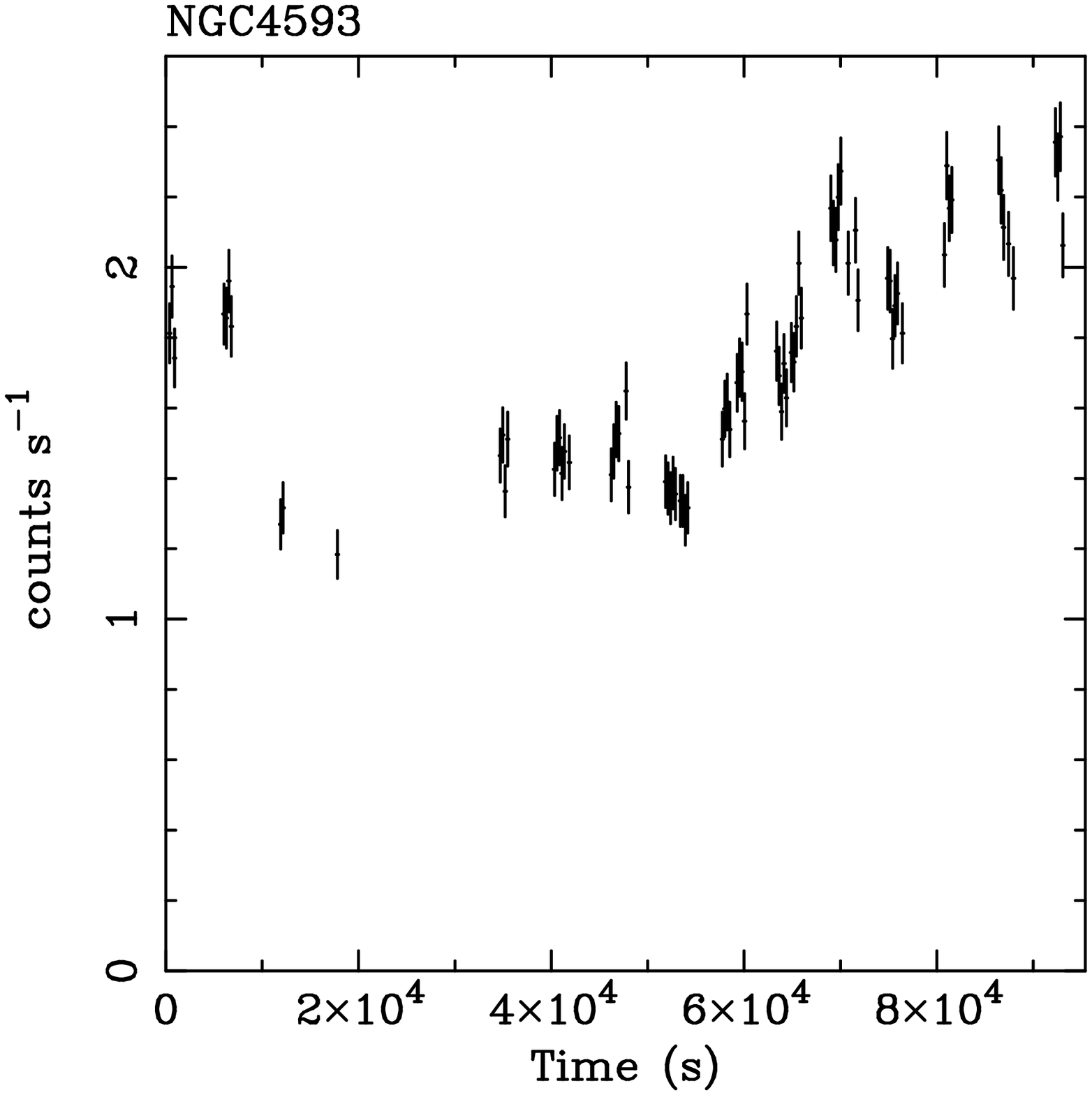,width=0.49\textwidth,height=0.23\textheight,angle=270}
\psfig{figure=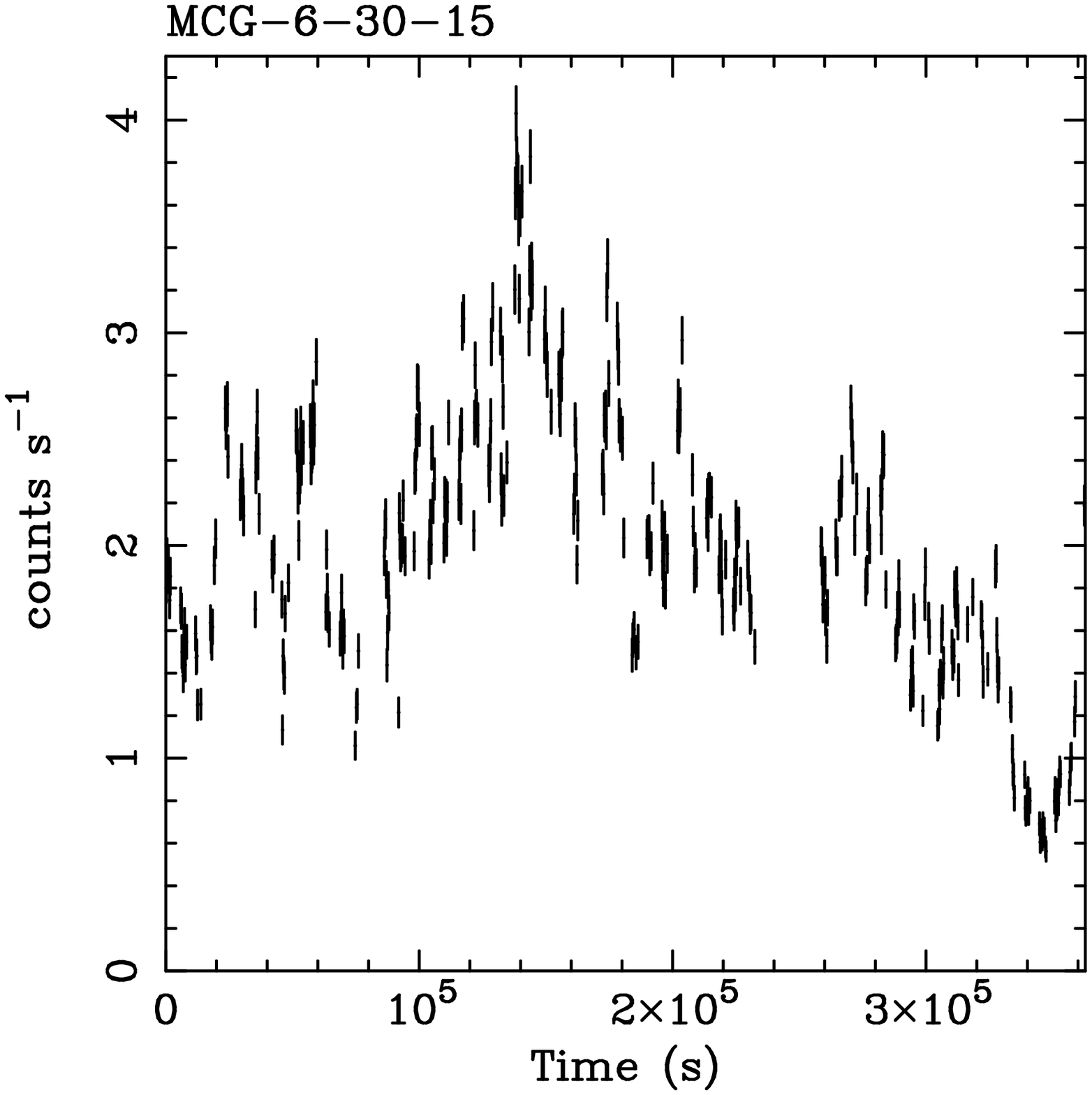,width=0.49\textwidth,height=0.23\textheight,angle=270}
}
\hbox{
\psfig{figure=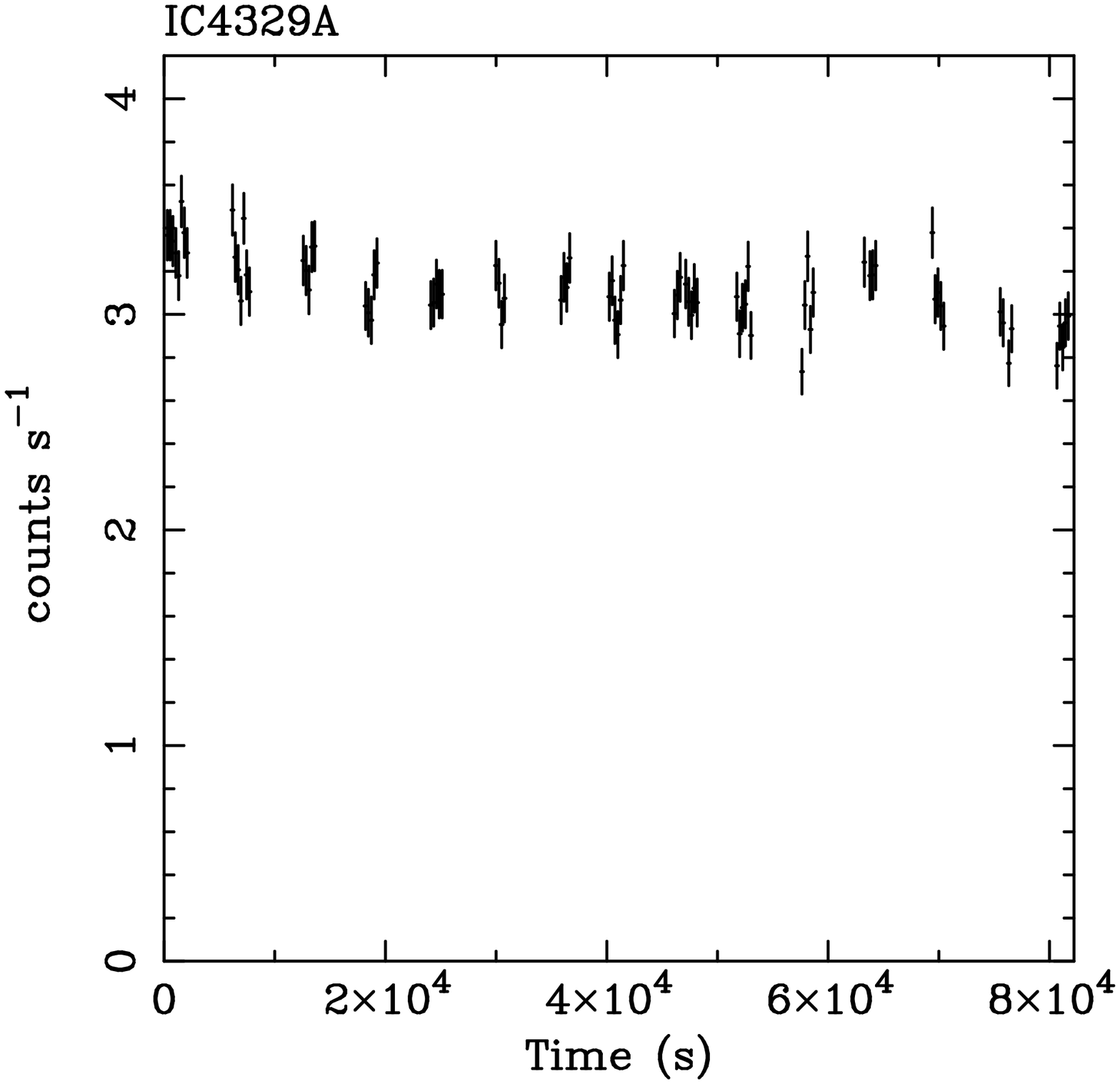,width=0.49\textwidth,height=0.23\textheight,angle=270}
\psfig{figure=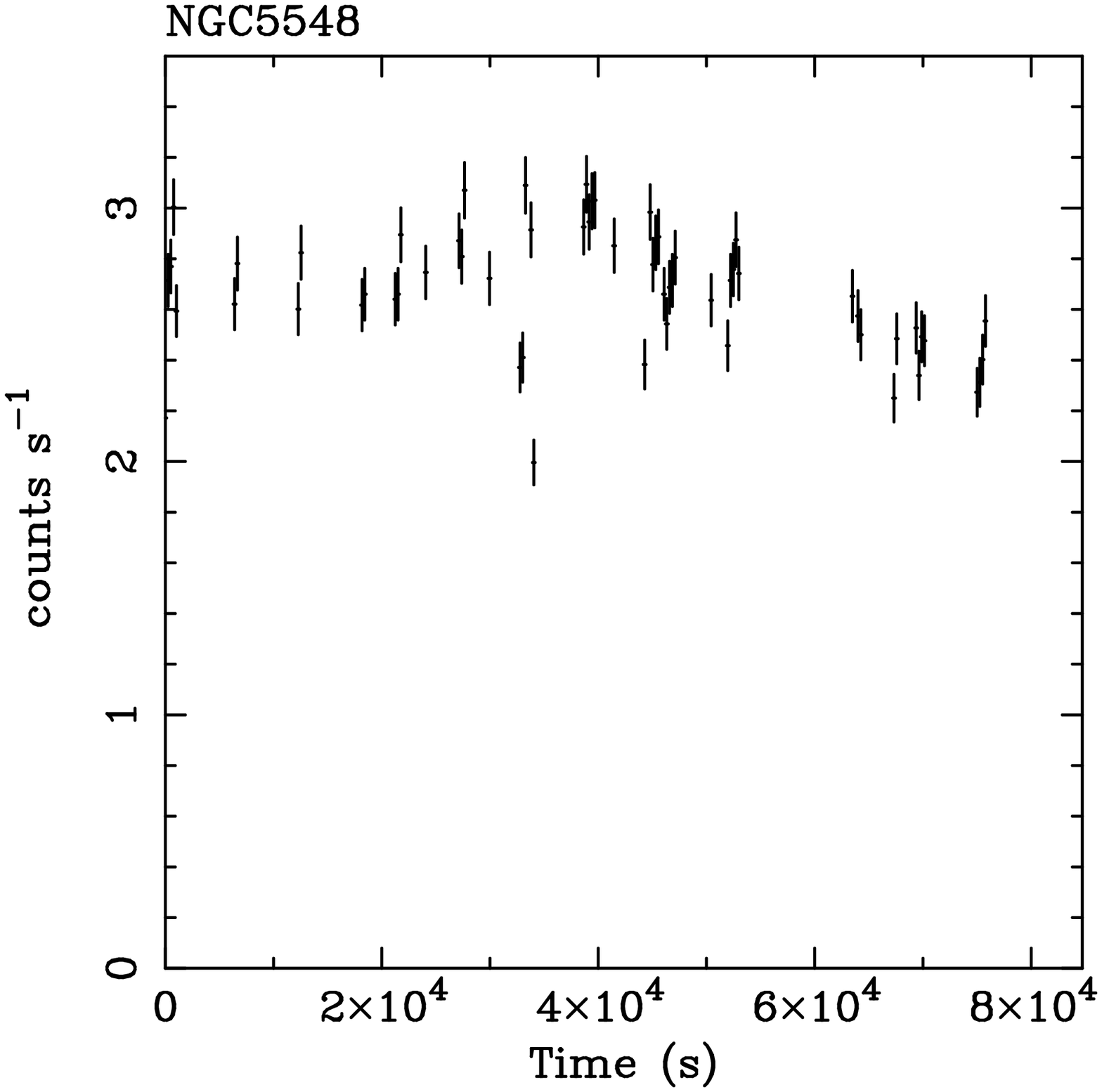,width=0.49\textwidth,height=0.23\textheight,angle=270}
}
\hbox{
\psfig{figure=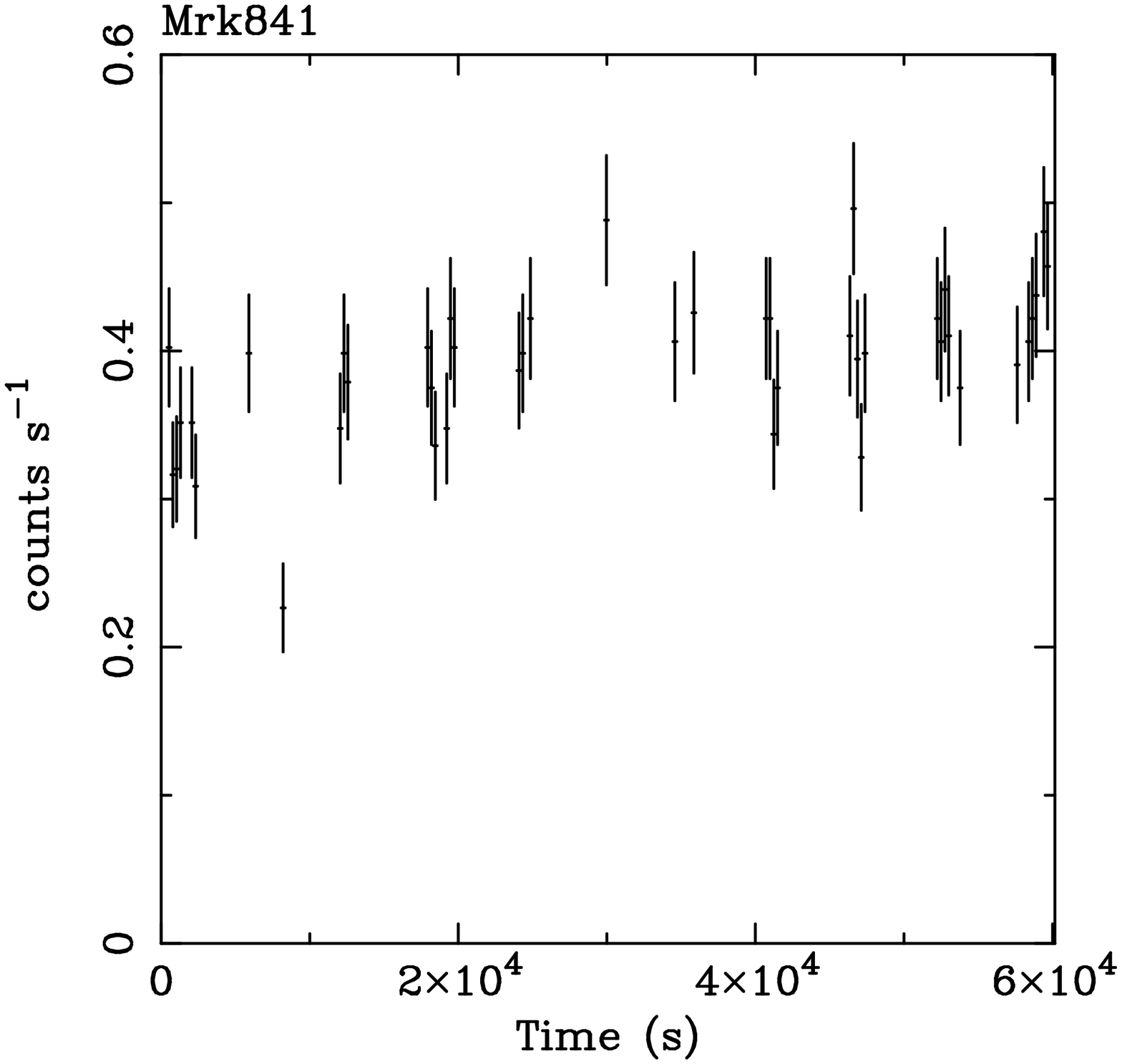,width=0.49\textwidth,height=0.23\textheight,angle=270}
\psfig{figure=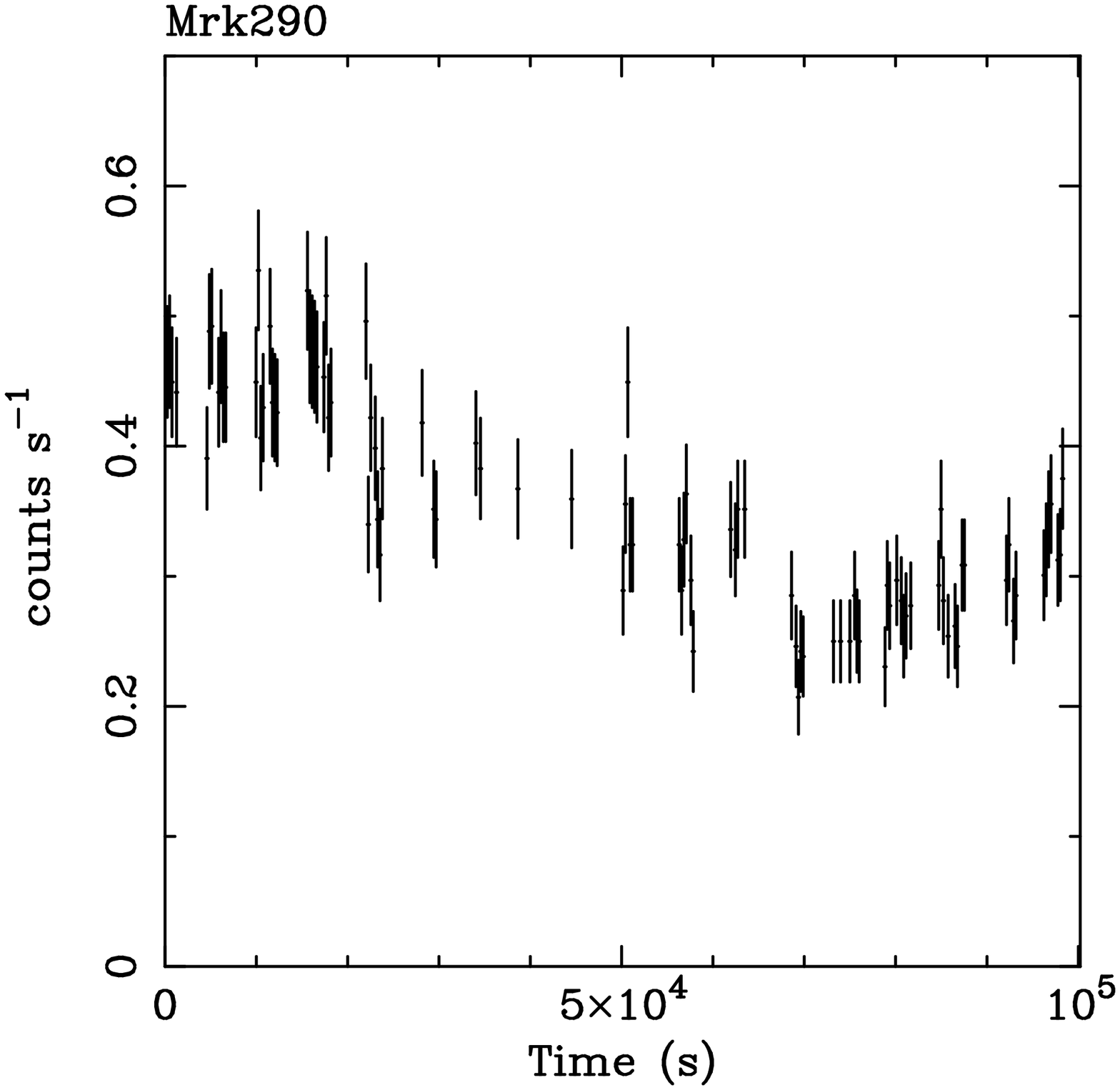,width=0.49\textwidth,height=0.23\textheight,angle=270}
}
\caption{cont.}
\end{figure*}

\addtocounter{figure}{-1}

\begin{figure*}
\hbox{
\psfig{figure=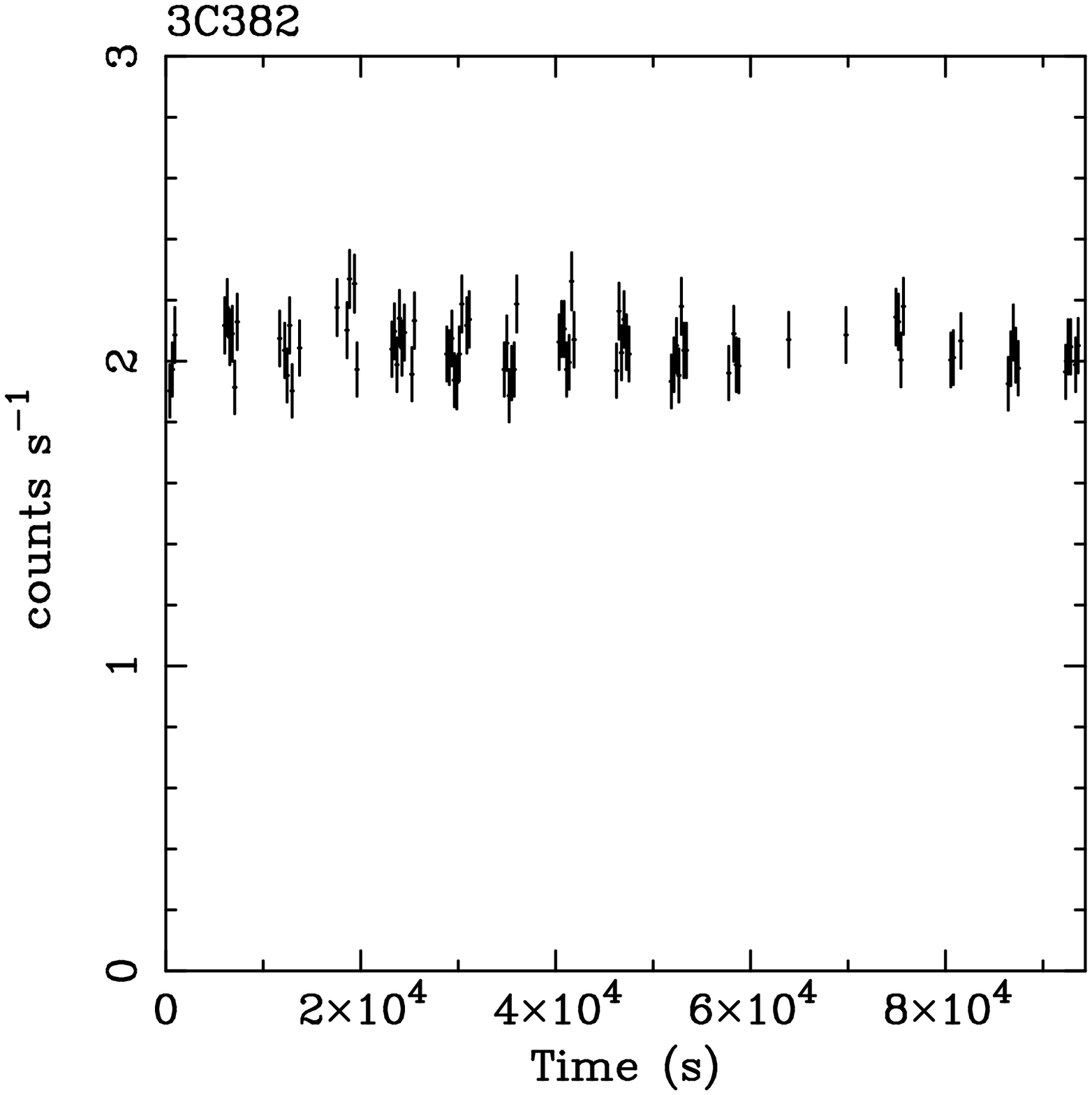,width=0.49\textwidth,height=0.23\textheight,angle=270}
\psfig{figure=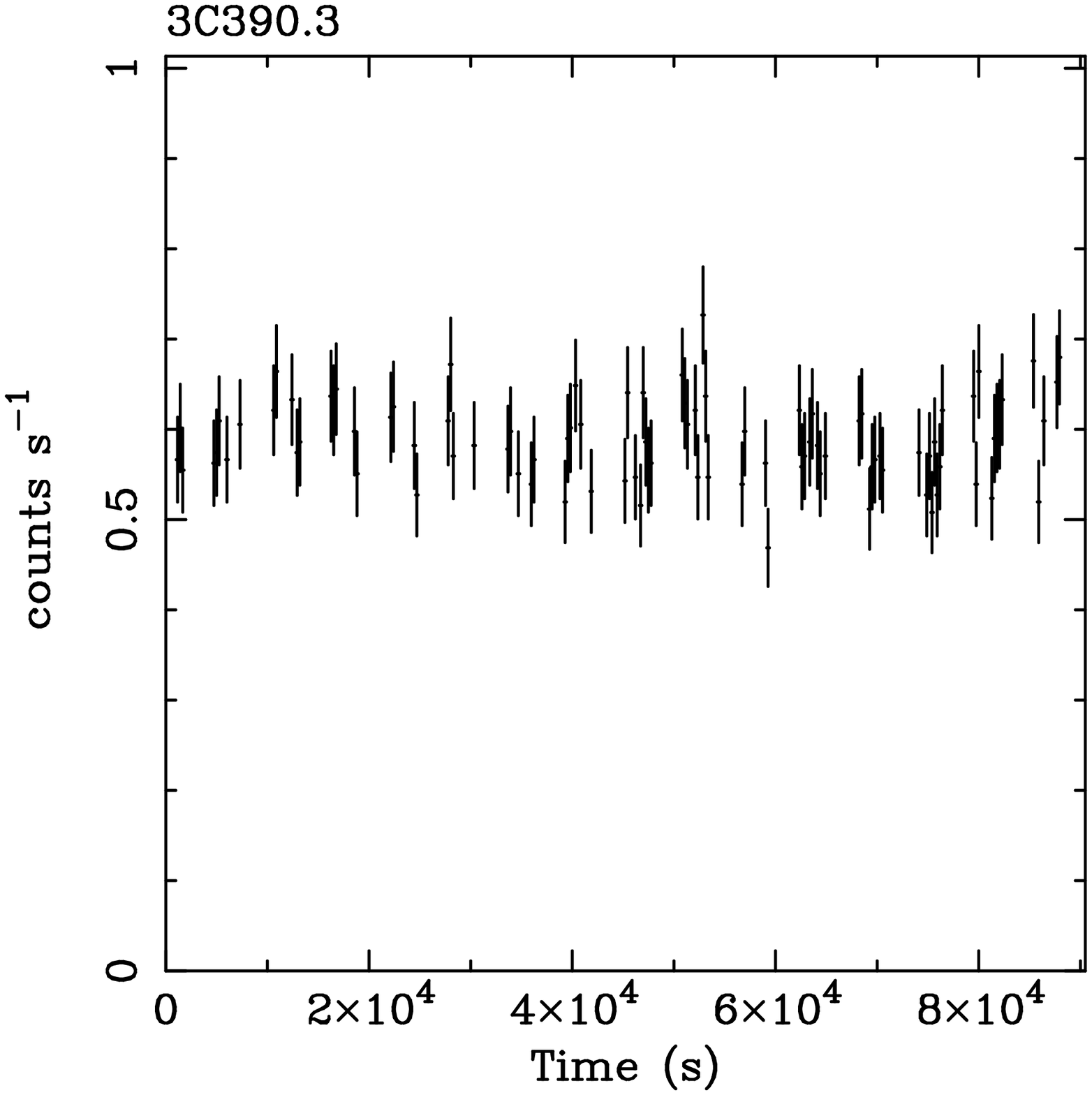,width=0.49\textwidth,height=0.23\textheight,angle=270}
}
\hbox{
\psfig{figure=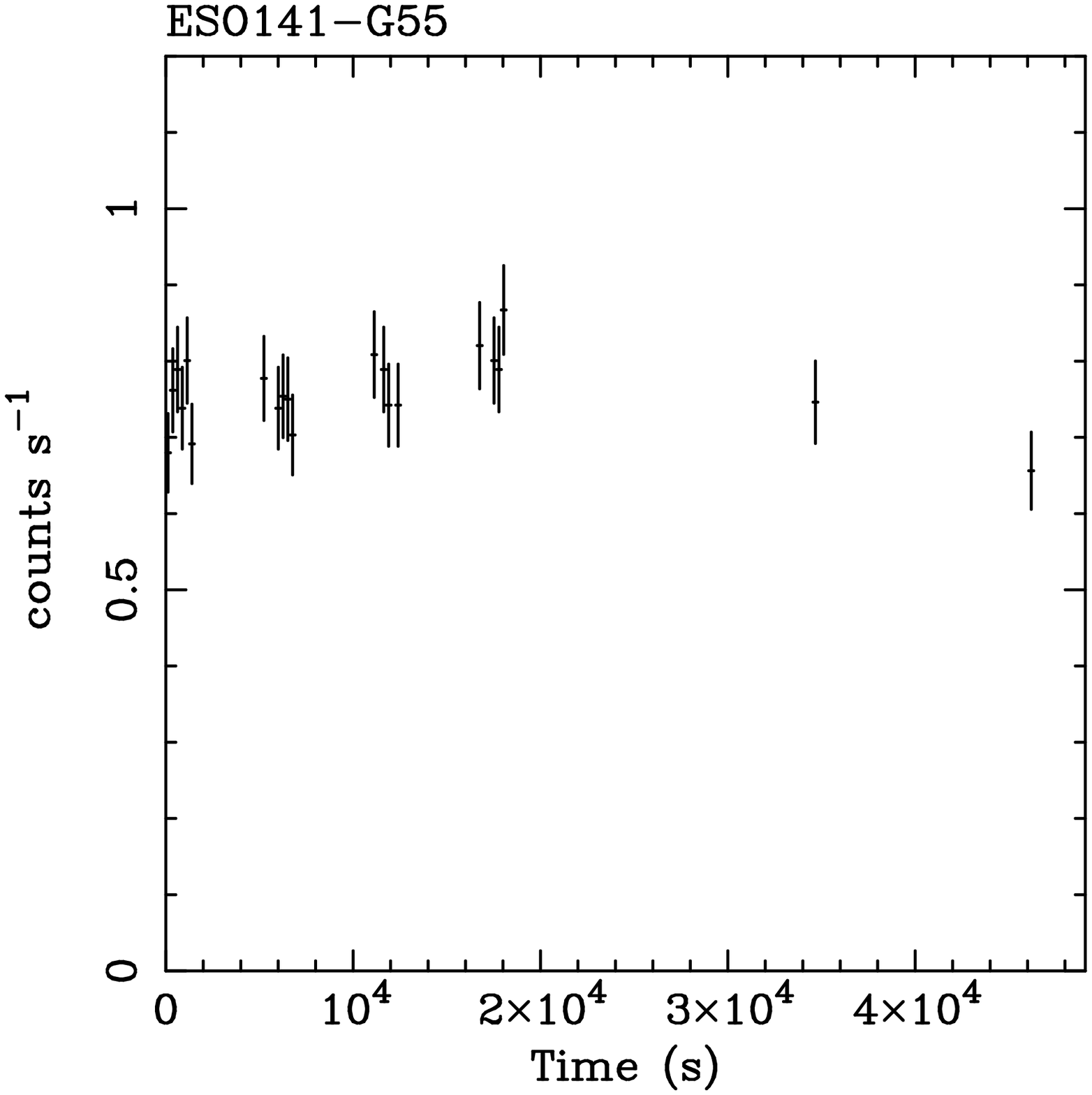,width=0.49\textwidth,height=0.23\textheight,angle=270}
\psfig{figure=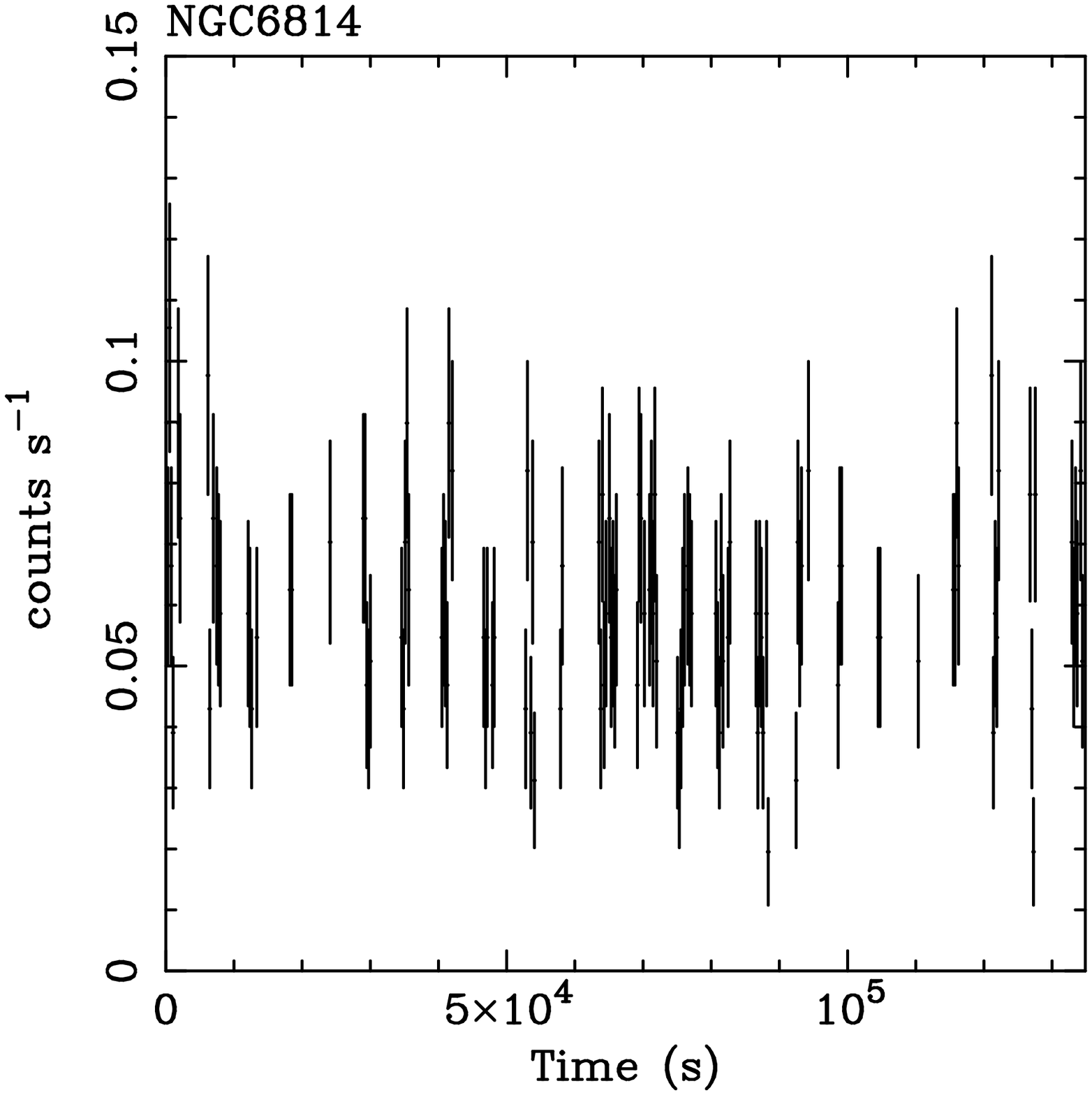,width=0.49\textwidth,height=0.23\textheight,angle=270}
}
\hbox{
\psfig{figure=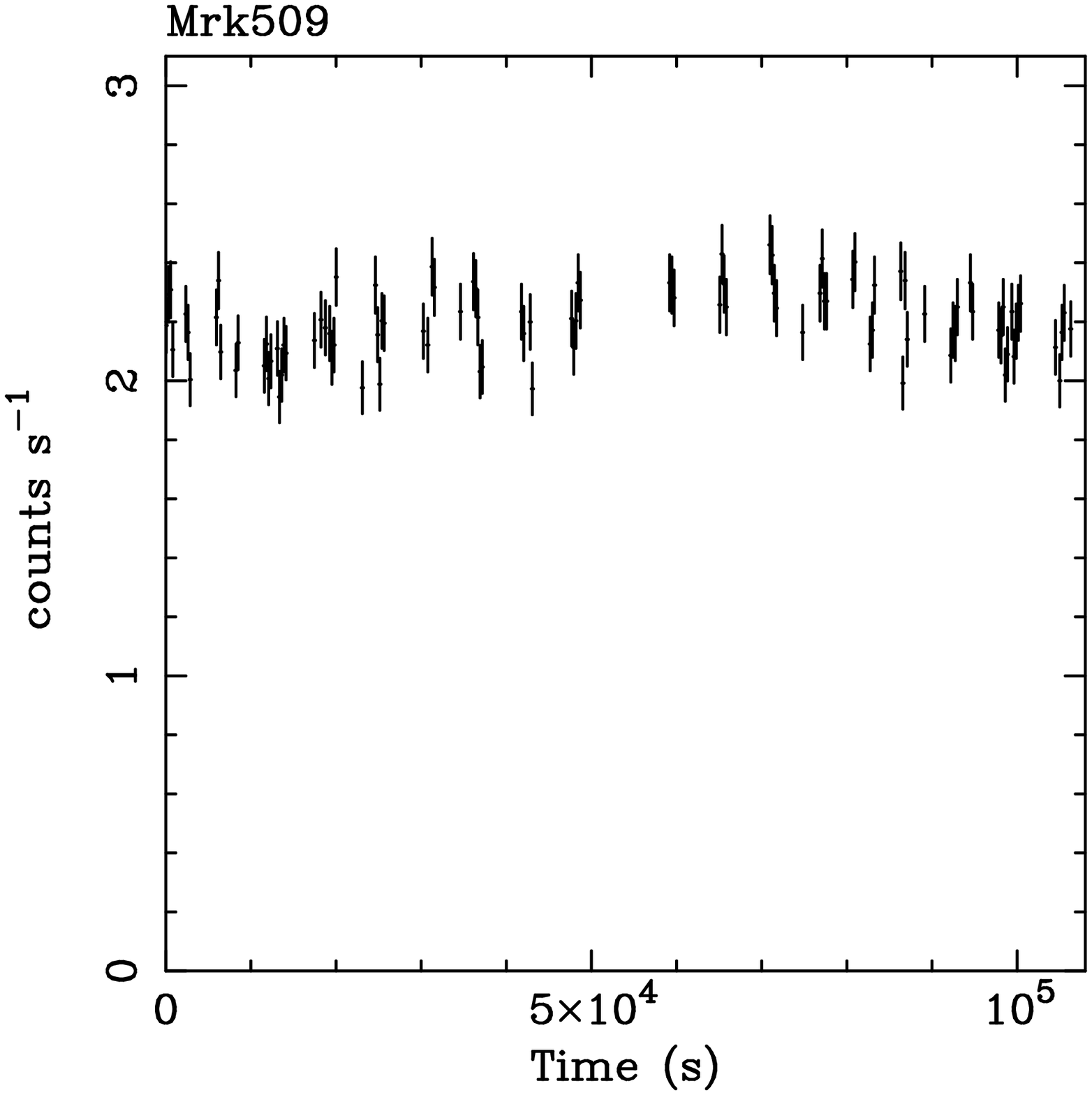,width=0.49\textwidth,height=0.23\textheight,angle=270}
\psfig{figure=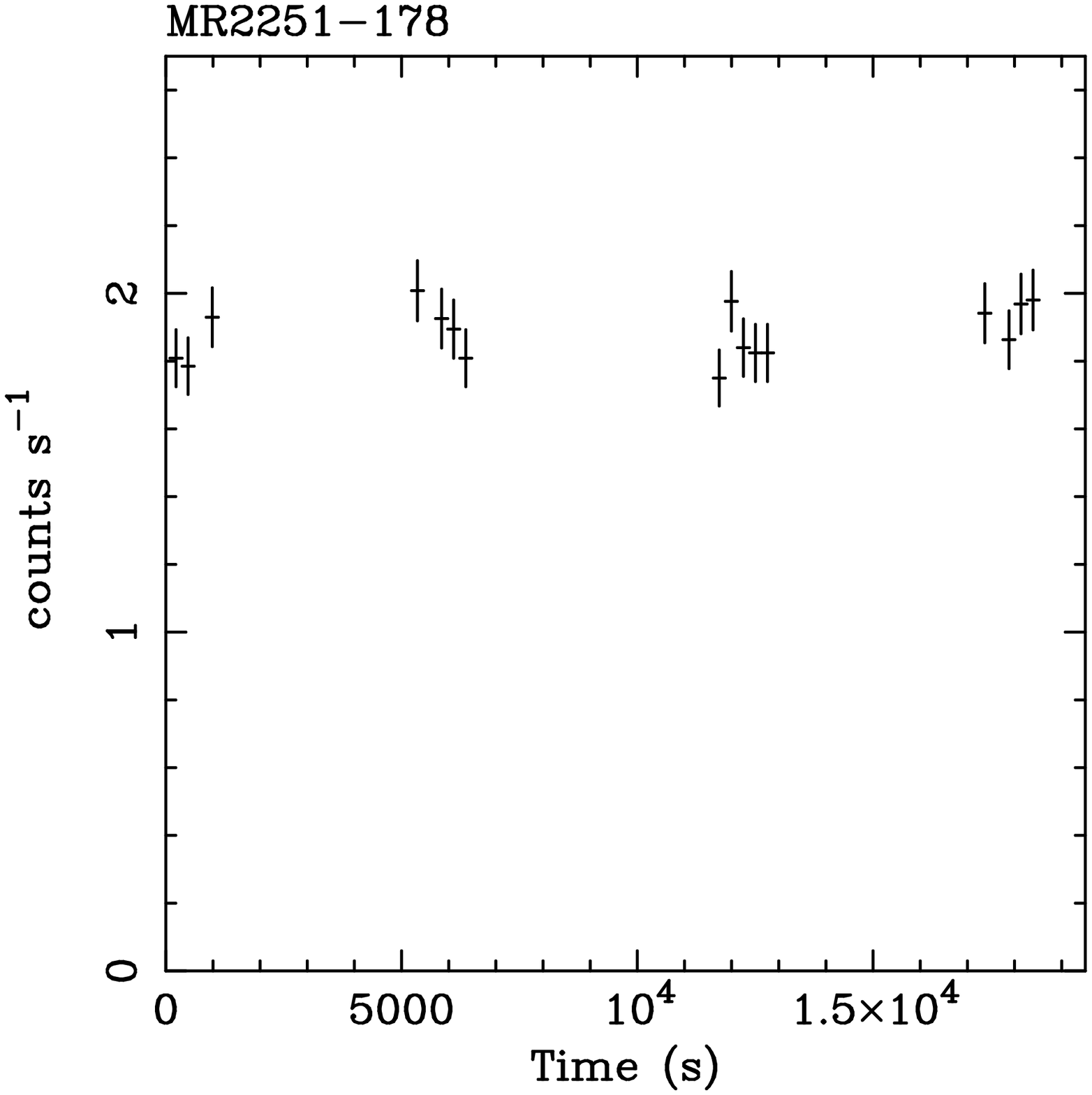,width=0.49\textwidth,height=0.23\textheight,angle=270}
}
\hbox{
\psfig{figure=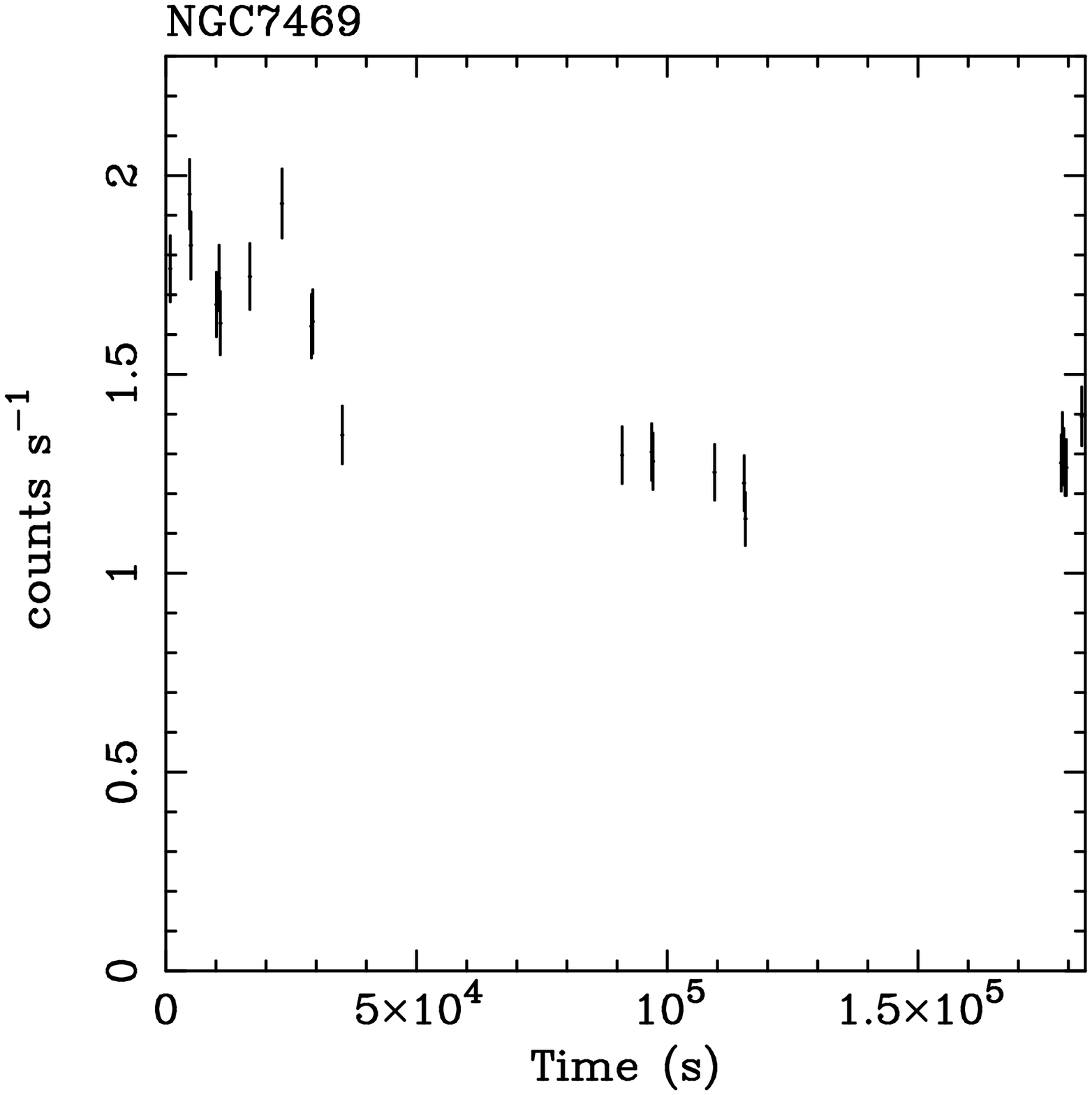,width=0.49\textwidth,height=0.23\textheight,angle=270}
\psfig{figure=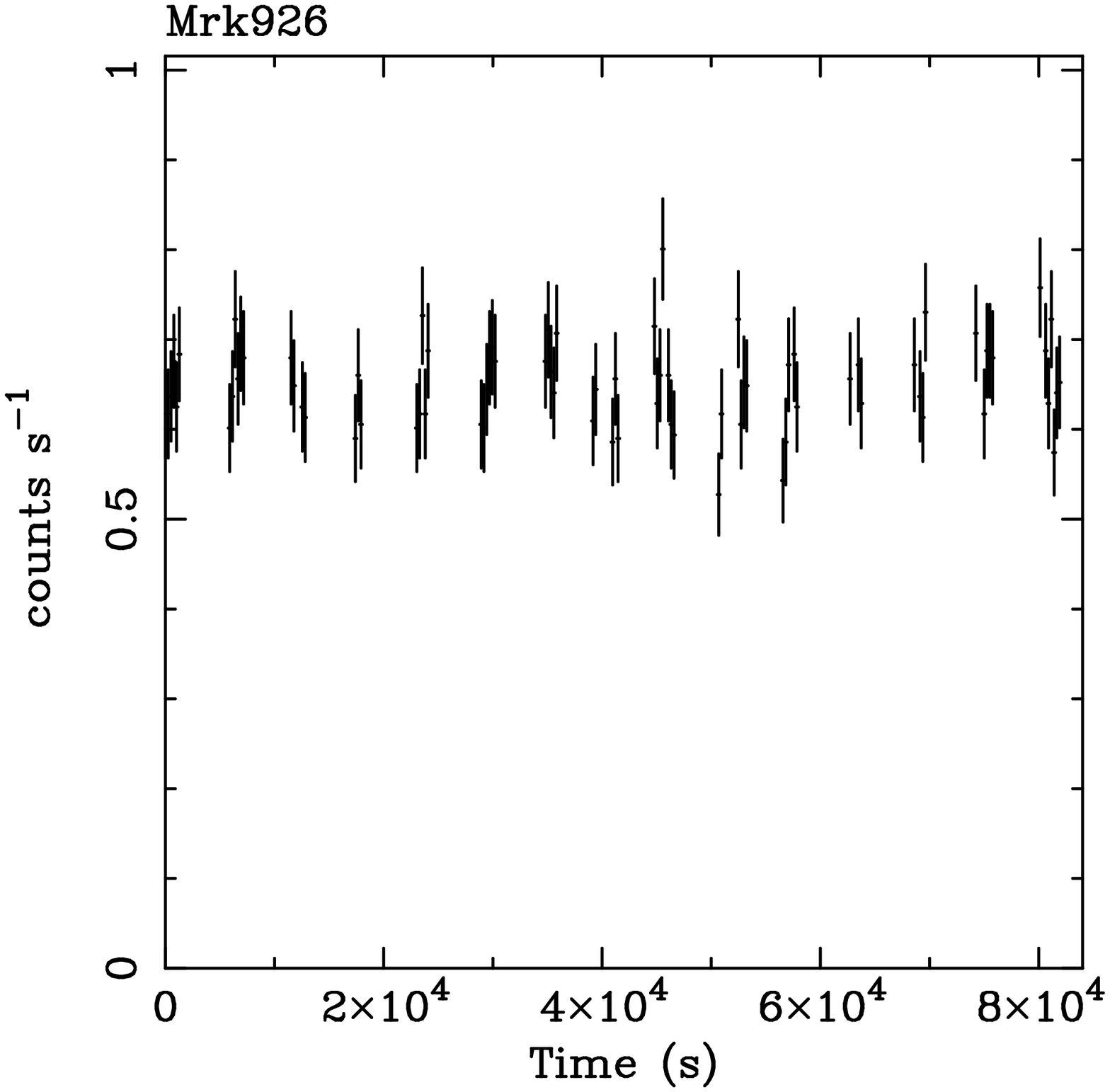,width=0.49\textwidth,height=0.23\textheight,angle=270}
}
\caption{cont.  Full-band SIS0 light curves for the 24 objects of this sample.  
Significant variability is detected in 16 out of 24 objects.  See the text for 
a detailed discussion.}
\end{figure*}

\subsection{Motivation}

The previous sections have considered the time-averaged spectra of
this AGN sample.  In this section, the variability properties are
briefly examined.  A detailed study of the continuum variability
properties is presented elsewhere (Nandra et al. 1996a).  Here I focus
on spectral variability that is relevant to a discussion of the warm
absorber.

The response of the photoionized absorber to changes in the primary
ionizing continuum provides an important probe of this plasma.  In a
simple one-zone model, the column density would be expected to remain
constant whilst the ionization paramater responds to changes in the
continuum flux on the recombination timescale [see Reynolds et
al. (1995) for a discussion of the relevant timescales.]  In practice,
the observed spectral variability in at least some objects tends to
disagree with this simple picture.  Under these circumstances, the
spectral variability is presumably providing information on the
structure of the complex absorbing regions.

\subsection{Rapid spectral variability}

A systematic search for rapid spectral variability (i.e.  within the
duration of a single observation) was performed using the hardness
ratio technique of Reynolds et al. (1995).  SIS0 light curves were
extracted for each object in the full {\it ASCA} band (0.4--10\,keV)
and two restricted energy bands: a soft band at 0.7--1.3\,keV (which
suffers from the effects of warm absorption) and a hard band at
2.5--5.0\,keV (representing unabsorbed continuum).  The hardness ratio
${\cal R}$ can then be defined as
\begin{equation}
{\cal R}=\frac{{\rm 2.5-5\,keV\ count\ rate}}{{\rm 0.7-1.3\,keV\ count\ rate}}
\end{equation}
For a fixed primary spectrum, changes in the warm absorber
parameters will lead to changes in the value of this hardness ratio.

Figure~8 displays the full band light curves for the sample objects.
Many of the objects show clear large amplitude variation during the
{\it ASCA} observation.  The null hypothesis of a constant count rate
is rejected at 90 per cent confidence (using a standard $\chi^2$ test;
Bevington 1969) in all except 8 of the sample sources (Fairall 9,
Mrk~1040, NGC~2992, 3C~382, 3C~390.3, ESO~141-G55, MR~2251-178 and
Mrk~926).  NGC~2992 and Mrk~1040 have small count rates (SIS0 count
rate is 0.082\,cts\,s$^{-1}$ and 0.12\,cts\,s$^{-1}$ respectively)
compared to all other objects in the sample except NGC~6814.  Thus,
only dramatic variability would be detectable during the observation
of these faint source.  The other sources for which primary
variability was not detected are all powerful sources with a
2--10\,keV luminosity, $L_{\rm 2-10}>10^{44}\ergps$.  In fact, only 3
out of 9 sources in the sample with $L_{\rm 2-10}>10^{44}\ergps$ show
statistical evidence for variability.  In contrast, 14 out of 15 less
powerful sources ($L_{\rm 2-10}<10^{44}\ergps$) show statistical
evidence for variability.  This is in agreement with the well
established luminosity-variability time scale relationship of AGN
(Green, McHardy \& Lehto 1993).

The hardness ratio ${\cal R}$ as a function of time along the light
curve was constructed in order to search for warm absorber
variability.  This study is severely hampered by photon statistics and
only the brightest and most variable sources might be expected to show
unambiguous short term variations of the hardness ratios.  Indeed, the
null hypothesis of a constant ${\cal R}$ could only be rejected with
90 per cent confidence in the case of 2 objects (NGC~3227 and
MCG$-6-30-15$).  I now discuss the variability of these two objects
individually.

\subsubsection{MCG$-6-30-15$}

Spectral variability of this object as observed by {\it ASCA} has been
previously reported by Fabian et al. (1994a), Reynolds et al. (1995),
Otani et al. (1996) and Iwasawa et al. (1996).  Otani et al. (using
the same dataset as that analysed here) find that the soft spectral
variability is characterised by an {\sc O\,vii} edge with a constant
depth and an {\sc O\,viii} edge whose depth anticorrelates with the
ionizing flux (which varies by a factor of 7 during the observation).
They show that this is {\it incompatible} with any one-zone
photoionization model and deduce that a two-zone (or multizone) model
is needed.

\subsubsection{NGC~3227}

Ptak et al. (1994) have performed a detailed analysis of the spectral
variability displayed by NGC~3227 during this {\it ASCA} observation.
The data are consistent with behaving in a similar manner to
MCG$-6-30-15$ (i.e.  constant {\sc O\,vii} edge and variable {\sc
O\,viii} edge).  However, it is difficult to reach unambiguous
conclusions due to poor photon statistics: in particular, the data
cannot rule out a variable primary power-law index modified by a
constant warm absorber.

\section{Notes on individual objects}

\subsection{Radio-Quiet sources}

\subsubsection{Mrk~335}

The Seyfert 1 galaxy Mrk~335 possesses a strong soft excess apparent
below 1\,keV which is readily observable both in the {\it ASCA} data
presented here and the BBXRT results of Turner et al. (1993b).  Turner
et al. suggest the presence of ionized absorption on the basis of a
joint {\it Ginga}/BBXRT analysis.  The absence of any obvious ionized
absorption in the present {\it ASCA} data suggests either that this
absorption is variable or that the supposed warm iron edge seen in the
{\it Ginga} data had some other origin.  One possibility is that the
sharp drop at $\sim$7--8\,keV that is characteristic of an iron
emission line from a relativistic disk may have been confused with K-shell
absorption from highly ionized iron in the {\it Ginga} data.

\subsubsection{Fairall-9}

\begin{figure}
\hspace{-1.5cm}
\hbox{
\psfig{figure=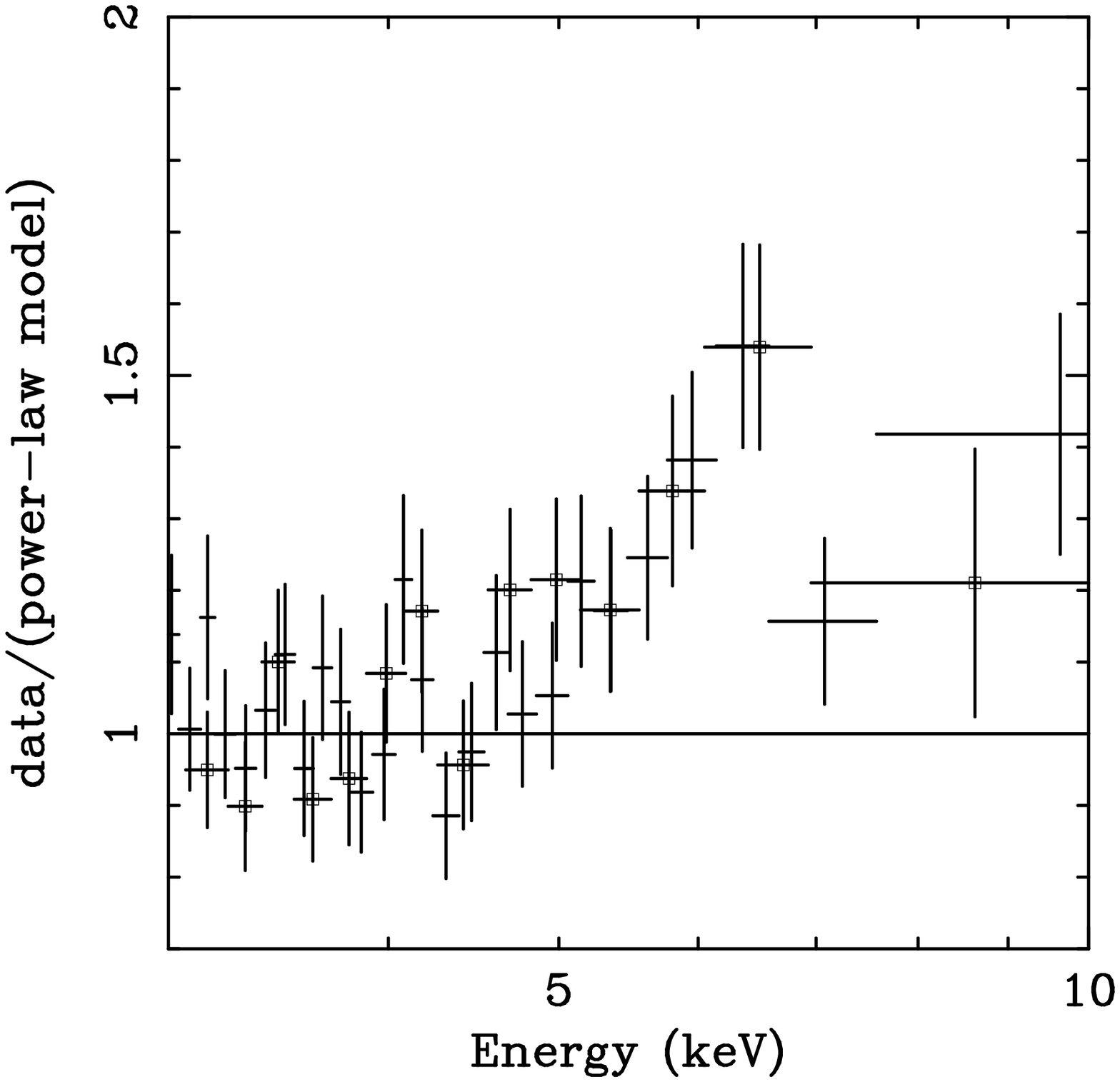,width=0.65\textwidth,angle=270}
}
\caption{Ratio of the hard-band {\it ASCA} data for Fairall-9 to the 
underlying primary continuum model (defined by fitting the data in the
1--4\,keV by a simple power-law spectrum).  The broad iron line can be
seen, as can a hard tail.}
\end{figure}

\noindent Fairall-9, a luminous Seyfert 1 galaxy, displays a strong 
and well detected broad fluorescent iron line (with $W_{\rm
Fe}=380^{+150}_{-100}\eV$) and a high-energy tail.  Figure~9
demonstrates these features which are well fitted by line emission
from a relativistic accretion disk (with inclination $27\pm
5\degmark$) and a strong Compton reflection continuum (Otani 1995).
The strength of the iron line can be plausibly explained as due to an
enhancement in iron abundance by a modest factor (George \& Fabian
1991; Reynolds, Fabian \& Inoue 1995).  Reverberation effects, most
likely associated with time-delayed reflection from the pututive
molecular torus, have to be invoked to reconcile the observed hard
tail with expectations from Compton reflection models.

\subsubsection{Mrk~1040}

The {\it ASCA} observation of Mrk~1040 has been presented previously
in Reynolds, Fabian \& Inoue (1995).  In the present work, we model
the soft complexity of this source as a combination of neutral
absorption and {\sc O\,vii} and {\sc O\,viii} edges in order to force
a comparison with the classical warm absorber sources.  However, due
to the small count rate and high Galactic line-of-sight absorption
($7.07\times 10^{20}\pcmsq$; Elvis, Lockman \& Wilkes 1989), the exact
nature of the soft complexity is unclear and several alternative
models are explored in Reynolds, Fabian \& Inoue (1995).  The
intrinsic neutral absorption is probably related to the ISM of the
host galaxy which possesses a large inclination (Amram et al. 1992).

\subsubsection{NGC~2992}

The 2--10\,keV flux of NGC~2992 has been found to have decreased by a
factor of $\sim 20$ during the past 16 years (Weaver et al. 1996).
During that time, its X-ray spectrum has changed from being that of a
typical Seyfert 1 galaxy to an extremely hard spectrum.  The {\it
ASCA} spectrum is well described by an absorbed power-law with photon
index $\Gamma=1.25$, very much harder than any other source in the
present sample, and an narrow (unresolved) neutral iron fluorescence
line with equivalent width $560\eV$.  Weaver et al. (1996) present a
detailed study of the {\it ASCA} spectrum and suggest this spectrum to
be dominated by a Compton reflection continuum that remains once the
intrinsically steep primary continuum has faded (preseumably due to
light travel time effects).  Significant neutral absorption and a
strong soft excess are also required to make this hypothesis
consistent with the {\it ASCA} data.  They identify the putative
molecular torus as a natural site for the origin of this time-delayed
reflection continuum.


\subsubsection{NGC~3227}

The rapid spectral variability of this object has been discussed in
Section 6.2.2.  As with MCG$-6-30-15$, such variability strongly
suggests a multi-zone warm absorber.

\subsubsection{NGC~3516}

Kriss et al. (1996a) present an analysis of a later {\it ASCA}
observation (1995 March 11/12) of the Seyfert 1 galaxy NGC~3516 than
that presented here (2 Apr 1994).  It is found that the source has a
similar luminosity in both epochs.  The inferred oxygen edge depths
are also consistent with not having changed between the two epochs.
Simultaneous observations with the Hopkins Ultraviolet Telescope (HUT)
were also performed (Kriss et al. 1996b).  Kriss et al. show that the
UV and X-ray data combined strongly argue for a multizone ionized
absorber containing a large range of ioniation states.  They also show
how detailed photoionization modelling of the 1995 {\it ASCA} data
hints at a multizone absorber.


\subsubsection{NGC~3783}

The Seyfert 1 galaxy NGC~3783 is one of the two objects displaying
deep, warm oxygen edges but little or no optical/UV reddening (the
other being NGC~3516).  The photoionized absorber has been discussed
in Section 5.1.  Alloin et al. (1995) have presented a snapshot of the
continuum emission from radio wavelengths through to $\gamma$-rays.
They show the existence of a classical big blue bump in the UV
(confirming the small UV extinction towards this source) and an IR
bump.  The IR bump is interpreted as originating from dust at
temperatures ranging from the sublimation temperature ($T\sim
1500\K$) down to relatively cool grains ($T\sim 200\K$).  These
observations support the hypothesis of the existence of dusty warm
(photoionized) plasma which, in this object, is out of the line of
sight to the central continuum source (see Sections 4.2.2 and 8.2.2).


\subsubsection{NGC~4051}

The soft X-ray properties of this Seyfert 1 galaxy are complex.  The
time-averaged soft {\it ASCA} PV spectrum (Mihara et al. 1994 and the
present work) can be explained as a combination of a primary
power-law, a black-body soft excess and the effects of ionized
absorption.  The present work completely agrees with that of Mihara et
al. (1994).  The black body soft excess has a best fitting temperature
$kT=0.12\pm 0.01\keV$ and a bolometric luminosity $L_{\rm B}=2.3\times
10^{41}\ergps$.

Complex spectral variability has been reported by Guainazzi et
al. (1996) during a later {\it ASCA} observation of NGC~4051.  These
authors find that in a two-edge parameterisation, both the best
fitting edge depth and edge energies increase when the source enters a
low state.  However, it is unclear how to interpret these results: a
complex interplay of a variable ionized absorber with a variable soft
excess could lead to misleading results when an oversimplified
spectral model is fitted.  An unidentified spectral feature at $\sim
0.9\keV$ may also affect conclusions about spectral variability.
Higher spectral resolution observations with future missions
(e.g. {\it ASTRO-E} and {\it AXAF}) will be invaluable in
understanding this complex source.

\subsubsection{MCG$-6-30-15$}

This objects has been discussed in the introduction and Sections 5.2
and 6.2.1.

\subsubsection{IC~4329a}

The Seyfert 1 galaxy IC~4329a displays a strong unmodelled high energy
excess after fitting with model-B.  Cappi et al. (1996) model this
hard excess as the onset of a very strong Compton reflection
continuum.  Time lag effects have to be invoked to obtain such a
reflection component.  As shown in Cappi et al., the iron line is weak
($W_{\rm Fe}\approx 90\pm 30\eV$) once the reflection continuum has
been included.  If this is the correct spectral model, such a weak
line requires an iron underabundance and/or an ionized disk.  This is
in contrast to many other well studied Seyfert galaxies.

\subsubsection{NGC~5548}

The oxygen edges of the warm absorber in the famous Seyfert 1 galaxy
NCG~5548 were first found by the {\it ROSAT} PSPC (Nandra et
al. 1993).  A single edge fit to the {\it ROSAT} data yields an edge
depth of $\tau\sim 0.4$, consistent with the values found here and
reported in Table~3.  The {\it ROSAT} data also finds a soft excess
component which, when modelled as a black body, has a best fitting
temperature of $kT=60^{+13}_{-20}\eV$.  This is rather softer than the
values found from the {\it ASCA} data ($kT=240^{+30}_{-20}\eV$) and
suggests either some spectral variability of the soft excess or a
complex soft excess.  For example, a blackbody soft excess with a hard
tail could reproduce these results due to the different bandpasses of
{\it ROSAT} and {\it ASCA}.  

The {\it ASCA} data presented here have been previously reported in
Fabian et al. (1994b).  These authors agree with the results obtained
here.

\subsubsection{MR~2251-178}

MR~2251-178 is the most luminous radio-quiet object of the current
sample.  It was the first object in which a warm absorber was
identified (Halpern 1984; Pan, Stewart \& Pounds 1990).

Historical spectral variability over the span of 6 {\it ASCA}
observations (all in AO-1) has been examined by Otani (1995).  He
finds no significant variation in the parameters of the warm absorber,
despite changes by almost a factor of two in the continuum X-ray flux.
All spectral variability can be attributed to changes in the continuum
photon index which is found to correlate with the flux (i.e. the
continuum gets steeper when the flux increases).  The warm absorber
parameters he derives are entirely consistent with those found here.

\subsubsection{NGC~7469}

These {\it ASCA} data on the Seyfert 1 galaxy NGC~7496 have been
previously presented in Guainazzi et al. (1994).  They found a weak
($W_{\rm Fe}\approx 120\eV$) narrow neutral iron emission line.  The
present work (which benefits from recently improved high-energy
calibration) also finds a very strong ($W_{\rm Fe}\sim 1\keV$) and
very broad ($\sigma\sim 1.2\keV$) emission feature at these energies,
although the narrow line can still be seen in the residuals of the
best fitting model-B.  The relativistic accretion disk emission line
model has difficulties explaining such a spectral feature.  A full
discussion of this will be presented in a future publication.

\subsection{Radio-Loud sources}

\subsubsection{3C~120}

As previously mentioned, the broad-line radio galaxy 3C~120 possesses
an extremely broad ($\sigma=1.5^{+0.6}_{-0.4}\keV$) and strong
($W_{\rm Fe}=960^{+520}_{-270}\eV$) spectral feature with a centroid
energy $E=6.43^{+0.23}_{-0.24}\keV$.  This is not consistent with
being iron fluorescence emission from the immediate vicinity of a
Schwarzschild black hole.  More investigation is required to determine
the origin of such features.

\subsubsection{3C~273}

The radio-loud quasar 3C~273 is the most luminous object of the current
sample by more than an order of magnitude.  The addition of a broad
Gaussian at energies characteristic of iron K$\alpha$ emission leads
to a change in the goodness of fit parameter by $\Delta\chi^2=32$ for
3 additional degrees of freedom (with best-fitting parameters reported
in Table~4).  The F-test shows this to be a significant improvement at
more than the 99 per cent level.  This is contrary to the result of
Yaqoob et al. (1994) who find no significant iron line emission.  This
discrepancy is understandable given small high-energy calibration
issues (associated with the X-ray mirror response) that affected the
early response matrices.  The analysis presented here should be more
reliable at detecting comparatively weak broad features at high
energies.

\subsubsection{3C~390.3}

A detailed analysis of these {\it ASCA} data on the broad line radio
galaxy 3C~390.3 has been reported by Eracleous, Halpern \& Livio
(1996).  These authors focus on the nature of the iron K$\alpha$
emission line.  In agreement with the present work, they show it to be
resolved and consistent with fluorescent emission from cold iron in
the accretion disk of this source.

\section{Discussion}

Prior to discussing some theoretical implications of this work, the
main observational results will be briefly reviewed.

\subsection{Summary of observational results}

\subsubsection{Ionized absorption}

Ionized (warm) absorption, characterised by the presence of {\sc
O\,vii} and {\sc O\,viii} K-shell absorption edges, is a common
feature in the soft X-ray spectra of this sample of objects.  The
inclusion of these absorption edges in the spectral fitting of these
objects (as in model-B) leads to a statistically significant
improvement in 12 out of the 24 objects.  Moreover, 9 of these objects
have a maximum optical depth greater than 0.2 (90 per cent confidence
level) for one or both of the absorption edges.  This is approximately
the level at which the ionized absorption is visually apparent in the
X-ray spectrum (e.g. see Fig.~1).  These results are robust to the
presence of a black-body soft excess.  They are also robust to any
currrently recognized SIS low-energy calibration issues.  

A striking observational result, and one which has received little
attention, is that many objects have ionized edge depths in the range
0.2--1.5.  There is no a priori reason for values so comparatively
close to unity unless some feedback mechanism related to the optical
depth regulates the ionized column density.

The present dataset is sufficiently large to examine general trends.
When the sample is split into high and low luminosity subsamples of
equal size, the low luminosity objects are found to be be more
susceptable to ionized absorption than the high luminosity objects.
However, the interpretation of this result is not straightforward
since all of the radio-loud objects fall into the high luminosity
subsample.  The data is also consistent with there being {\it no}
luminosity dependence and, instead, a dependence on the radio
properties of the objects (with radio-loud AGN having weak ionized
absorbers in comparison with radio-quiet AGN).

Optical/UV reddening is also related to the properties of the ionized
absorber.  At the current level of investigation, the trend is best
characterised by dividing the sample into three types: unreddened
objects with only small {\sc O\,vii} edges, reddened objects with
appreciable {\sc O\,vii} edges and unreddened objects with very large
{\sc O\,vii} edges.  The formal statistical significance of this
relation is shown in Section 4.2.2.  The {\sc O\,viii} edge may
weakly reflect this trend but with much less significance.

\subsubsection{Primary continuum}

Model-B seems to adequately describe the X-ray reprocessing phenomena
relevant in the {\it ASCA} band.  Therefore, the photon index $\Gamma$
derived from fitting with this model should represent the slope of the
primary non-thermal continuum of the central engine.  The distribution
of photon indices for the full sample has mean
$\langle\Gamma\rangle=1.81$ and dispersion $\sigma=0.21$.  No
significant difference between the photon index distributions of
radio-loud and radio-quiet objects is found, although the number of
radio-loud objects is too small for this to be addressed fully.

\subsubsection{Iron lines}

Most objects show spectral complexity at energies characteristic of
K$\alpha$ emission lines of iron (6--7\,keV).  A detailed treatment of
this emission in a similar sample of radio-quiet objects is presented
in Nandra et al. (1996b).

An interesting point to arise from the current work is a difference
between the high energy complexity in the radio-loud and radio-quiet
subsamples.  The radio-quiet objects generally display cold iron
fluorescent emission lines with profiles that agree well with the
hypothesis of originating from a relativistic accretion disk.  In
constrast, the radio-loud objects generally possess significantly
broader emission features which are poorly fit by standard
relativistic accretion disk models.  

\subsection{AGN models and ionized absorption}

Modern unified AGN models aim to explain the wide diversity of AGN
properties in terms of a few physical principles (see Antonucci 1993
for a recent review).  Whilst the radio-loud/radio-quiet dichotomy is
clearly a fundamental one, most other observable properties can be
explained as an orientation effect.

In the spirit of these unification schemes, I hypothesize that all
radio-quiet AGN have optically-thin photoionized plasma in the
vicinity of the central engine.  The frequency of occurence of
observed ionized absorbers would then suggest this material to cover a
fraction of the sky $f_{\rm c}\sim 0.5$ as seen by the primary X-ray
source.  As will be demonstrated below, the quantities of photoionized
plasma involved are significant as are the mass flow rates of this
plasma.  Thus, this material represents an important component of the
central engine environment.  Within such a scheme, any observed
luminosity dependence of the warm absorption features would correspond
to a luminosity dependence of the column density, ionization state
and/or covering fraction of warm material.

\subsubsection{Location of the warm absorber}

The best studied warm absorber, that in MCG$-6-30-15$, is strongly
believed to be comprised of at least two main zones on the basis of a
detailed study of its spectral variability (Otani et al. 1996).  Other
well studied objects also show evidence for a multizone absorber
(NGC~3227: Ptak et al. 1994, NGC~3516: Kriss et al. 1996a,b).  Thus,
it seems most likely that all objects have, to some extent, a
multizone absorber if viewed from the correct orientation.  In a
complete physical picture, these various absorbing regions will
represent different parts of some global flow pattern within the AGN.
However, for now I simplify the discussion into that of an `inner'
absorber and an `outer' absorber.

\subsubsection{The outer absorber}

In the case of MCG$-6-30-15$, the outer absorber is the less ionized
one and is responsible for much of the {\sc O\,vii} edge.
Photoionization models characterise this zone as having a column
density $N_{\rm W}=4.6\times 10^{21}\pcmsq$ and ionization parameter
$\xi=17.4\,{\rm erg}\,{\rm cm}\,{\rm s}^{-1}$ (Otani et al. 1996).
The constancy of this edge, coupled with recombination timescale
arguments, imply that this material is tenuous ($n<2\times
10^5\pcmcu$) and at radii $R>1\pc$.  These are radii characteristic of
the putative molecular torus, the Seyfert 2 scattering medium and
narrow line region (NLR).

This material is a significant component of the AGN environment.
Using a one-zone approximation to model the outer absorber, the mass
of this ionized plasma is
\begin{eqnarray}
M&=&4\pi R^2 N m_{\rm p} f_{\rm c}\\
&\approx &1000\,R_{\rm pc}^2 N_{22} f_{\rm c}\,\Msun
\end{eqnarray}
where $R=R_{\rm pc}\pc$ is the radial distance of the absorbing
region, $N=10^{22}N_{22}\pcmsq$ is the column density and $f_{\rm c}$
is the covering fraction.  Evaluating for the parameters relevant to
MCG$-6-30-15$ gives $M\approxgt 500\,f_{\rm c}\,\Msun$.  If the
central engine is accreting at more than a few per cent of the
Eddington limit, radiation pressure will dominate the dynamics of this
material and drive an outflow (Murray \& Chiang 1995; Reynolds \&
Fabian 1995).  Suppose the outflow has a velocity $v=10^3\,v_8\kmps$,
number density $n$ and volume filling factor $f_{\rm v}$ at radius
$R$.  Further, suppose the flow occurs over a solid angle $\Omega$.
Note that $\Omega/4\pi\approxgt f_{\rm c}$ since the outflow itself
may be cloudy.  The mass flow rate is then given by
\begin{eqnarray}
\Mdot_{\rm abs} &=& \Omega R^2 n m_{\rm p} v f_{\rm v}\\
&=& \Omega R N m_{\rm p} v\\
&\approx & 1 R_{\rm pc} N_{22} v_8 \left(\Omega/4\pi\right) \Msunpyr
\end{eqnarray}
Evaluating using the parameters of MCG$-6-30-15$ gives $\Mdot_{\rm
abs}\approxgt 0.5\,(\Omega/4\pi)v_8\Msunpyr$.  This mass flow rate
greatly exceeds the accretion rate within Seyfert galaxies (thought to
be $\Mdot_{\rm acc}\sim 0.01\Msunpyr$ assuming an accretion efficiency
of $\eta\sim 0.1$).

If the warm absorber is dusty, as suggested in the cases of
MCG$-6-30-15$ and IRAS\,13349$+$2438, the dust must reside in this
outer absorber: dust would rapidly sublime if present in the intensely
irradiated inner absorber.  The observed correspondence between the
depth of the {\sc O\,vii} edge and the optical reddening in the
present sample of objects is further circumstancial evidence for dust
coincident with the {\sc O\,vii} absorbing region.  The observed
amount of ionized absorption and optically reddening in radio-quiet
AGN would then depend solely on orientation.  Some lines-of-sight may
have no intervening ionized absorber or dust whereas other
lines-of-sight may have a dusty ionized absorber.  The two objects
displaying deep ionized absorption edges and no optical reddening
(NGC~3783 and NGC~3516) may be intrinsically unusual or suggest that
there are lines-of-sight with a large column of dust free ionized
absorption.

The presence of dust in the outer ionized absorber provides clues as
to its origin.  It is unlikely to originate by cooling from a hot
phase (at the Compton temperature of $T\sim 10^6$--$10^7\K$, say).
The hot phase would be completely dust free (any dust would be rapidly
sputtered) and dust formation from metals in the warm gas ($T\sim
10^5\K$) would be prevented since the gas temperature very much
exceeds the sublimation temperature; dust grains could never assemble
at such temperatures.  Therefore, the dusty outer absorber is likely
to originate via heating of previously cold, dusty material.  The
classical molecular torus of the standard AGN unification schemes
provides an obvious source for such cold dusty material.  Thus, the
outer absorber may represent a radiatively-driven dusty wind from the
torus.  The fact that the ionized edge depth is often observed to be
within an order of magnitude of unity may also be suggestive of the
importance of radiation pressure.

A detailed understanding of such outflows is intimately linked with a
physical understanding of molecular tori, a full discussion of which
is beyond the scope of the present paper.  However, some general
considerations hint at the ingredients of a complete model.  The
dichotomy between type-1 (broad line) and type-2 (narrow line) AGN
suggests the existence of a well defined obscuring torus.  The outer
ionized absorber may constitute material that forms an ionized skin on
the inner edge of this torus and is then driven along radial paths by
the action of radiation pressure.  The ionized dusty material would
possess a conical shell-like structure.  The opening angle of the flow
in the wall of this conical shell (and thus the solid angle subtended
by the flow) would depend on the depth of the ionized torus skin.  The
observations presented here find that $\sim 20$ per cent (4 out of 20)
radio-quiet type-1 AGN possess significant ionized absorption {\it
and} significant optical reddening.  Taking into account the solid
angle in which an observer would see a type-2 object ($2\pi$--$3\pi$
steradians; Antonucci 1993), the dusty ionized outflow covers $\sim
5$--10 per cent of the total solid angle as seen from the central
source.  This implies either a thick ionized torus skin ($\sim
0.1$--0.3\,pc thick in order to subtend this solid angle at the
primary source) or that the wind is non-radial.

Ionized material may be caused not to follow a radial path and thus be
streamed into the line of sight towards the nucleus via the action of
magnetic fields.  For example, in the centrifugally driven wind model
(Blandford \& Payne 1982) matter is driven off the surface of an
accretion disk and made to stream along a poloidal magnetic field by
the action of centrifugal forces and radiation pressure.  Any dust
present in the disk material will be carried along in this flow and,
beyond the sublimination radius, can survive to form a dusty obscuring
`torus' (Konigl \& Kartje 1994).  Observers at small inclination
(i.e. face-on) will be free from any obscuring matter and see bare AGN
emission.  As the inclination is increased, the line of sight may
intercept ionized material (corresponding to the outer warm absorber).
For still larger inclinations, dusty ionized material may obscure the
central engine.  Finally, there would be some critical inclination
beyond which a very large column of gas (both ionized and cold) and
dust blocks the central engine thereby producing a type-2 source.
Konigl \& Kartje (1994) suggest that the opening angle of the
obscuration-free cone increases with luminosity due to the action of
radiation pressure on material streaming along the field lines.  Such
a mechanism may provide a luminosity dependence to the ionized
absorption of the type suggested in Section 4.2.1.

Dusty warm plasma is the most likely the source of strong hot
infrared bumps seen in the spectra of many AGN.  IRAS~13349$+$2438,
probably the best known individual case of a dust warm absorber to
date (Brandt, Fabian \& Pounds 1996), has a prominent near IR bump
with grain temperatures ranging up to $\sim 500\K$ (Beichman et
al. 1986; Barvainis 1987; Wills et al. 1992).

\subsubsection{The inner absorber}

The inner absorber of MCG$-6-30-15$ possesses a higher ionization
parameter than the outer absorber and is responsible for much of the
{\sc O\,viii} edge seen when the source is in a low flux state.  A
one-zone approximation of this absorber yields parameters $N_{\rm
W}=1.3\times 10^{22}\pcmsq$ and $\xi=74\,{\rm erg}\,{\rm cm}\,{\rm
s}^{-1}$.  The short recombination timescale of the warm plasma leads
to the rapid spectral variability of this object and implies it to be
at distances characteristic of the BLR ($R<10^{17}\cm$).  The rapid
spectral variability seen in NGC~3227 (and possibly NGC~4051) suggests
that MCG$-6-30-15$ is not unique and such inner absorbers are common.

The presence of an optically-thin component within BLRs has been
suggested by detailed examination of the optical/UV broad line
properties.  Zheng \& O`Brien (1990) examined changes in the line
profiles and relative strengths of the Ly$\alpha$, {\sc C\,iv} and
Mg\,{\sc ii} broad lines in the Seyfert 1 galaxy Fairall-9.  They
argued that some of the Ly$\alpha$ emission and practically all of the
{\sc C\,iv} emission originates from optically-thin clouds in the
inner BLR which are heated by an anisotropic radiation field.
Ferland, Korista \& Peterson (1990) also investigated an
optically-thin inner BLR (the `very broad line region') and suggested
that it might account for the UV continuum of Seyfert galaxies.  These
authors linked such a component to warm absorption features.  More
recently, a detailed study of the (rest-frame) UV emission line
profiles of the radio-quiet quasar Q0207$-$398 reveal a highly-ionized
and outflowing optical-thin plasma within the BLR (Baldwin et
al. 1996).  The photoionization models examined by these authors show
a clear link between these optically-thin BLR clouds and ionized X-ray
absorption.

Using the expressions of Section 8.2.2 and assuming a distance of
$R\sim R_{\rm BLR}\sim 10^{17}\cm$ suggests that the mass of the inner
absorber is $M\sim 1\,f_{\rm c}\Msun$ for the parameters of Seyfert
galaxies.  This is comparable with the total mass of the classical
(optically-thick) broad emission line clouds (although this is a very
model dependent quantity\footnote{A lower limit on the mass of the
classical broad line clouds can be set by considering H$\beta$
emission.  For MCG$-$6-30-15, the luminosity in the broad H$\beta$
line is $L_{{\rm H}\beta}\approx 3\times 10^{40}\ergps$ (Pineda et
al. 1980).  If the line emitting material has volume $V$ and density
$n=10^9n_9\pcmcu$, we can write
\begin{equation}
L_{{\rm H}\beta}=V n^2 \alpha_{{\rm H}\beta} h \nu_{{\rm H}\beta}
\end{equation}
where $\alpha_{{\rm H}\beta}$ is the H$\beta$ effective recombination
coefficient ($\alpha_{{\rm H}\beta}=2.0\times 10^{-14}\cmcu\ps$;
Osterbrock 1989) and $\nu_{{\rm H}\beta}$ is the frequency of H$\beta$
line emission.  Rewriting this is terms of the mass $M_{\rm BLR}$ of
emitting material and evaluating for MCG$-$6-30-15 gives $M_{\rm
BLR}=0.2n_9^{-1}\Msun$}), suggesting that this material is an
important component of the BLR.

If the inner X-ray absorber is truly related to the BLR, an
understanding of its origin and physical state is intricately linked
to an understanding of the BLR generally.  Different BLR models would
lead to a dramatically different interpretation for the optically-thin
material (see discussion in Reynolds \& Fabian 1995).  A detailed
discussion of such models is beyond the scope of this paper.

\section{Conclusions}

The primary aim of this work is to study the X-ray spectral properties
of an unbiased sample of type-1 AGN using the unprecedented spectral
capability of {\it ASCA}.  The sample contains both radio-quiet
objects (18 Seyfert 1 galaxies and 2 RQQ) and radio-loud objects (3
BLRG and 1 RLQ) and thus comparisons can be drawn between these two
broad classes of source.  The main results and conclusions are listed
below.

(i) Spectral complexity, defined as deviations from the primary
power-law spectrum, is common within this sample.  In particular, at
least 12 out of 24 sources display absorption features due to
photoionized material (the so-called warm absorber) implying that the
covering fraction of the warm material is $f_{\rm c}\sim 0.5$.  In
addition, almost all sources show an excess at energies characteristic
of iron K$\alpha$ emission.

(ii) The effect of the warm absorber in the {\it ASCA} band is well
described by two K-shell absorption edges due to {\sc O\,vii} and {\sc
O\,viii} at 0.74\,keV and 0.87\,keV respectively.  Fully consistent
one-zone photoionization models show {\sc O\,vi} and Ne\,{\sc ix}
K-shell edges to be also observable, although the oversimplified
assumptions within these models (especially the assumption of a
one-zone absorber) leads to certain failures and limits the use of
such models.  Thus, I focus on the phenomenological two-edge
model of the warm absorber.  Unmodelled spectral features could lead
to false (unphysical) shifts of the edge threshold energies if these
energies were left as free parameters.  For example, an unmodelled
{\sc O\,vi} edge (at 0.68\,keV) could produce a small redshift of the
{\sc O\,vii} edge.  To avoid this, the edge energies have been fixed
at the physical values.

(iii) There is a trend for there to be less ionized absorption
(i.e. edge depths are smaller) in either more luminous objects and/or
radio-loud objects.  The present data cannot distinguish these two
possibilities because the radio-loud objects are all amongst the most
luminous sources in the sample.  It must be noted that there {\it are}
at least two luminous radio-loud objects known, 3C~212 (Mathur 1994)
and 3C~351 (Mathur et al. 1994), which display warm absorbers.  The
existence of this apparent exceptions highlights the need for further
study with larger samples.

(iv) Sources displaying significant optical reddening ($X>1$) also
display deep {\sc O\,vii} edges.  Coupled with other studies of
particular objects (MCG$-6-30-15$ and IRAS~13349$+$2438), this
supports the hypothesis that there is warm photoionized plasma
containing dust in thermal equilibrium with the primary radiation
field.  Within the context of unified AGN models, a radiatively driven
wind originating from the molecular torus is a plausible
identification of such this material.  The large covering fraction of
this outflow (as inferred from the frequency of occurence in this
sample) impies that either the torus has a thick ionized skin or some
mechanism (e.g. magnetic field) forces a non-radial flow.

(v) Optically thin clouds/filaments within the BLR are also likely to
be the source of some warm absorption features.  The presence of such
material has been suggested by recent detailed optical/UV emission
line studies as well as X-ray spectral variability.  This material is
likely to be an important component of the BLR.

(vi) The iron K$\alpha$ emission is thought to be due to fluorescence
within optically-thick cold material irradiated by the primary X-ray
continuum.  16 out of 20 radio-quiet objects show resolved broad lines
which are consistent with originating from the inner parts of an
accretion disk.  The corresponding spectral feature in radio-loud
objects is found to be significantly broader and poorly fit by
standard relativistic disk models.  An interesting possibility is that
the presence of a relativistic jet is affecting the nature of this
emission.

(vii) Once all reprocessing mechanisms have been modelled, the primary
continuum can be studied.  The photon index distribution has mean
$\langle\Gamma\rangle=1.81$ and standard deviation
$\sigma_{\Gamma}=0.21$.  There is no significant difference between
radio-quiet and radio-loud objects (although the number of radio-loud
objects is small).  

\section*{Acknowledgements}

I thank Andy Fabian and Kazushi Iwasawa for many useful discussions
throughout the course of this work.  The referee is thanked for a
careful reading of the original manuscript.  This research has made
use of the NASA/IPAC Extragalactic Database (NED) which is operated by
the Jet Propulsion Laboratory, California Institute of Technology,
under contract with the National Aeronautics and Space Administration.
This work has also made extensive use of data obtained through the
High Energy Astrophysics Science Archive Research Center (HEASARC)
Online Service, provided by the NASA-Goddard Space Flight Center.  I
acknowledge the Particle Physics and Astronomy Research Council
(PPARC) for support.

\end{document}